\documentclass[letterpaper,11pt,twoside]{article}

\usepackage[margin=1.3in]{geometry}

\usepackage[numbers,square,sort&compress]{natbib}

\usepackage{amsmath}
\usepackage{amssymb}
\usepackage{amsthm}

\usepackage[
  english 
  , ruled 
  , linesnumbered 
  , vlined 
]{algorithm2e}
\newcommand{\assign}{\ensuremath{\leftarrow}\xspace}
\usepackage{graphicx} 
\usepackage{xcolor,xspace}
\definecolor{RoyalBlue}{cmyk}{1, 0.50, 0, 0}
\definecolor{ForestGreen}{cmyk}{0.864, 0.0, 0.429, 0.396}
\definecolor{DarkGreen}{rgb}{0.076, 0.379, 0.306}
\definecolor{Brown}{cmyk}{0.0,0.692,0.925,0.529}
\definecolor{LightPurple}{rgb}{0.550,0.394,0.664}
\definecolor{DarkPurple}{rgb}{0.433,0.172,0.569}
\definecolor{Orange}{rgb}{0.93,0.478,0.121}
\definecolor{MyGreen}{cmyk}{0.95, 0.05, 0.95, 0.05}
\definecolor{MyPurple1}{rgb}{0.45,0.353,0.963}
\definecolor{MyPurple2}{rgb}{0.63,0.4,0.63}
\definecolor{MyYellow}{rgb}{0.901,0.547,0.0}
\definecolor{CambridgeRed}{rgb}{0.797, 0.000, 0.001}
\definecolor{DarkOrchid}{rgb}{0.6, 0.2, 0.8}
\definecolor{Purple}{rgb}{0.5, 0.0, 0.5}
\definecolor{cornellred}{rgb}{0.7, 0.11, 0.11}
\definecolor{eggplant}{rgb}{0.38, 0.25, 0.32}
\definecolor{eggshell}{rgb}{0.94, 0.92, 0.84}
\definecolor{mediumlavendermagenta}{rgb}{0.8, 0.6, 0.8}
\definecolor{mayablue}{rgb}{0.45, 0.76, 0.98}
\definecolor{mediumelectricblue}{rgb}{0.01, 0.31, 0.59}
\newif\ifcomments
\commentstrue
\newcommand{\Dnote}[1]{\ifcomments{\color{Orange}[Dim: #1]}\xspace\fi}

\usepackage[
   colorlinks%
   ,plainpages=false
   ,hypertexnames=true
   ,naturalnames
   ,hyperindex
   ,backref=page
   ,citecolor=MyGreen 
   ,urlcolor=blue
   ,pdfauthor={}
   ,pdftitle={}
   ,pdfsubject={}
   ,breaklinks=true
]{hyperref}
\usepackage{breakurl}   
\usepackage{amsmath}
\usepackage{amsthm}
\usepackage{nicefrac,psfrag}
\usepackage{upgreek,marvosym,subfigure}
\usepackage{thm-restate}
\usepackage{bm}
\usepackage{multirow}
\usepackage{nicefrac}
\usepackage{comment}
\usepackage{paralist}
\usepackage{upgreek}
\usepackage{booktabs}
\usepackage{pstricks}
\psset{unit=0.45cm}
\psset{linewidth=0.05}%
\psset{fillstyle=solid}%
\usepackage{amsmath}

\DeclareMathOperator*{\arginf}{arg\,inf}
\usepackage{enumitem}
\newcommand{\getsample}{\ensuremath{\underset{m}{\leftarrow}}\xspace}
\newcommand{\DD}{\ensuremath{D}\xspace} 
\newcommand{\DDset}{\ensuremath{\mathcal D}\xspace} 

\newcommand{\myprob}[1]{\ensuremath{\mathbb{P}\left( #1 \right)}\xspace}
\newcommand{\probsub}[2]{\ensuremath{\mathbb{P}_{#1}\left( #2 \right)}\xspace}
\newcommand{\expected}[1]{\ensuremath{\mathbb{E}\left[#1\right]}\xspace}
\newcommand{\expectedsub}[2]{\ensuremath{\mathbb{E}_{#1}\left[#2\right]}\xspace}

\newcommand{\naturalexp}[1]{\ensuremath{\mbox{\ensuremath{\mathrm{exp}}}\left(#1\right)}\xspace}

\providecommand{\norm}[1]{\left\lVert#1\right\rVert}
\newcommand{\abs}[1]{\left\lvert #1\right\rvert}
\newcommand{\heads}{\ensuremath{\mbox{\texttt{H}}}\xspace}
\newcommand{\tails}{\ensuremath{\mbox{\texttt{T}}}\xspace}
\newcommand{\set}[1]{\ensuremath{\left\{#1\right\}}}

\newcommand*{\boldone}{\text{\usefont{U}{bbold}{m}{n}1}}
\newcommand{\charfunc}[1]{\ensuremath{\boldone\left\{#1\right\}}\xspace}

\newcommand{\posex}{\ensuremath{1}\xspace}
\newcommand{\negex}{\ensuremath{0}\xspace}
\newcommand{\XX}{\ensuremath{\mathcal{X}}\xspace} 
\newcommand{\YY}{\ensuremath{\mathcal{Y}}\xspace} 
\newcommand{\Hh}{\ensuremath{\mathcal H}\xspace} 
\newcommand{\Cc}{\ensuremath{\mathcal C}\xspace}

\newcommand{\sample}{\ensuremath{S}\xspace}

\newcommand{\arisk}[2]{\ensuremath{R_{#1}\left(#2\right)}\xspace}
\newcommand{\aemprisk}[2]{\ensuremath{\widehat{R}_{#1}\left(#2\right)}\xspace}
\newcommand{\oracle}{\ensuremath{\textsl{EX}\left(c, \DD\right)}\xspace}
\newcommand{\genericfaultyoracle}{\ensuremath{\textsl{O}_{\textcolor{red}{\textsl{N}}}}\xspace}
\newcommand{\customoracle}[1]{\ensuremath{\textsl{EX}\left(c, #1\right)}\xspace}
\newcommand{\rcnoracle}{\ensuremath{\textsl{EX}_{\textsl{\textcolor{DarkOrchid}{RCN}}}^{\eta}\left(c, \DD\right)}\xspace}
\newcommand{\mcnoracle}{\ensuremath{\textsl{EX}_{\textsl{\textcolor{red}{MCN}}}^{\eta}\left(c, \DD\right)}\xspace}

\newcommand{\massoracle}{\ensuremath{\textsl{EX}_{\textsl{\textcolor{MyGreen}{Mass}}}^{\eta(\bm{x})}\left(c, \DD\right)}\xspace}
\newcommand{\algo}{\ensuremath{\mathcal A}\xspace}

\newcommand{\losszo}{\ensuremath{\ell^{\text{0-1}}}\xspace}
\newcommand{\OO}[1]{\ensuremath{\mathcal O\left(#1\right)}\xspace}

\newcommand{\Reals}{\ensuremath{\mathbb{R}}\xspace}
\newcommand{\RR}{\ensuremath{\Reals}\xspace}

\usepackage{pifont}
\newcommand{\xmark}{\ding{55}}%
\usepackage{fontawesome5}
\usepackage{skull}
\usepackage{txfonts} 
\usepackage{tikz}
\usetikzlibrary{arrows.meta,positioning,shapes,fit,calc}
\usepackage{forest}
\usepackage{booktabs}
\usepackage{tabularx}
\usepackage{array}
\usepackage{rotating}

\newcolumntype{Y}{>{\raggedright\arraybackslash}X}

\usepackage{multido}
\usepackage{float}

\newtheorem{theo}{Theorem}[section]

\newtheorem{prop}[theo]{Proposition}

\newtheorem{cla}[theo]{Claim}

\theoremstyle{definition}
\newtheorem{defi}[theo]{Definition}

\newtheorem{examp}[theo]{Example}
\newtheorem{rem}[theo]{Remark}

\def\emptyset{\varnothing}

\usepackage[most]{tcolorbox}
\tcolorboxenvironment{defi}{
  breakable, 
  colback=mayablue!15!white,
  colframe=mediumelectricblue!75!mediumelectricblue,
  fonttitle=\bfseries,
  boxrule=0.8mm,
  before upper={\parindent=\dimexpr\fontdimen6},
  sharp corners
}
\tcolorboxenvironment{theo}{
  breakable,
  colback=green!5!white,       
  colframe=green!55!black,     
  fonttitle=\bfseries,
  boxrule=0.8mm,
  before upper={\parindent=\dimexpr\fontdimen6},
  sharp corners
}
\tcolorboxenvironment{lem}{
  breakable,
  colback=green!5!white,       
  colframe=green!55!black,     
  fonttitle=\bfseries,
  boxrule=0.8mm,
  before upper={\parindent=\dimexpr\fontdimen6},
  sharp corners
}
\tcolorboxenvironment{prop}{
  breakable,
  colback=green!5!white,       
  colframe=green!55!black,     
  fonttitle=\bfseries,
  boxrule=0.8mm,
  before upper={\parindent=\dimexpr\fontdimen6},
  sharp corners
}
\tcolorboxenvironment{cor}{
  breakable,
  colback=green!5!white,       
  colframe=green!55!black,     
  fonttitle=\bfseries,
  boxrule=0.8mm,
  before upper={\parindent=\dimexpr\fontdimen6},
  sharp corners
}
\tcolorboxenvironment{examp}{
  breakable,
  colback=mediumlavendermagenta!10!white,      
  colframe=mediumlavendermagenta!60!DarkPurple,     
  fonttitle=\bfseries,
  boxrule=0.8mm,                
  before upper={\parindent=\dimexpr\fontdimen6},
  sharp corners
}
\tcolorboxenvironment{rem}{
  breakable,
  colback=eggshell!40!white,      
  colframe=eggshell!60!brown,     
  fonttitle=\bfseries,
  boxrule=0.8mm,                
  before upper={\parindent=\dimexpr\fontdimen6},
  sharp corners
}

\begin{document}

\title{Fundamentals for Binary Classification Under Label Noise\thanks{%
This is a preprint of the following chapter: Dimitrios I. Diochnos,
\emph{Fundamentals for Binary Classification Under Label Noise}, published in
\emph{Mathematics and Artificial Intelligence}, edited by Martin Charles Golumbic
and Frederick Hoffman, 2027, Springer. It is the version of the author's
manuscript prior to acceptance for publication and has not undergone editorial
and/or peer review on behalf of the Publisher (where applicable). The final
authenticated version is available online at:
\href{https://doi.org/10.1007/978-3-032-39161-2_3}{https://doi.org/10.1007/978-3-032-39161-2\_3}}}
\author{Dimitrios I.~Diochnos\\
The University of Oklahoma\\ 
Norman, OK, USA\\
\href{mailto:diochnos@ou.edu}{diochnos@ou.edu}}
\date{November 18, 2026}
%
%
\maketitle


\begin{abstract}
This chapter surveys fundamental results for binary classification under label noise.
Binary classification under label noise involves training models on datasets where class labels are partially incorrect, leading to degraded performance and misclassification. We distinguish poisoning attacks from noise processes and formalize random classification noise, Massart noise, and malicious classification noise within the probably approximately correct (PAC) framework. These models capture different adversarial strengths and data-generation realities.
We then highlight some core ideas and representative results. In particular, we explain why random classification noise preserves the expected ordering of hypotheses and thus supports learning using the empirical risk minimization (ERM) principle, how sample complexity depends on the noise rate, and why learning algorithms often require an upper bound on the corruption level, as well as how such a bound may be estimated from data in the case of random classification noise. We also discuss computational aspects of ERM under noise and present a canonical negative result technique---the method of induced distributions---within the malicious noise model where the adversary has more power and may affect not only the label of an example but also the instance. We conclude with pointers to related work and a taxonomy that organizes classical label-noise models using the NCAR/NAR/NNAR lens.
\end{abstract}

\newpage
\setcounter{tocdepth}{3}
\tableofcontents

\newpage

\section{Introduction}
This chapter deals with learning in the presence of noise, an important topic in decision-making under uncertainty. The simplest approach to machine learning assumes that the information that is provided to learning algorithms is honest.  However, in the real world this is hardly ever the case.
For this reason there is a need to design learning algorithms that are \emph{robust} and provide good machine learning solutions, even when the data are corrupted. 

Perhaps the most well-established paradigm within machine learning is that of \emph{supervised learning}, where the learning algorithm is given \emph{examples} of some underlying phenomenon that we study and is supposed to learn a useful approximation of the phenomenon that we study. 
These examples come in the form of \emph{instance-label} pairs and in the case of \emph{classification} the labels typically correspond to unordered categories.  As a simple illustrative example for classification, one may want to provide images of cats and dogs to a machine learning algorithm, where the label `cat' or `dog' expresses the fact as to whether a particular image (instance) corresponds to a cat or dog, respectively.
The other approach to supervised learning is \emph{regression}, where the labels are real numbers; e.g., they may correspond to weather temperature given some weather characteristics that we see in the instance.  Our discussion in this chapter focuses on classification tasks and in particular in the case where a label can be one of two possible categories; we refer to this framework as \emph{binary classification}.

More generally, labels may be corrupted by human error, ambiguity, faulty annotation pipelines, or sensor failures. 
Moreover, in higher-stakes settings, such as security, finance, and online platforms, training data may even be deliberately manipulated through adversarial or strategic behavior. As a result, learning using data that may not faithfully reproduce the reality of an underlying phenomenon is not an edge case but a central theme in the theory and practice of machine learning. Understanding how different types of data manipulation affect learnability, sample complexity, generalization, and algorithm design is therefore essential both for theoretical foundations and for reliable real-world deployment.

\emph{Poisoning}~\citep{Surveys:Poisoning:Tsipras,Surveys:Poisoning:DL,Surveys:Poisoning:WildPatterns} and \emph{evasion attacks}~\citep{AdvEgs:Survey,Surveys:AdvEgs:Opportunities,Surveys:AdvEgs:LongLists,AdvEgs:Intriguing} are mature topics at this point, with many important techniques and results within \emph{adversarial machine learning}~\citep{Book:AdvML1,Book:AdvML2,Book:DatasetShift,Book:Laird}, which in turn falls under the umbrella of \emph{trustworthy machine learning}~\citep{Book:Trustworthy}. However, a perhaps simpler approach within adversarial learning is learning in the presence of \emph{noise}. 
Over time several noise models have been introduced, such as 
random classification noise, class-conditional noise, instance-dependent noise, noise that depends on the margin, hybrid and partition-based noise frameworks, and fully adversarial noise frameworks such as malicious and nasty noise. 
Each noise framework captures a different data-generation reality and may lead to different statistical and computational guarantees.

%
In this chapter, marking the 35th anniversary of the International Symposium on Artificial Intelligence and Mathematics (ISAIM), we discuss topics and results on label noise that I consider fundamental for students interested in provable guarantees for machine learning algorithms when the data we have access to are corrupted by noise. 
The content is based on notes and slides that I use
when discussing noise in a computational learning theory course that I teach. 
I hope this material will help future students as well as others who may be interested in dealing with noise in machine learning.


\subsection{Scope}
This chapter is necessarily selective.  The focus is on binary classification under the $0$--$1$ loss, batch learning, and classical PAC-style statistical and computational guarantees. Multiclass classification under label noise, noisy regression, active and online learning in the presence of noise, and modern noise-mitigation methods for training deep neural networks with noisy labels are outside our main scope. Nevertheless, several of the fundamental distinctions and ideas discussed in this chapter extend, at least conceptually, to some of these broader settings.

\subsection{Roadmap}
The material covered is organized as follows.
Section~\ref{sec:background} provides notation, preliminaries, and background information on the probably approximately correct (PAC) model of learning~\citep{PAC:Valiant} when the learner has access to honest data labeled by a deterministic ground truth function.
In Section~\ref{sec:NoiseIntro} we separate poisoning attacks from noise, define a basic mechanism for noise models considered in this chapter, and discuss the \emph{method of induced distributions}, a popular technique for proving negative learnability results.
We illustrate this technique under \emph{arbitrary malicious noise}, which helps put into perspective the positive results that follow when corruption is restricted to labels.
Section~\ref{sec:classification-noise} defines formally label-noise models and presents the chapter's main results, assuming, apart from some initial remarks, that \emph{realizability} holds in the no-noise setting.
Under this assumption, Section~\ref{sec:RCN-ERM} examines the statistical complexity of empirical risk minimization under the random classification noise model of Angluin and Laird~\citep{PAC:Noise:Angluin}.
This result assumes that the learner knows an upper bound $\eta_b$ on the noise rate $\eta$; Section~\ref{sec:rcn-upper-bound} shows how such a bound can instead be estimated from a separate set of examples.
Section~\ref{sec:ERM-Noise-Computation} discusses the computational efficiency of ERM through one tractable and one computationally hard example.
In Section~\ref{sec:ERM-Beyond} we study the statistical complexity of ERM under malicious label noise and show that it is not harder than under random classification noise.
We conclude the section with a simple example showing that an adversary with a fairly modest \emph{poisoning} budget may cause ERM to produce a catastrophically bad hypothesis.
Section~\ref{sec:related} discusses related work, provides pointers for further study, and presents a taxonomy connecting theoretical noise models to three high-level categories commonly used in applied discussions of label noise.
Finally, Section~\ref{sec:conclusion} concludes the presentation and outlines directions for future work.

\section{Preliminaries and Background}\label{sec:background}
We start by formalizing notation.
%
Real numbers are denoted by \Reals and non-negative real numbers by $\Reals_{\ge 0}$.
Vectors are denoted with boldface; e.g., $\bm{a}\in\RR^n$ is a vector in $\RR^n$.
The transpose of a vector $\bm{a}$ is denoted as $\bm{a}^T$.
We use $\charfunc{\cdot}$ for an indicator function.
That is, 
$\charfunc{P}$ returns 1 when the predicate $P$ is true; otherwise 0.
For a random variable $X$, we denote its expected value by $\expected{X}$.
To make some natural-base exponents easier to read, we write $\naturalexp{z}$ to denote $e^z$.
Furthermore, we use $\ln(z)$ to denote the natural logarithm of $z$, while $\lg(z)$ is the logarithm of $z$ in base 2.

We use 
\XX to denote the \emph{set of instances} and \YY to denote the \emph{set of labels}. 
By default, an individual instance $\bm{x}\in\XX$ has $n$ different \emph{attributes} (or, \emph{features}) and is thus of the form $\bm{x} = (x_1, \ldots, x_n)^T$. 
Hence, in a collection of instances $\set{\bm{x_i}}_{i=1}^m$, $x_{ij}$ denotes the value of the $j$-th feature in the $i$-th instance. 
In the case of binary classification we have $\YY = \set{\negex, \posex}$, where $\negex$ (resp.~$\posex$) corresponds to the negative (resp.~positive) class. 
%
%
We use capital non-calligraphic letters for particular distributions (e.g., \DD), 
while we use capital calligraphic letters for sets; e.g., \DDset for a set of distributions. 

The simplest setup for \emph{distribution-free} PAC learning assumes that there is an unknown but fixed distribution \DD over the instance space \XX.  We indicate that an instance $\bm{x}$ is sampled from \XX according to some distribution \DD with $\bm{x}\sim\DD$.
A label $y = c(\bm{x})$ is assigned to such an instance by some deterministic ground truth function $c\in\Cc$, thus formulating a labeled instance $(\bm{x}, y)$, which we call \emph{example}.\footnote{Notice that the term is fitting as the pair $(\bm{x}, y)$ \emph{exemplifies} the input-output relationship of the unknown function $c$ that we are trying to learn.}
The class of functions \Cc from which the ground truth function $c$ is drawn, is called the \emph{concept class}.  
The function $h$ that we develop to approximate $c$ is called a \emph{hypothesis} and is one of the plausible functions that one can find in what we call a \emph{hypothesis space}, denoted by \Hh.

A tool that we will use for calculating concentration is Hoeffding's inequality, shown below.
\begin{prop}[Hoeffding's Inequality; \citep{Basics:Hoeffding}]\label{prop:Hoeffding}
Let $\sample = (X_{1},\cdots, X_{k})$ be a sequence of $k$ independent random variables over a distribution $\DD$ such that $X_{i}\in [\alpha,\beta]$ for all $i=1,\cdots, k$. Let 
$\bar{X}_{k}:=\frac{1}{k}\sum_{i=1}^{k}X_{i}$. For any $\upxi>0$, we have
\begin{displaymath}
\probsub{\sample\sim\DD^k}{\abs{\bar{X}_{k}-\expected{\bar{X}_{k}}}\geq \upxi}\leq 2\naturalexp{\frac{{-2k\upxi^{2}}}{(\beta-\alpha)^{2}}}\,.  
\end{displaymath}
\end{prop}

\subsection{Access to Oracles}\label{sec:prelim:oracles}
A convenient tool for formulating situations and simplifying arguments is to allow the learner to have access to some \emph{oracle} that returns examples that can be used for learning.
Normally such examples respect the ground truth $c$ and the underlying distribution \DD on \XX.  Furthermore, it is easy to define different noise models and ``hide'' the mechanics of these noise models inside the different oracles. In turn this will make it much easier for us to discuss situations where we have access to examples that have certain properties due to the oracle that we will be using; e.g., a particular noise model has tampered with the clean example, and we will not have to explain again how the particular example has been modified by noise as this will be clear by the specific oracle that is used in a particular context.

In the simplest possible form, we will be using the oracle \oracle to denote an oracle that returns clean data to the learner.  That is, requesting an example from the examples oracle \oracle implies the following steps:
\begin{enumerate}
    \item draw an instance $\bm{x}\sim\DD$,
    \item set $y = c(\bm{x})$, and
    \item return the example $(\bm{x}, y)$.
\end{enumerate}
Therefore, the oracle \oracle returns honest examples of the phenomenon of interest that we are trying to learn, in the sense that both the instances respect the underlying distribution \DD on \XX and moreover all the associated labels returned by the oracle can be trusted to be correct.

In situations where our data are corrupted by noise, we talk about \emph{faulty} oracles and we will use appropriate notation to indicate that later below. 
Regardless of the case, with $(\bm{x}, y)\sim O$ we denote an example sampled from an oracle $O$, which may be faulty or not.
Having said that, it is also convenient to introduce the notation $\sample \getsample O$ where one obtains a sample of size $\abs{\sample} = m$ from an oracle $O$ in a \emph{single call}.  This notation can be equivalent to $\sample \sim O^m$, where one obtains $m$ independent and identically distributed (i.i.d.)~examples, say through $m$ calls to the oracle $O$.  However, more importantly, the $\sample \getsample O$ notation is handy because it allows the sequence of $m$ examples that have been generated by $O$ for the formation of the sample \sample \emph{to depend on each other}. 
The notation thus allows the treatment of later examples by a faulty oracle to depend on earlier examples and corruptions.
We will see and clarify this issue further later on.
One could also use this notation for generating non-i.i.d.~honest examples; however, this idea is beyond our scope.

\subsection{Measuring the Quality of the Learned Functions}
%
Loss functions have the form $\ell \colon \Hh \times \XX \times \YY \rightarrow \Reals_{\ge 0}$ and allow us to argue about the quality of different hypotheses as shown 
below. 
By default we will be using the 0-1 loss, which is a very `natural' loss function for classification.  
The 0-1 loss function is defined to be $\losszo\left(h, \bm{x}, y\right) = \charfunc{h(\bm{x}) \neq y}$.  
%
\begin{defi}[Risk]\label{def:risk}
Let \DD be a distribution over \XX and $c\in\Cc$ be the concept that we are trying to learn.
Given an oracle \oracle and a hypothesis $h\in\Hh$, the \emph{risk} of $h$ is defined by 
\begin{displaymath}
\arisk{\oracle}{h} = \expectedsub{(\bm{x}, y)\sim\oracle}{\losszo\left(h, \bm{x}, y\right)}\,.
\end{displaymath}
\end{defi}

Note that in Definition~\ref{def:risk} we have used the oracle \oracle which returns honest examples; i.e., it holds that $y = c(\bm{x})$ as it was explained in Section~\ref{sec:prelim:oracles}.
However, if this oracle is faulty, instead of measuring the true risk of the function $h\in\Hh$, we are measuring the \emph{disagreement rate} that this function has with respect to the examples returned by the faulty oracle.  In other words, we have the following.
\begin{defi}[Disagreement Rate]\label{def:disagreement-rate}
Given a, potentially faulty, oracle $O$ and a hypothesis $h\in\Hh$, the \emph{disagreement rate} of $h$ with respect to $O$ is defined by 
\begin{displaymath}
\arisk{O}{h} = \expectedsub{(\bm{x}, y)\sim O}{\losszo\left(h, \bm{x}, y\right)}\,.
\end{displaymath}
\end{defi}

Therefore, if we know that we are dealing with a non-faulty oracle $O$, we will use the term of \emph{risk} of $h$ w.r.t.~$O$ to stress the fact that we are calculating how good the predictions of $h$ are with respect to the true function that is being learned as these examples are returned from an honest oracle $O$.
When we do not know if the oracle $O$ that we have access to returns examples that have been tampered by some form of noise, or because we may in fact know that examples are being drawn from a faulty oracle $O$, then we will be talking about the \emph{disagreement rate} that $h$ has with this oracle $O$.

\begin{defi}[Empirical Risk, Empirical Disagreement Rate]\label{def:emprisk}
Given a hypothesis $h\in\Hh$ 
and a sequence of examples 
$\sample = \left\{(\bm{x_i}, y_i)\right\}_{i=1}^m$, 
the \emph{empirical risk} (or, the \emph{empirical disagreement rate}) of $h$, is defined by
\begin{displaymath}
\aemprisk{\sample}{h} = \dfrac{1}{m}\cdot\sum_{i=1}^m \losszo\left(h, \bm{x_i}, y_i\right)\,.    
\end{displaymath}
\end{defi}
Notice that Definition~\ref{def:emprisk} does not involve an oracle, but rather looks at a sequence of examples \sample and calculates via $\losszo(\cdot)$ the average number of times that $h$ made an incorrect prediction according to \sample.

\begin{rem}\label{rem:non-iid-generation}
The traditional approach for generating a dataset \sample is by making $m$ calls to some oracle $O$ and this is denoted by $\sample\sim O^m$.  
However, as we explained near the end of Section~\ref{sec:prelim:oracles} when we study formally noise models, in the general case, it is convenient to consider the notation $\sample \getsample O$ that we introduced, so that we can allow dependencies on tampered examples as these examples are being generated by a faulty oracle in a contaminated sample returned by the oracle through a \emph{single call}.

In Definition~\ref{def:pac-label-noise} we will define formally the learning process and the goal for learning under some label noise models and this point will be clearer as the notation $\sample \getsample O$ will elucidate the fact that the hypothesis $h$ that is the output of some learning algorithm \algo has been obtained using information from a sample that has been generated with potential adversarial dependencies 
among its examples. 
\end{rem}

\begin{rem}
Should we be certain that we are measuring the goodness of $h$ w.r.t.~a set of examples \sample obtained i.i.d.~from an \emph{honest} oracle \oracle, we will try to use the term \emph{empirical risk} to make this relationship clearer.  Otherwise we will be using the term \emph{empirical disagreement rate}.    
\end{rem}


\subsection{Empirical Risk Minimization and the Goal for Learning}

The most important paradigm for learning is that of \emph{empirical risk minimization (ERM)}; that is, a learning algorithm reduces to a \emph{search problem} among hypotheses in \Hh, so that it identifies one that minimizes the empirical risk on a given dataset. 
In other words, given a sequence of $m$ examples \sample, an ERM solution is obtained when the learning algorithm returns a hypothesis $h^\star\in\Hh$ such that $h^\star\in\arginf_{h\in\Hh}\aemprisk{S}{h}$.

The hope is that, using a sufficiently large sample \sample, an ERM solution $h^\star\in\Hh$ should be close to the best possible $h^{\varheartsuit}\in\Hh$ that one can \emph{truly} achieve among all the possible hypotheses in $\Hh$.
The idea relies on the following two components:
\begin{enumerate}[label=(\arabic*)] 
    \item On one hand, given a sample \sample 
    obtained from an oracle $O$ under conditions that permit appropriate concentration (e.g., obtained i.i.d.), the empirical disagreement rate $\aemprisk{\sample}{h}$ is used as a proxy for the true (expected) disagreement rate $\arisk{O}{h}$ that the hypothesis $h$ has against examples obtained from $O$.

    \item On the other hand, the hypothesis $h^\star$ obtained by the application of the ERM principle will not behave much worse compared to the best possible $h\in\Hh$, denoted as $h^\varheartsuit$, since the sample \sample was sufficiently large and thus allowed the learning algorithm to make a fairly good decision on the selection of $h^{\star}$.
\end{enumerate}

For a moment, sticking to the idealized situation where the examples oracle $O$ returns honest examples; i.e., it holds that $O = \oracle$, the learning algorithm wants to find, with probability at least $1-\delta$, a hypothesis $h^\star$ that satisfies:
\begin{equation}\label{eq:agnostic-pac-learning}
\arisk{\oracle}{h^{\star}} 
< 
\arisk{\oracle}{h^{\varheartsuit}} 
+ 
\varepsilon\,.
\end{equation}

Hence, $h^\star$ is a \emph{probably approximately correct (PAC)}~\citep{PAC:Valiant} hypothesis. 
In other words, when working in PAC learning, the goal of the learner is to identify a hypothesis $h^\star\in\Hh$ that is not much worse compared to $h^{\varheartsuit}$, where $h^{\varheartsuit}\in\Hh$ 
denotes a hypothesis having the lowest possible \emph{true risk} $\arisk{\oracle}{h}$ among all hypotheses $h\in\Hh$.
Also, quite often the goal of the learning algorithm stated in (\ref{eq:agnostic-pac-learning}) may involve a loose inequality rather than the strict inequality that we chose to use.  There is not much of a difference in the results in these cases and people may use whatever is more mathematically convenient for the results that they are trying to present.  We use a strict inequality in (\ref{eq:agnostic-pac-learning}) since all the results that we discuss can satisfy this slightly stricter criterion and make the discussion in at least one case a little bit simpler.
Regardless, we will simplify further and formalize this discussion soon.

\subsubsection{Representation}
There are two requirements that are assumed in our discussion.
First,  
we assume that there is some \emph{reasonable} representation for every hypothesis $h\in\Hh$ that on one hand does not take up too much space and on the other hand allows the calculation of $h(\bm{x})$ efficiently.
More concretely, the space required to represent a hypothesis $h$ (respectively, a concept $c$) is denoted by $\mathrm{size}(h)$ (respectively, $\mathrm{size}(c)$).
Regarding the second requirement, we say that a hypothesis space is \emph{polynomially evaluatable} if the value $h(\bm{x})$ can be computed in time polynomial in the dimension $n$ and $\mathrm{size}(h)$.

\begin{examp}
As a simple example 
consider the class of 
disjunctions over $n$ Boolean variables (i.e., no negated variables allowed). There are $\abs{\Hh} = 2^n$ such functions and one simple representation would be a Boolean row vector of size $n$ where if the $i$-th entry is a 1 this means that the variable $x_i$ is included in the representation, otherwise it is not.  As an example, the function $h = x_2 \vee x_5$ can be represented as $h = (0, 1, 0, 0, 1, 0, 0, 0)$ in a situation where the dimension of the instances is $n=8$.  Moreover, notice that evaluating any $h\in\Hh$ on some arbitrary input $\bm{x}\in\set{0, 1}^8$ also takes polynomial time in $n$.  In fact, a single pass over $\bm{x}$ and over the vector representation of $h$ simultaneously, allows for the calculation of the output $h(\bm{x})$.
\end{examp}

\subsubsection{Proper Learning, Improper Learning, and Realizability}

Oftentimes we may know the concept class \Cc of functions from where the ground truth function $c$ is drawn.
The knowledge of \Cc typically 
allows us to 
select a hypothesis space \Hh such that some useful relationship between \Hh and \Cc holds. 
In particular, we have \emph{proper learning} when the hypothesis class contains only functions from the concept class; i.e.,~$\Hh \subseteq \Cc$.
The most common situation for proper learning is when $\Hh = \Cc$, but $\varepsilon$-covers as discussed in Section~\ref{sec:distribution-specific} provide the characteristic example of situations where proper learning can occur and it is not the case that all functions in \Cc can be found in the output of some learning algorithm \algo.
When the learning algorithm is not restricted to output a function that appears in the concept class \Cc, then 
we have \emph{improper} or \emph{representation-independent} learning. 
%

\smallskip

Furthermore, 
for a fixed target concept $c\in\Cc$ and a fixed distribution \DD over \XX, we say that learning occurs in the \emph{realizable case} when there exists some $h_{c, \DD}\in\Hh$ for which it holds $\arisk{\oracle}{h_{c, \DD}} = 0$; see, e.g.,~\citep{Book:Understanding}.
This assumption implies that with probability~1 over any random sample $\sample\sim\left(\oracle\right)^m$ we have $\aemprisk{\sample}{h_{c, \DD}} = 0$.
Notice however that realizability for a fixed pair $(c, \DD)$ does not necessarily imply $c\in\Hh$.
Indeed, a hypothesis may agree with the concept $c$ everywhere in the support of \DD, while differing outside of the support of \DD. Put it differently, the hypothesis $h_{c, \DD}\in\Hh$ that achieves realizability \emph{may depend} on both $c$ and \DD. 
Conceptually, learning a particular concept class $\Cc$ in the realizable case under a particular family of distributions $\DDset$ corresponds to requiring:
\begin{equation}\label{eq:realizability:dist-specific}
(\forall c\in\Cc)(\forall \DD\in\DDset)(\exists h_{c, \DD}\in\Hh)\left[\arisk{\oracle}{h_{c, \DD}} = 0\right]\,.
\end{equation}
While the formulation in~(\ref{eq:realizability:dist-specific}) allows $\Hh$ to be designed for a particular concept class $\Cc$ and family of distributions $\DDset$, the resulting hypothesis space remains fixed, whereas the zero-risk witness $h_{c,D}\in\Hh$ may depend on the particular pair $(c,D)$.
In this book chapter, however, we adopt 
a simpler but stronger distribution-independent convention for learning in the realizable case as explained below.

We are interested in studying and providing good learning guarantees for all possible concepts $c$ from some concept class $\Cc$ and for  (unrestricted) distribution-free learning we want to do so regardless of the underlying distribution \DD that may govern the instance-generating process.
Hence, in this book chapter, 
we will say that realizability holds whenever the hypothesis space \Hh is at least as expressive as the concept class \Cc, in the sense that 
for every $c\in\Cc$ there exists at least one $h_c\in\Hh$ such that regardless of the underlying distribution \DD, it holds $\arisk{\oracle}{h_c} = 0$. 
In other words, the following relationship holds:
\begin{equation}\label{eq:realizability:dist-free}
(\forall c\in\Cc)(\exists h_c\in\Hh)(\forall \DD \text{ over } \XX)\left[\arisk{\oracle}{h_c} = 0\right]\,.
\end{equation}
Notice that due to the order of quantifiers in (\ref{eq:realizability:dist-free}) the hypothesis $h_c\in\Hh$ may now depend \emph{only on} $c\in\Cc$ and \emph{not on} the particular distribution \DD.
A simpler statement for this relationship is to exclude \DD from the definition of realizability as follows:
\begin{equation}\label{eq:realizability:instance-space}
(\forall c\in\Cc)(\exists h_c\in\Hh)(\forall \bm{x}\in\XX)\left[h_c(\bm{x}) = c(\bm{x})\right]\,.
\end{equation}
The equivalence between~(\ref{eq:realizability:dist-free}) and~(\ref{eq:realizability:instance-space}) follows because unrestricted distribution-free learning includes distributions with support only on a few points; e.g., point-mass distributions: if a particular hypothesis $h_c\in\Hh$ satisfies (\ref{eq:realizability:dist-free}) for a particular concept $c$ to be learned and this is true for all possible distributions over \XX, then that $h_c$ cannot disagree with $c$ at some point $\bm{x}\in\XX$.
Put differently, if $h_c$ and $c$ disagreed at some $\bm{x}\in\XX$, then the distribution concentrated on $\bm{x}$ would give $h_c$ positive risk contradicting the definition in~(\ref{eq:realizability:dist-free}).
We can further simplify (\ref{eq:realizability:instance-space}) to the following definition that we adopt in this chapter.
\begin{defi}[Realizability Assumption]\label{def:realizability}
Let \Cc be a concept class and \Hh a hypothesis space.  We say that learning occurs in the realizable case if $\Cc \subseteq \Hh$ holds.
\end{defi}
%
Definition~\ref{def:realizability}
includes the typical proper learning variant where $\Hh = \Cc$, as well as situations where one works with a strictly richer set of functions \Hh compared to \Cc.
Notably, when realizability holds, then $\arisk{\oracle}{h^{\varheartsuit}} = 0$ in (\ref{eq:agnostic-pac-learning}) 
for the simple reason that one can use $h^{\varheartsuit} = c$.
Therefore, the PAC criterion reduces to the simpler version of identifying an $h^{\star}\in\Hh$ such that 
$\arisk{\oracle}{h^{\star}} 
<
\varepsilon$. 

\subsubsection{Working with Non-Trivial Concept Classes}\label{sec:distinct}
Unless stated otherwise, for the discussion and results that follow we are interested in learning non-trivial concept classes.  This means that such concept classes have at least some level of basic richness regarding the functions that they contain.  In particular, a concept class \Cc is called \emph{distinct} if there are concepts $c_1, c_2 \in \Cc$ and instances $\bm{u}, \bm{v}$ in the domain of the concepts $c_1$ and $c_2$, such that $c_1(\bm{u}) = c_2(\bm{u})$ and
$c_1(\bm{v}) \neq c_2(\bm{v})$~\citep{PAC:Noise:Malicious,PAC:Noise:FourTypes:Sloan}.

When one restricts the attention to the domain $\XX = \set{0, 1}^n$ then 
a concept class \Cc is called 
\emph{instance distinct}~\citep{PAC:Noise:FourTypes:Sloan} 
if there are concepts $c_1, c_2\in\Cc$ and an instance $\bm{u}\in\XX$ such that $c_1(\bm{u}) \neq c_2(\bm{u})$ and for all $\bm{v}\neq\bm{u}$ it holds $c_1(\bm{v}) = c_2(\bm{v})$.
In other words, a concept class \Cc is called instance distinct if there exist two concepts that disagree on a single instance in their domain. 

\subsubsection{Probably Approximately Correct (PAC) Learning}
We can now define formally the goal for learning when having access to honest examples.

\begin{defi}[PAC Learning]\label{def:pac-learning}
Let \Hh be a hypothesis space and \Cc be a concept class from where the ground truth function $c\in\Cc$ is drawn.
We say that \algo is a \emph{probably approximately correct (PAC) learning algorithm} 
if there exists a polynomial function $\mbox{poly}(\cdot, \cdot, \cdot, \cdot)$
that satisfies the following:
for every distribution \DD over \XX, for every $c\in\Cc$, for every $0 < \varepsilon, \delta < \nicefrac{1}{2}$, if \algo is given access to the inputs $n, \varepsilon, \delta$, and $\text{size}(c)$, and an (idealized) oracle \oracle, 
then \algo, 
using at most $\mbox{poly}(1/\varepsilon, 1/\delta, n, \text{size}(c))$ examples from the oracle,
outputs an $h^\star\in\Hh$ such that with probability at least $1-\delta$ satisfies
\begin{equation}\label{eq:pac-learning}
\arisk{\oracle}{h^{\star}} 
<
\varepsilon\,.
\end{equation}
The probability is over the random examples that are drawn from \oracle and over any possible randomization that \algo may have.
Furthermore, algorithm \algo is an \emph{efficient} PAC-learning algorithm if \algo runs in $\mbox{poly}(1/\varepsilon, 1/\delta, n, \text{size}(c))$ time.
\end{defi}

Note that the oracle that is used for generating the training examples for \algo is the same as the one that is used for testing the hypothesis $h$ that \algo learns from this data. This is a typical assumption in machine learning in that the distribution \DD between training and deployment of a machine learning model does not change.  Furthermore, this assumption is usually crucial for establishing rigorous results that satisfy Definition~\ref{def:pac-learning}.

\paragraph{Agnostic Learning.}

As another minor remark here, one can have a more general definition compared to Definition~\ref{def:pac-learning} in the sense that one does not need to know \Cc and moreover the error criterion is along the lines of (\ref{eq:agnostic-pac-learning}) rather than (\ref{eq:pac-learning}).  In such situations the distribution \DD is usually defined over $\XX\times\YY$ thus encapsulating the ground truth function entirely in \DD.  
Notice also that in this situation the ground truth can be randomized and return different labels for the same instance, or this can be a way of modeling noise along the labels.
Moreover, the learning algorithm is trying to do the best possible using hypotheses in \Hh without any knowledge about \Cc and thus the whole learning process takes place in the \emph{agnostic} model of learning.
We will not follow this route.  In our case \DD will be defined over \XX and a deterministic concept $c$ will provide the label associated with each particular input $\bm{x}$.  The implications of noise will be well defined but hidden inside the mechanism that governs a faulty oracle that adheres to a particular noise model.  
We will deal with a variation of Definition~\ref{def:pac-learning} 
and accommodate three noise models appropriately, later in Definition~\ref{def:pac-label-noise}.


\subsection{Distribution-Independent and Distribution-Specific Learning}\label{sec:distribution-specific}
Definition~\ref{def:pac-learning} refers to the \emph{distribution-free} (or, \emph{distribution-independent}) model of learning.  This means that the learning algorithm will produce a hypothesis that satisfies the PAC criterion regardless of the distribution \DD that is governing \XX. 
In fact, the algorithm is agnostic as to what distribution is generating the instances.  
This has often been a weak spot for PAC learning as it allows one to come up with some artificial but ``problematic'' (in some sense) distribution $\DD_p$ and show negative results (e.g., one cannot perform learning when the set of training examples has size below a certain threshold) since a learning algorithm \algo that would succeed under any possible distribution, unfortunately fails to acquire some desired property under $\DD_p$ and therefore, in general, the required property cannot be achieved by \algo in a distribution-free sense, since, in particular, \algo fails under $\DD_p$.

Partly because of the above, and partly because we may be interested in the behavior of algorithms under specific distributions, or finally because it is the case that we can have provable mathematical results for certain algorithms only when we study the said algorithms under a particular distribution, or sets of distributions, has led to a variant of PAC learning that is about \emph{distribution-specific} learning. 
This line of work was initiated in~\citep{PAC:Fixed:BenedekItai}. 
Work along the lines of distribution-specific learning
has generated a plethora of results over the 
last 30 years 
and distribution-specific results are a commonplace for situations where robustness properties are investigated; e.g.,~\citep{AdvEgs:Diochnos:Neurips,AdvEgs:Diochnos:AAAI,Adversarial:EvasionBackdoor:Diochnos,Gourdeau:RobustHardness,Gourdeau:Robustness:DL,Urner:Robustness:NotAtOdds,Urner:UAI,PAC:Robustness:Mohri}.
%
%

Note that one can have at least two different variants of distribution-specific learning.  In one case one may receive as a hint the underlying distribution and then design a learning algorithm that adapts to the particular distribution.  However, a different idea is to have a single learning algorithm (i.e., uniform setup that exhibits no adaptation on particular distributions) that can operate over a range of distributions but not for all possible distributions.  Therefore, in a sense, in this latter case we have \emph{distribution-independent learning, but for a range of distributions}.
See for example the work of~\citet{Evolvability:JAIR} for an example of such a result in combination with the idea of an $\varepsilon$-cover that is discussed below.

\paragraph{$\varepsilon$-Covers.}
As a final remark, several distribution-specific learning results are obtained using \emph{$\varepsilon$-covers}.
In these situations the learning algorithm is tasked with the goal of identifying a PAC solution satisfying Definition~\ref{def:pac-learning} but without realizability holding.  For example, \Hh can be a strict subset of \Cc, but the functions inside \Hh are selected in such a way so that no matter which $c\in\Cc$ is the ground truth, there is at least some $h\in\Hh$ that is $\varepsilon$-close to $c$ under the distribution \DD (or set of distributions \DDset) where learning will occur. Some examples of this approach have been studied inside evolvability~\citep{Evolvability:EuroGP,Evolvability:ALT,Evolvability:JAIR}.
In particular, \citet{Evolvability:JAIR} provides \emph{one mechanism that is adaptive} to the exact underlying distribution in the sense that the distribution is used as a hint for the representation of the hypothesis that is being evolved and \emph{one more mechanism that is not adaptive} to the exact underlying distribution but rather evolves a hypothesis with specific representational prescription regardless of the underlying distribution, as long as the distribution is one of a permissible set of distributions.  Hence, the first mechanism corresponds to standard distribution-specific learning, whereas the second mechanism corresponds to distribution-independent learning restricted to a particular family of distributions as explained above.

%
%
%

\section{Adversaries and Noise Models}\label{sec:NoiseIntro}
In Definition~\ref{def:pac-learning} 
all the examples are known to the learner ahead of time.  This is known as the \emph{batch} model of learning.  In this case the study of learning is decomposed largely into two phases: the \emph{training} phase and the \emph{inference} (or \emph{test}) phase. 
The identification of a good hypothesis $h^\star$ after having access to some training data happens during training.
However, in general, adversaries may manipulate the data in either one of the two phases.

Inference-time attacks are known as \emph{evasion} attacks~\citep{Adversarial:Evasion:Biggio,Book:AdvML1}, or more commonly as \emph{adversarial examples}~\citep{AdvEgs:ExplainHarness,AdvEgs:Madry}; see also,~\citep{AdvEgs:Diochnos:Neurips,Gourdeau:RobustHardness} for implications and discussion on definitional aspects around this topic. Such attacks fall under the umbrella of \emph{exploratory} attacks~\citep{Book:AdvML2} and while this is a very interesting line of work, nevertheless, it is outside the scope of our discussion.

Training-time attacks are known as \emph{causative} attacks~\citep{Adversarial:Barreno,Book:AdvML2}.
Among other taxonomies, training-time attacks can be decomposed into \emph{poisoning} attacks and \emph{noise} affecting the data-generating process.
Our focus is on noise and we will largely ignore poisoning attacks in our discussion.
For uniformity, we sometimes personify a noise corruption mechanism as an \emph{opportunistic adversary}. However, under purely random models the modification is prescribed rather than strategically chosen.
This idea will be clarified further below and within the different noise models we can argue about the power that such an opportunistic adversary has in order to manipulate the data.

\subsection{Separating Noise Models from Poisoning Attacks}\label{sec:poisoning-vs-noise}

%
Formal lines of work on noise models\footnote{We use the terms \emph{``noise model''} and \emph{``noise framework''} interchangeably.} 
were developed before adversarial machine learning was the rich field that it is today.
In some of the early lines of work we see that poisoning attacks were sometimes treated as exceptionally adversarial cases of particular noise models.
To simplify our discussion we take the approach explained below in order to separate 
work and results that fall under what we mean by the term ``noise'' as opposed to what we mean by the term ``poisoning attack''.

We view \emph{poisoning attacks} as situations where the adversary 
may inject new examples or alter existing ones, subject to a prescribed
\emph{budget}, and may also \emph{choose which examples to target}.
Furthermore, poisoning attacks occur primarily as an \emph{offline} vulnerability model that the learning algorithm may have to deal with.  

On the other hand, we view \emph{noise} as the situation where the corruption occurs
as part of the data-generation process.
As each clean example is generated, a stochastic mechanism determines whether the adversary is granted a tampering opportunity. There is no prescribed budget and what may happen when an opportunity occurs depends on the particular noise model.

\begin{tcolorbox}[colback=cornellred!10!white,colframe=cornellred!50!eggplant,title=\faLightbulb \ Important: Noise vs.~Poisoning Attacks]
Therefore, 
the two crucial differences between noise and poisoning attacks as we treat them in this chapter are:
\begin{enumerate}[label=(\roman*)]

\item 
under noise, the adversary does not select the datapoints at which
tampering opportunities arise, whereas the adversary has such flexibility under poisoning attacks, and

\item 
under poisoning attacks the adversary is given a  budget ahead of time that they can use for their alterations, whereas under noise, the adversary acts in an opportunistic sense whenever they are allowed to do so.
\end{enumerate}
\end{tcolorbox}
Figure~\ref{fig:noise:mechanism} illustrates the generic corruption-opportunity mechanism used for the noise models considered in this chapter. It shows which examples receive tampering opportunities, with the locations of these opportunities determined by an underlying stochastic process.

\begin{figure}[ht]
\begin{center}
\psset{unit=0.65cm}
\begin{pspicture*}(-11.3,1.5)(11.5,6)
%
%
\rput(0,5.5){\textcolor{black}{$\bm{x_1}$}}
\rput(1,5.5){\textcolor{black}{$\bm{x_2}$}}
\rput(2,5.5){\textcolor{black}{$\bm{x_3}$}}
\rput(3,5.5){\textcolor{black}{$\bm{x_4}$}}
\rput(4,5.5){\textcolor{black}{$\bm{x_5}$}}
\rput(5,5.5){\textcolor{black}{$\bm{x_6}$}}
\rput(6,5.5){\textcolor{black}{$\bm{x_7}$}}
\rput(7,5.5){\textcolor{black}{$\bm{x_8}$}}
\rput(8,5.5){\textcolor{black}{$\bm{x_9}$}}
\rput(9,5.5){\textcolor{black}{$\bm{x_{10}}$}}
\rput(10,5.5){\textcolor{black}{$\bm{x_{11}}$}}
\rput(11,5.5){\textcolor{black}{$\bm{x_{12}}$}}
\pscircle[linecolor=black,fillcolor=white](0,5){2pt}
\pscircle[linecolor=black,fillcolor=white](1,5){2pt}
\pscircle[linecolor=black,fillcolor=white](2,5){2pt}
\pscircle[linecolor=black,fillcolor=white](3,5){2pt}
\pscircle[linecolor=black,fillcolor=white](4,5){2pt}
\pscircle[linecolor=black,fillcolor=white](5,5){2pt}
\pscircle[linecolor=black,fillcolor=white](6,5){2pt}
\pscircle[linecolor=black,fillcolor=white](7,5){2pt}
\pscircle[linecolor=black,fillcolor=white](8,5){2pt}
\pscircle[linecolor=black,fillcolor=white](9,5){2pt}
\pscircle[linecolor=black,fillcolor=white](10,5){2pt}
\pscircle[linecolor=black,fillcolor=white](11,5){2pt}
%
\rput[r](-0.5,4){\textcolor{black}{\small $c(\bm{x_i}) \rightarrow$}}
\rput[r](-0.5,5){\textcolor{black}{\small instance $\bm{x_i} \rightarrow$}}
\rput[r](-0.5,3){\textcolor{black}{\small \textcolor{magenta}{tampering-opportunity indicator $\mathcal{T}_i$} $\rightarrow$}}
\rput(0,4){\textcolor{black}{1}}
\rput(1,4){\textcolor{black}{1}}
\rput(2,4){\textcolor{black}{0}}
\rput(3,4){\textcolor{black}{0}}
\rput(4,4){\textcolor{black}{0}}
\rput(5,4){\textcolor{black}{1}}
\rput(6,4){\textcolor{black}{0}}
\rput(7,4){\textcolor{black}{1}}
\rput(8,4){\textcolor{black}{0}}
\rput(9,4){\textcolor{black}{1}}
\rput(10,4){\textcolor{black}{1}}
\rput(11,4){\textcolor{black}{0}}
%
\rput(0,3){\textcolor{black}{0}}
\rput(1,3){\textcolor{black}{0}}
\rput(2,3){\textcolor{black}{0}}
\rput(3,3){\textcolor{magenta}{1}}
\rput(4,3){\textcolor{magenta}{1}}
\rput(5,3){\textcolor{magenta}{1}}
\rput(6,3){\textcolor{black}{0}}
\rput(7,3){\textcolor{black}{0}}
\rput(8,3){\textcolor{black}{0}}
\rput(9,3){\textcolor{magenta}{1}}
\rput(10,3){\textcolor{black}{0}}
\rput(11,3){\textcolor{black}{0}}
%
\rput[r](-0.5,2){adversary \textcolor{magenta}{\small may modify} the example $\rightarrow$}
%
\rput(0,2){\textcolor{black}{\xmark}}
\rput(1,2){\textcolor{black}{\xmark}}
\rput(2,2){\textcolor{black}{\xmark}}
\rput(3,2){\textcolor{magenta}{\checkmark}}
\rput(4,2){\textcolor{magenta}{\checkmark}}
\rput(5,2){\textcolor{magenta}{\checkmark}}
\rput(6,2){\textcolor{black}{\xmark}}
\rput(7,2){\textcolor{black}{\xmark}}
\rput(8,2){\textcolor{black}{\xmark}}
\rput(9,2){\textcolor{magenta}{\checkmark}}
\rput(10,2){\textcolor{black}{\xmark}}
\rput(11,2){\textcolor{black}{\xmark}}
\end{pspicture*}
\end{center}
\vspace*{-0.3cm}
\caption{Generic corruption-opportunity mechanism used in this
chapter. For each clean example $(\bm{x_i},c(\bm{x_i}))$, with $\bm{x_i}\sim \DD$, that is generated, the noise mechanism generates a binary tampering-opportunity indicator 
$\mathcal{T}_i\in\{0,1\}$.  
If $\mathcal{T}_i = 0$, the example remains unchanged; if $\mathcal{T}_i = 1$, the noise model permits a modification of the corresponding example.
What happens when $\mathcal{T}_i = 1$ depends on the noise model: the modification may be prescribed, or chosen adversarially, possibly using the preceding history. 
In the illustration, tampering opportunities occur at indices $i\in\{4,5,6,10\}$ and for clarity each opportunity is also marked by a checkmark \textcolor{magenta}{\checkmark} in the final row.}
\label{fig:noise:mechanism}
\end{figure}

As a reminder, in adversarial models, the modification made at a granted position may depend on the preceding history.  This motivates the notation $\sample \getsample O$ introduced in Section~\ref{sec:prelim:oracles} and discussed in Remark~\ref{rem:non-iid-generation}, under which a faulty oracle generates and returns the complete sample as one object.


\subsection{Classification Noise: Random and Malicious}\label{sec:label-noise-basics}
When an oracle is ``suffering'' from \emph{random classification noise (RCN)} with noise rate $\eta < 1/2$, the following happens.
Upon request of an example by the learner from the (noisy) examples oracle, 
the oracle samples an $\bm{x}\sim\DD$ and returns: 
\begin{equation}\label{eq:RCNmodel}
\left\{
\begin{array}{rl}
    (\bm{x}, c(\bm{x})), & \mbox{with probability $1 - \eta$}, \\
    (\bm{x}, 1 - c(\bm{x})), & \mbox{with probability $\eta$}.
\end{array}
\right.
\end{equation}
In other words, such a faulty oracle will flip the label of an example with probability $\eta$, and the oracle will be denoted as \rcnoracle.
It is often convenient to think that an example $(\bm{x}, y')$ returned by the oracle \rcnoracle is obtained by first drawing an example $(\bm{x}, y)$ from \oracle and then obtaining $y'$ from $y$ by following rule~(\ref{eq:RCNmodel}).

\smallskip

In the case of \emph{malicious classification noise (MCN)} the adversary may choose to flip or not the label of the example when 
the tampering opportunity arises.
In other words, the rule is similar to~(\ref{eq:RCNmodel}), but with probability $\eta$ in the second branch of the rule, the adversary may return any label that they want.

\medskip

Tampering opportunities in these two noise models arise by tossing a coin that lands heads (\heads) with constant probability $\eta < 1/2$ for each generated example.

\begin{examp}
Consider a situation similar to Figure~\ref{fig:noise:mechanism}. 
This time the tampering opportunity for corrupting the data is determined by tossing a coin that lands heads (\heads) with constant probability $\eta < 1/2$ for each example generated.

Then, Figure~\ref{fig:ex:noise-application} 
shows the labels that are obtained under RCN and MCN.
\begin{figure}[H]
\begin{center}
\psset{unit=0.65cm}
\begin{pspicture*}(-6,0)(11.5,6)
%
%
\rput(0,5.5){\textcolor{black}{$\bm{x_1}$}}
\rput(1,5.5){\textcolor{black}{$\bm{x_2}$}}
\rput(2,5.5){\textcolor{black}{$\bm{x_3}$}}
\rput(3,5.5){\textcolor{black}{$\bm{x_4}$}}
\rput(4,5.5){\textcolor{black}{$\bm{x_5}$}}
\rput(5,5.5){\textcolor{black}{$\bm{x_6}$}}
\rput(6,5.5){\textcolor{black}{$\bm{x_7}$}}
\rput(7,5.5){\textcolor{black}{$\bm{x_8}$}}\rput(8,5.5){\textcolor{black}{$\bm{x_9}$}}
\rput(9,5.5){\textcolor{black}{$\bm{x_{10}}$}}
\rput(10,5.5){\textcolor{black}{$\bm{x_{11}}$}}
\rput(11,5.5){\textcolor{black}{$\bm{x_{12}}$}}
\pscircle[linecolor=black,fillcolor=white](0,5){2pt}
\pscircle[linecolor=black,fillcolor=white](1,5){2pt}
\pscircle[linecolor=black,fillcolor=white](2,5){2pt}
\pscircle[linecolor=black,fillcolor=white](3,5){2pt}
\pscircle[linecolor=black,fillcolor=white](4,5){2pt}
\pscircle[linecolor=black,fillcolor=white](5,5){2pt}
\pscircle[linecolor=black,fillcolor=white](6,5){2pt}
\pscircle[linecolor=black,fillcolor=white](7,5){2pt}
\pscircle[linecolor=black,fillcolor=white](8,5){2pt}
\pscircle[linecolor=black,fillcolor=white](9,5){2pt}
\pscircle[linecolor=black,fillcolor=white](10,5){2pt}
\pscircle[linecolor=black,fillcolor=white](11,5){2pt}
%
\rput[r](-0.5,4){\textcolor{black}{\small $c(\bm{x}_i) \rightarrow$}}
\rput[r](-0.5,5){\textcolor{black}{\small instance $\bm{x_i} \rightarrow$}}
\rput[r](-0.5,3){\textcolor{black}{\small coin toss outcome $\rightarrow$}}
\rput(0,4){\textcolor{black}{1}}
\rput(1,4){\textcolor{black}{1}}
\rput(2,4){\textcolor{black}{0}}
\rput(3,4){\textcolor{black}{0}}
\rput(4,4){\textcolor{black}{0}}
\rput(5,4){\textcolor{black}{1}}
\rput(6,4){\textcolor{black}{0}}
\rput(7,4){\textcolor{black}{1}}
\rput(8,4){\textcolor{black}{0}}
\rput(9,4){\textcolor{black}{1}}
\rput(10,4){\textcolor{black}{1}}
\rput(11,4){\textcolor{black}{0}}
%
\rput(0,3){\textcolor{black}{\tails}}
\rput(1,3){\textcolor{black}{\tails}}
\rput(2,3){\textcolor{black}{\tails}}
\rput(3,3){\textcolor{magenta}{\heads}}
\rput(4,3){\textcolor{magenta}{\heads}}
\rput(5,3){\textcolor{magenta}{\heads}}
\rput(6,3){\textcolor{black}{\tails}}
\rput(7,3){\textcolor{black}{\tails}}
\rput(8,3){\textcolor{black}{\tails}}
\rput(9,3){\textcolor{magenta}{\heads}}
\rput(10,3){\textcolor{black}{\tails}}
\rput(11,3){\textcolor{black}{\tails}}
%
\rput[r](-0.5,2){\textcolor{DarkOrchid}{\small RCN label $\rightarrow$}}
\rput[r](-0.5,1){\textcolor{red}{\small MCN label $\rightarrow$}}
\rput(0,2){\textcolor{black}{1}}
\rput(1,2){\textcolor{black}{1}}
\rput(2,2){\textcolor{black}{0}}
\rput(3,2){\textcolor{DarkOrchid}{1}}
\rput(4,2){\textcolor{DarkOrchid}{1}}
\rput(5,2){\textcolor{DarkOrchid}{0}}
\rput(6,2){\textcolor{black}{0}}
\rput(7,2){\textcolor{black}{1}}
\rput(8,2){\textcolor{black}{0}}
\rput(9,2){\textcolor{DarkOrchid}{0}}
\rput(10,2){\textcolor{black}{1}}
\rput(11,2){\textcolor{black}{0}}
\rput(0,1){\textcolor{black}{1}}
\rput(1,1){\textcolor{black}{1}}
\rput(2,1){\textcolor{black}{0}}
\rput(3,1){\textcolor{red}{?}}
\rput(4,1){\textcolor{red}{?}}
\rput(5,1){\textcolor{red}{?}}
\rput(6,1){\textcolor{black}{0}}
\rput(7,1){\textcolor{black}{1}}
\rput(8,1){\textcolor{black}{0}}
\rput(9,1){\textcolor{red}{?}}
\rput(10,1){\textcolor{black}{1}}
\rput(11,1){\textcolor{black}{0}}
\end{pspicture*}
\end{center}
\vspace*{-0.6cm}
\caption{Example application of random and malicious classification noise. Tampering opportunities in these two noise models arise by tossing a coin that lands heads (\heads) with constant probability $\eta < 1/2$.}\label{fig:ex:noise-application}
\end{figure}

As can be seen from the figure, under RCN the learner obtains the misclassified examples
$(\bm{x_4}, 1), (\bm{x_5}, 1), (\bm{x_6}, 0),$ and $(\bm{x_{10}}, 0)$.
In the case of MCN the adversary is allowed to assign either label to these four instances and this is shown with a question mark in the figure. 
The adversary's label choices may depend on preceding examples and on earlier label-manipulation decisions.
\end{examp}

\subsection{Method of Induced Distributions}\label{sec:induced-distributions}
A popular technique for proving impossibility results for learning under noise 
is called \emph{the method of induced distributions}.
Similar to the idea that was discussed in Section~\ref{sec:distribution-specific}, 
we will be studying a particular \emph{pathological} distribution where some property of the learning algorithm will fail.  This time, with the method of induced distributions the failure will be in a worst-case sense in that the learning process will be destined to fail under the pathological  distribution that we will construct, no matter how large of a sample size the learner will be allowed to have access to for the learning process!
This will happen because an adversary can  set up the induced distribution in such a way so that a learner will be unable to distinguish between two competing concepts, where one of them is the ground truth and the other one has prohibitively large risk under the PAC criterion that we would like to accomplish.

We will see such an example application of the method of induced distributions below for malicious noise which leads to a classic result showing that no learning algorithm can tolerate malicious noise at rate $\varepsilon/(1+\varepsilon)$ or higher---a threshold slightly below the permissible PAC error rate $\varepsilon$.
This will be in sharp contrast to some results that we will see for label noise later where the noise rate will be allowed to be arbitrarily close to $1/2$.  
Therefore such a result will allow us to put things into perspective and appreciate more the positive results that can be obtained under label noise and will be discussed in sequence.

\subsection{Application of Method of Induced Distributions on  Malicious Noise}\label{sec:arbitrary-malicious}
As promised above 
we will detour slightly and show a pessimistic result about learning in the presence of noise, using the method of induced distributions that was mentioned in Section~\ref{sec:induced-distributions}.

In particular,~\citet[Theorem 1]{PAC:Noise:Malicious} showed that the noise rate that can be tolerated by PAC learning algorithms in the malicious noise model is upper bounded by  $\varepsilon$; actually, slightly smaller as we will see in Theorem~\ref{thm:KearnsLi} below.
They did so by allowing the learner to have access to two oracles; one that returns only positive examples and another one that returns only negative examples. This model of PAC learning (with two oracles) was a popular variant at the time. However, in our presentation we investigate the situation where the learning algorithm has access to a single oracle and for this reason below we will see a proof using just one oracle that returns both positive and negative examples at the same time, respecting some underlying distribution \DD that governs all instances. 
Furthermore, the result is simpler to state if we require that the learner generates a hypothesis that has error rate \emph{strictly less} than $\varepsilon$; i.e., as in Definition~\ref{def:pac-learning} we want to generate a hypothesis $h$ that satisfies $\arisk{\oracle}{h} < \varepsilon$, rather than $\arisk{\oracle}{h} \le \varepsilon$.

\smallskip

In the arbitrary malicious noise model (i.e., not just malicious classification noise as it was discussed above in Section~\ref{sec:label-noise-basics}), the adversary has the opportunity to tamper with each example $(\bm{x}, y)$ that is being generated honestly, at a rate $\eta$, following the basic mechanism of Figure~\ref{fig:noise:mechanism}.
Hence, in such situations the learner has access to some arbitrary example $(\bm{z}, \ell)$ from the noisy oracle.  
That is, both the instance and the label can be changed when the adversary has such an opportunity.  Moreover, the adversary's tampering decision may depend on the decisions that they have made earlier in the data-generating process, similar to what was mentioned earlier in Sections~\ref{sec:poisoning-vs-noise} and~\ref{sec:label-noise-basics} and also connecting to our discussion in Remark~\ref{rem:non-iid-generation}.

\smallskip

We are now ready to see the theorem.

\begin{theo}[Theorem~1 in~\citet{PAC:Noise:Malicious}]\label{thm:KearnsLi}
Let \Cc be a concept class that is distinct (see Section~\ref{sec:distinct}). Let $\varepsilon < 1/2$ and assume that we work with the strict PAC objective of Definition~\ref{def:pac-learning}.
Then, while learning this concept class \Cc, no learning algorithm can tolerate malicious noise at a rate of  $\dfrac{\varepsilon}{1+\varepsilon}$, or higher.
\end{theo}

\begin{proof}
Let $\varepsilon < 1/2$. 
Let the concept class $\Cc$ be \emph{distinct}; that is, there are at least two concepts $c_1, c_2\in\Cc$ such that 
on two instances $\bm{u}, \bm{v}\in\XX$ it holds 
$c_1(\bm{u}) = c_2(\bm{u})$ and $c_1(\bm{v}) \neq c_2(\bm{v})$. 
In other words, 
the concepts $c_1$ and $c_2$ agree in one of the instances of the domain and disagree in another one.
Furthermore, allow the distribution \DD over \XX to be concentrated on these two instances as shown below:
\begin{equation}
\left\{
\begin{array}{rcl}
    \DD(\bm{u}) &=& 1 - \varepsilon \\
    \DD(\bm{v}) &=& \varepsilon 
\end{array}
\right.,
\end{equation}
and to make things more concrete say we have
\begin{equation}
\left\{
\begin{array}{rcl}
    c_1(\bm{u}) &=& c_2(\bm{u}) = 1  \\
    c_1(\bm{v}) &=& 1 \\
    c_2(\bm{v}) &=& 0
\end{array}
\right.\,.
\end{equation}

Now notice that regardless of whether $c_1$ or $c_2$ is the ground truth, any hypothesis that disagrees with the ground truth on even one of these two points has error at least $\varepsilon$ and thus it is not accepted as a solution satisfying the PAC criterion (with strict inequality).

\smallskip

The argument is symmetric in $c_1$ and $c_2$, so either concept may be taken as the ground truth $c$.
In the end it will be clear why this is the case as the adversary's approach is precisely to confuse the learning algorithm \algo and make it impossible for \algo to distinguish between these two functions.
Now, let's come to the adversary's strategy and how the induced distribution looks like from where the learning algorithm will obtain noisy examples.

\smallskip

\paragraph*{Adversary's Strategy.}
During the course of learning, the adversary presents a point drawn from \DD with probability $1-\eta$. Furthermore, with probability $\eta$ the adversary may tamper with the data and in that case will return $\bm{v}$ with the opposite label of the target function that we want to learn.
%
%

\paragraph*{Induced Distribution $\DD'$.}
The induced distribution $\DD'$ is shown below.
\begin{center}
\begin{tabular}{r|cc|lccl}
\toprule
\multicolumn{1}{c|}{} & $c_1$ & $c_2$ & \multicolumn{3}{c}{$\DD'$} \\
\midrule
$\bm{u}$ & 1 & 1 & $(1-\eta)\cdot(1-\varepsilon)$ & $+$ & $0$ & $\leftarrow$ always return label 1 \\
$\bm{v}$ & 1 & 0 & $(1-\eta)\cdot\varepsilon$ & $+$ & $\eta$ & $\leftarrow$ return both labels \\
\bottomrule
\multicolumn{3}{c}{} & \multicolumn{1}{c}{\ensuremath{\uparrow}} & \phantom{dimis} & 
\multicolumn{1}{c}{\ensuremath{\uparrow}} & 
\\
\multicolumn{3}{c}{} & \multicolumn{1}{c}{\ensuremath{\myprob{\mbox{honest label}}}} & & \multicolumn{1}{c}{\ensuremath{\myprob{\mbox{opposite label}}}} & 
\end{tabular}
\end{center}
In other words, the malicious oracle returns the following examples:
\begin{equation}\label{eq:malicious:confusion}
\left\{
\begin{array}{rl}
    (\bm{u}, 1), & \mbox{with probability \ } (1-\eta)(1-\varepsilon) \\
    (\bm{v}, c(\bm{v})), & \mbox{with probability \ } (1-\eta)\varepsilon \\
    (\bm{v}, 1 - c(\bm{v})), & \mbox{with probability \ } \eta \\
\end{array}
\right.\,.
\end{equation}

\paragraph*{Confusing the Learning Algorithm.}
If we require now $(1-\eta)\varepsilon = \eta \Leftrightarrow \eta = \frac{\varepsilon}{1+\varepsilon}$, then it follows that the instance $\bm{v}$ is returned to the learner with both labels at the same rate $\varepsilon/(1+\varepsilon)$ in (\ref{eq:malicious:confusion}).
Furthermore, the same distribution $\DD'$ is obtained \emph{regardless} of which function among the $c_1$ or $c_2$ is the true ground truth function $c$ that we are trying to learn. Therefore, any algorithm that produces an $\varepsilon$-\emph{good} hypothesis with probability at least $1-\delta$ when the target function is $c_1$, then with the same probability that same algorithm must produce an $\varepsilon$-\emph{bad} hypothesis when the target is $c_2$ (and vice-versa). 
Hence, the malicious noise rate that can potentially be tolerated is strictly less than $\varepsilon/(1+\varepsilon)$ as desired by the statement of the theorem.
\end{proof}

\begin{rem}
If the adversary has more power for tampering with the data at a rate larger than $\varepsilon/(1+\varepsilon)$, then they can always return an honest example for the excess part of the probability above this threshold and now repeat the above argument.
\end{rem}

\begin{rem}
The requirement that the proof of Theorem~\ref{thm:KearnsLi} contains instances $\bm{u}$ and $\bm{v}$ is in fact meaningful. For example, let $\bm{u} = (1, \ldots, 1)^T$ and $\bm{v} = (1, \ldots, 1, 0)^T$ and the two concepts be 
\begin{displaymath}
\left\{
\begin{array}{rcl}
c_1 &=& x_1\wedge x_2\wedge \ldots\wedge x_{n-1} \\
c_2 &=& x_1\wedge x_2\wedge \ldots \wedge x_{n-1} \wedge x_{n}
\end{array}
\right.
\end{displaymath}
Then, $c_1(\bm{u}) = c_2(\bm{u}) = 1$, $c_1(\bm{v}) = 1$, and $c_2(\bm{v}) = 0$ as desired.
\end{rem}

\begin{rem}
As a result of the above, when the learning algorithm wants to produce a better hypothesis (i.e., having risk bounded by a smaller value of $\varepsilon$) in the distribution-free setting, then the malicious noise that can be tolerated has to have rate $\eta$ that satisfies $\eta < \varepsilon/(1+\varepsilon)$.
That is, the better the solution that we want to produce in the general case, the smaller the malicious noise rate that we can tolerate.
\end{rem}

\section{Some Important Results and Ideas on Classification Noise}\label{sec:classification-noise}
We discussed the basic mechanism for RCN and MCN in Section~\ref{sec:label-noise-basics} but below we will formalize them in the context of PAC learning.
In addition, a closely related noise model is that of Massart noise~\citep{PAC:Noise:Massart} and we will formalize it too in the same definition.
Hence, our task is to modify Definition~\ref{def:pac-learning} so that we can accommodate the above noise models in the learning process.

\begin{defi}[PAC Learning under Random, Massart, or Malicious Classification Noise]\label{def:pac-label-noise}
A concept class $\mathcal{C}$ is \emph{probably approximately correct (PAC)}-learnable by the hypothesis space $\mathcal{H}$ in the presence of random classification noise, 
or Massart noise, or malicious classification noise, 
if there exists a learning algorithm $\algo$ and 
a polynomial function $m(\cdot, \cdot, \cdot, \cdot, \cdot)$ with the property that
for every $c\in\Cc$, every distribution \DD over \XX, every $\varepsilon, \delta \in (0, 1/2)$, 
every noise-rate function 
$\eta \colon \XX \to [0, 1/2)$
satisfying the conditions of the corresponding noise model, and every bound $\eta_b$ such that 
$\sup_{\bm{x}\in\XX} \eta(\bm{x})\leq\eta_b<1/2$,
if \algo is given inputs $n, \mathrm{size}(c), \varepsilon, \delta, \eta_b$, 
and access to a noisy oracle \genericfaultyoracle with behavior as specified below, 
when \algo receives a random sample \sample of size
at least 
$m(\nicefrac{1}{\varepsilon}, \nicefrac{1}{\delta}, \nicefrac{1}{(1-2\eta_b)}, n,\mathrm{size}(c))$
drawn from 
\genericfaultyoracle, 
then \algo will produce a hypothesis $h^\star\in\Hh$
that satisfies 
\begin{equation}\label{eq:pac-noise-criterion}
\probsub{\sample \getsample \genericfaultyoracle, h^\star\assign \algo(\sample)}{
\arisk{\oracle}{h^\star} 
< 
\varepsilon
} \ge 1 - \delta\,,
\end{equation}
where \oracle is the oracle that returns honest examples not affected by noise.
The probability is over the clean instance draws, the corruption opportunities, any randomization used by the adversary, and any internal randomization of \algo.

\begin{itemize}[leftmargin=0.3cm]
    \item 
    When $\genericfaultyoracle = \rcnoracle$ we have \emph{random classification noise (RCN)} and it holds that $\eta(\bm{x}) = \eta, \forall \bm{x}\in\XX$. Thus, according to the general mechanism scheme that was discussed in Figure~\ref{fig:noise:mechanism} and was clarified further in Figure~\ref{fig:ex:noise-application} for this noise model, for every instance $x\in\XX$ a coin is being tossed that lands heads (\heads) with probability $\eta \le \eta_b < 1/2$.  Whenever an \heads is the outcome of the coin toss, then the label of the honest example is flipped. Notice that the noise rate is \emph{instance-independent} in this case.
    For this reason in (\ref{eq:pac-noise-criterion}) one can view $\sample \getsample \genericfaultyoracle$ as $\sample\sim\left(\genericfaultyoracle\right)^m$ due to the fact that mislabelings occur independently of other mislabelings. 


    \item 
    When $\genericfaultyoracle = \massoracle$,
    we have the case of \emph{Massart noise} and 
    the noise rate \emph{is} in fact now \emph{instance-dependent} as every instance $\bm{x}\in\XX$ has a different rate $\eta(\bm{x})$ by which 
    its label is flipped.
    In other words, under Massart noise, 
    whenever the tampering-opportunity indicator in 
    Figure~\ref{fig:noise:mechanism} satisfies $\mathcal{T}_i = 1$, 
    the label \emph{has to be flipped},
    similar to the case of RCN.  
    However, unlike RCN, the coin that is being tossed in order to determine if the label will be flipped or not, lands heads (\heads) with different probability $\eta(\bm{x})$ for every instance $\bm{x}$. 
    This \emph{non-uniform} classification noise rate allows for a more realistic noise setting than RCN; for example, human experts may find it harder to label certain instances compared to others and hence a variable noise rate across \XX can be justified.
    Regardless, note that again the noise is \emph{bounded} as we have $\eta(\bm{x}) \le \eta_b < 1/2$.
    Similar to the case of RCN, for Massart noise 
    in (\ref{eq:pac-noise-criterion}) one can view $\sample \getsample \genericfaultyoracle$ as $\sample\sim\left(\genericfaultyoracle\right)^m$ due to the fact that mislabelings occur independently of other mislabelings (even if a different coin with rate $\eta(\bm{x})$ is being flipped internally in the oracle in order to decide about the label flip). 

    \item 
    When $\genericfaultyoracle = \mcnoracle$ we have the case of \emph{malicious classification noise (MCN)}.
    Notice that, similar to RCN, in this case the classification noise rate is instance-independent; that is, $\eta(\bm{x}) = \eta$ for all $x\in\XX$.
    However, contrary to the case of RCN, in the case of MCN the adversary \emph{gets to choose} whether they will flip a label or not, whenever they have the chance (i.e., whenever the coin toss is an \heads that would allow the adversary to tamper with the data). This choice of the label flip may also depend on examples generated earlier and potential label flips that the adversary chose to perform. 
    For this reason, in the case of MCN, in (\ref{eq:pac-noise-criterion}) one should \emph{not} substitute $\sample\getsample\genericfaultyoracle$ with $\sample\sim\left(\genericfaultyoracle\right)^m$ anymore, as the draws are no longer i.i.d.~from \genericfaultyoracle.
    As a further consequence of this, under MCN a learning algorithm really obtains a sample \sample of size $m$ with a \emph{single push of a button} on the oracle \mcnoracle and receives all $m$ examples as a single entity, rather than one by one.
\end{itemize}

Finally, if furthermore \algo runs in time $m(\nicefrac{1}{\varepsilon}, \nicefrac{1}{\delta}, \nicefrac{1}{(1-2\eta_b)}, n, \mathrm{size}(c))$, then it is said to be an \emph{efficient} PAC-learning algorithm in the presence of random classification noise, Massart noise, or malicious classification noise, depending on the context of the faulty oracle \genericfaultyoracle.
\end{defi}

\begin{rem}
Notice that while the training examples are obtained from the noisy oracle \genericfaultyoracle, the PAC criterion in (\ref{eq:pac-noise-criterion}) requires the hypothesis $h^\star$ identified by \algo to have risk not $\varepsilon$ or more against an example drawn from the \emph{honest} oracle \oracle.
\end{rem}

\begin{rem}
Also note that as $\eta_b$ approaches $\nicefrac{1}{2}$, then the learning algorithm is allowed to use more examples during the learning process.
\end{rem}

\subsection{Empirical Risk Minimization under Random Classification Noise}\label{sec:RCN-ERM}
The largest part of our discussion is concentrated around RCN, which was proposed in 1988 by~\citet{PAC:Noise:Angluin}.  In the absence of noise, ERM translates to finding a hypothesis that is consistent with the training data; i.e., has zero empirical risk. 
However, under noise it could very well be the case that no hypothesis in \Hh is consistent (makes no mistakes) with the training data.
Nevertheless, the idea of ERM under noise translates to finding a hypothesis $h\in\Hh$ that has the smallest disagreement rate (see Definitions~\ref{def:disagreement-rate} and~\ref{def:emprisk}) with the training examples available to the learning algorithm.

\subsubsection{Hypothesis-Ordering Preservation under Random Classification Noise}\label{sec:RCN-ordering-of-hypotheses}

The following simple result is central to the study of RCN.

\begin{prop}[Noise-Invariant Hypothesis Ordering 
under RCN;~\citep{PAC:Noise:Angluin}]\label{prop:RCN-hypotheses-ordering} 
In expectation, 
the ordering of the hypotheses w.r.t.~their goodness of fit to the unknown target function $c$ that is being learned is not distorted under RCN, regardless of the RCN rate $\eta < 1/2$ that governs the underlying noise process.
This is justified because for every $h\in\Hh$ it holds that
\begin{displaymath}
\arisk{\rcnoracle}{h} = \eta + \arisk{\oracle}{h}(1-2\eta)\,.
\end{displaymath}
\end{prop}
\begin{proof}
We can calculate the rate by which a hypothesis $h_i\in\Hh$ will disagree with an example obtained from the examples oracle \rcnoracle.
\begin{itemize}[leftmargin=0.3cm] 
    \item The first case is if \rcnoracle returns an honest example and $h_i$ makes a mistake. This will happen with probability $(1-\eta)\cdot \arisk{\oracle}{h_i}$.

    \item The second case is if \rcnoracle returns a mislabeled example, but $h_i$ predicts the correct label. This will happen with probability 
    $\eta(1-\arisk{\oracle}{h_i})$.
\end{itemize}
Hence we have,
\begin{displaymath}
\arisk{\rcnoracle}{h_i} = (1-\eta)\cdot \arisk{\oracle}{h_i} + \eta\cdot (1-\arisk{\oracle}{h_i})
\end{displaymath}
Or, by re-arranging, we obtain the desired equation.
\end{proof}
In other words, when the hypotheses are considered in the no-noise case in terms of their (true) goodness of fit against $c$, then that ordering is preserved even in the presence of RCN!
Furthermore, Proposition~\ref{prop:RCN-hypotheses-ordering} holds even when the realizability assumption does not hold! Hence, if 
\begin{displaymath}
h^\varheartsuit\in\arginf_{h\in\Hh} \arisk{\oracle}{h}\,,    
\end{displaymath}
then Proposition~\ref{prop:RCN-hypotheses-ordering} implies that $h^\varheartsuit$ is also the solution to the problem of minimizing disagreements against a noisy oracle; i.e., a solution to 
\begin{displaymath}
\arginf_{h\in\Hh} \arisk{\rcnoracle}{h}\,.
\end{displaymath}
It is also easy to see that the converse is also true in Proposition~\ref{prop:RCN-hypotheses-ordering}; 
that is, 
\begin{displaymath}
h^\varheartsuit\in\arginf_{h\in\Hh}\arisk{\rcnoracle}{h}
\Longrightarrow
h^\varheartsuit\in\arginf_{h\in\Hh}\arisk{\oracle}{h}\,.
\end{displaymath}

In particular, if realizability holds, then 
$\arisk{\oracle}{h^\varheartsuit} = 0$.
Thus, using  Proposition~\ref{prop:RCN-hypotheses-ordering} we know that the following holds:
\begin{equation}\label{eq:pacwnoise-minimizer}
\left\{
\begin{array}{l}
    \Cc \subseteq \Hh \\
     h^\varheartsuit\in\arginf_{h\in\Hh} \arisk{\rcnoracle}{h}
\end{array}
\right.
\Longrightarrow
\arisk{\rcnoracle}{h^\varheartsuit} = \eta\,.
\end{equation}
This is another extremely important idea under RCN, since if we have enough data we can approximate sufficiently well the noise rate $\eta$ that is governing the noise process by simply looking at the empirical disagreement rate that the best hypothesis $h\in\Hh$ accomplishes against the (potentially mislabeled) training data. We will revisit this idea in Section~\ref{sec:rcn-upper-bound}.


\subsubsection{The Statistical Complexity of Empirical Risk Minimization under Random Classification Noise}\label{sec:RCN-ERM-sample-complexity}
We will now prove the theorem for the sample complexity of learning algorithms that rely on ERM under RCN.

\begin{theo}[\citet{PAC:Noise:Angluin}; \citet{Book:Laird}]\label{thm:RCN-ERM-sample-complexity}
Let $\Hh$ be a finite hypothesis space 
and 
$\Cc$ be a concept class, both over $\XX$, such that $\Cc\subseteq\Hh$. 
Let $0 < \eta \le \eta_b < 1/2$.
If
\begin{displaymath}
m\geq 2\varepsilon^{-2}(1-2\eta_{b})^{-2}\ln{\left(2\abs{\Hh}/\delta\right)}\,,
\end{displaymath}
then 
for any $h\in\Hh$ that minimizes the disagreements (i.e., $h(\bm{x_{i}})\neq y_i$) with the sequence 
of examples $\sample=\{(\bm{x_{i}},y_i)\}_{i=1}^{m}$ obtained by the noisy oracle \rcnoracle, it holds that 
\begin{displaymath}
\probsub{\sample\sim\left(\rcnoracle\right)^m}{\arisk{\oracle}{h} < \varepsilon} \geq 1-\delta\,.
\end{displaymath}
That is, whenever $\log\abs{\Hh}$ is polynomially bounded in the representation parameters, Definition~\ref{def:pac-label-noise} is satisfied for RCN and therefore $\Cc$ is PAC-learnable by $\Hh$ over $\XX$ 
having access to examples from a noisy oracle \rcnoracle via a method that minimizes disagreements with the sequence of training examples \sample.
\end{theo}
\begin{proof}[Sketch of the Proof]
The proof is similar to the proof for agnostic PAC learning using a finite hypothesis space.  That is, we will draw enough examples to approximate sufficiently well the disagreement rate of each hypothesis against the noisy oracle \rcnoracle.  Then, due to Proposition~\ref{prop:RCN-hypotheses-ordering} it follows that if we perform ERM over \Hh using these empirical disagreement rates, we will identify a hypothesis~$h^\star$ that is less than $\varepsilon$-worse compared to the best possible hypothesis~$h^\varheartsuit\in\Hh$.
\end{proof}
\begin{proof}
First we prove the correctness of the claim.  Then we will calculate the sample size (i.e., the statistical complexity) that is needed to support the claim.

\paragraph{Correctness of the Claim.}
Figure~\ref{fig:ERM-RCN-proof} provides a pictorial representation of the different hypotheses, as far as their true (expected) disagreement rate 
is concerned with respect to examples obtained from the noisy oracle \rcnoracle.
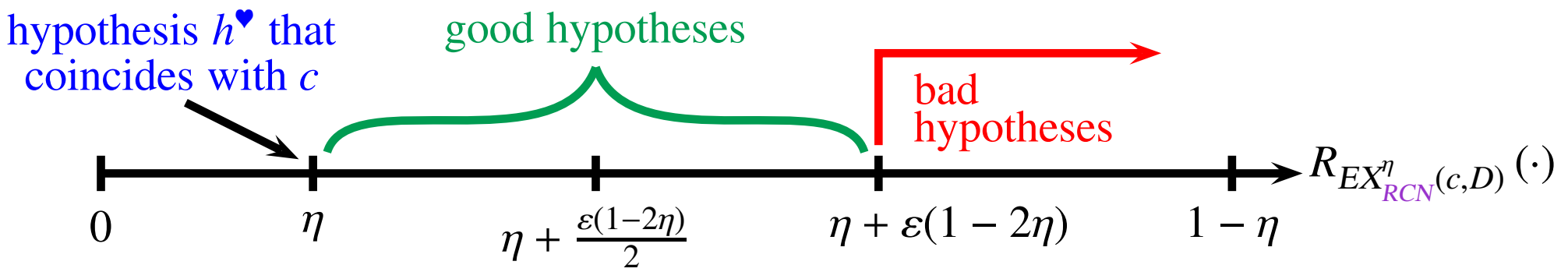
\begin{figure}[ht]
\begin{center}
\psset{unit=0.6cm}
\begin{pspicture*}(-1.5,-1.5)(21,2.4)
%
%
%
%
\psline[linecolor=black,linewidth=2pt]{->}(0,0)(17,0)
\psline[linecolor=black,linewidth=2pt]{-}(0,-0.25)(0,0.25)
\psline[linecolor=black,linewidth=2pt]{-}(16,-0.25)(16,0.25)
\psline[linecolor=black,linewidth=2pt]{-}(3,-0.25)(3,0.25)
\psline[linecolor=black,linewidth=2pt]{-}(7,-0.25)(7,0.25)
\psline[linecolor=black,linewidth=2pt]{-}(11,-0.25)(11,0.25)
%
\rput[l](17.1,0){\textcolor{black}{$\arisk{\rcnoracle}{\cdot}$}}
\rput[c](0,-0.75){\textcolor{black}{0}}
\rput[c](16,-0.75){\textcolor{black}{$1-\eta$}}
\rput[c](3,-0.75){\textcolor{black}{$\eta$}}
\rput[c](7,-0.9){\textcolor{black}{$\eta + \frac{\varepsilon(1-2\eta)}{2}$}}
\rput[c](12,-0.75){\textcolor{black}{$\eta + \varepsilon(1-2\eta)$}}
%
\rput[c](1,2){\textcolor{blue}{hypothesis $h^\varheartsuit$ that}}
\rput[c](1,1.4){\textcolor{blue}{coincides with $c$}}
\rput[c](7,2){\textcolor{MyGreen}{good hypotheses}}
\psline[linecolor=black,linewidth=2pt]{->}(1.2,1.0)(2.8,0.2)
\psbezier[linecolor=MyGreen,linewidth=2pt,showpoints=false]{-}(3.2,0.3)(3.5,1.3)(6.5,0.1)(7,1.5)
\psbezier[linecolor=MyGreen,linewidth=2pt,showpoints=false]{-}(7,1.5)(7.5,0.1)(10.5,1.3)(10.8,0.3)
\psline[linewidth=2pt,linecolor=red]{->}(11,0.4)(11,1.7)(15,1.7)
\rput[l](11.5,1.2){\textcolor{red}{bad}}
\rput[l](11.5,0.6){\textcolor{red}{hypotheses}}
\end{pspicture*}
\end{center}
\caption{Schematic representation regarding the relationship of the different hypotheses as far as their true (expected) disagreement rate is concerned, with respect to examples obtained from the noisy oracle \rcnoracle. 
A hypothesis $h\in\Hh$ is called \emph{bad} if $\arisk{\oracle}{h} \ge \varepsilon$; otherwise, it is called \emph{good}.
}\label{fig:ERM-RCN-proof}
\end{figure}
As it was explained in (\ref{eq:pacwnoise-minimizer}), the hypothesis $h^\varheartsuit$ that coincides with the target concept $c$ that we want to learn has the smallest true disagreement rate among all $h\in\Hh$ and in fact it holds that $\arisk{\rcnoracle}{h^\varheartsuit} = \arisk{\rcnoracle}{c} = \eta$.
Furthermore, this is the hypothesis that has zero true risk in the no-noise setting and thus the ideal solution 
in the no-noise setting.  
Therefore, good hypotheses in the sense of satisfying the PAC learning criterion under RCN (Definition~\ref{def:pac-label-noise}) are all these hypotheses $h\in\Hh$ such that 
$\arisk{\oracle}{h} 
< \varepsilon$.
Due to Proposition~\ref{prop:RCN-hypotheses-ordering} all these ``good'' hypotheses have true (expected) disagreement rate $\arisk{\rcnoracle}{h} 
<
\eta + \varepsilon(1-2\eta)$.  This corresponds to the green half-open interval $[\eta, \eta + \varepsilon(1-2\eta))$ in Figure~\ref{fig:ERM-RCN-proof}.

Now all that we really want to accomplish is to select a sufficiently large dataset~\sample for estimating this disagreement rate of every one of these hypotheses in such a way that the ERM principle would never select one of the ``bad'' hypotheses; i.e., the hypotheses that have prohibitively high true risk in the no-noise setting.  An easy way out from this situation would be to approximate the disagreement rate $\arisk{\rcnoracle}{h_i}$ of all the hypotheses $h_i\in\Hh$ within half (or less) of the length of the green interval in Figure~\ref{fig:ERM-RCN-proof}; that is, within $\upxi = \varepsilon(1-2\eta)/2$ of their true value.  
To see this, note that should this guarantee be satisfied, then no bad hypothesis will have empirical disagreement rate against such a sample $\sample \sim \left(\rcnoracle\right)^m$ that has $m$ i.i.d.~examples drawn from \rcnoracle, to the left of the midpoint of the green interval that contains all the good hypotheses.
In other words, for such a sample $\sample$, for any bad hypothesis $h^{\mathrm{bad}}$, with high probability, it will hold that 
\begin{equation}\label{eq:RCN-ERM-proof-1}
\aemprisk{S}{h^{\mathrm{bad}}} 
>
\eta + \varepsilon(1-2\eta)/2\,.    
\end{equation}
In addition, for the same sample $\sample$, the hypothesis $h^\varheartsuit = c$ will have, with high probability, empirical disagreement rate
\begin{equation}\label{eq:RCN-ERM-proof-2}
\aemprisk{S}{h^\varheartsuit} < \eta + \varepsilon(1-2\eta)/2\,.
\end{equation}
Therefore, if the sample \sample is sufficiently large so that (\ref{eq:RCN-ERM-proof-1}) and (\ref{eq:RCN-ERM-proof-2}) are satisfied (with high probability), then no ERM learning algorithm will ever select a bad hypothesis in the output, as it holds that $\aemprisk{\sample}{h^\varheartsuit} < \aemprisk{\sample}{h^{\mathrm{bad}}}$ for every hypothesis $h^{\mathrm{bad}}\in\Hh$ that has prohibitively large risk in the no noise setting; i.e., for every hypothesis $h^{\mathrm{bad}}$ for which it holds $\arisk{\oracle}{h^{\mathrm{bad}}} \ge \varepsilon$.
Of course it can be the case that an ERM algorithm will select an $h^\star\in\Hh$ that is different from $h^\varheartsuit$.  This is fine, because all such hypotheses that can confuse the ERM process are all good, in the sense that in the no-noise setting for all of them it holds that $\arisk{\oracle}{h} < \varepsilon$.

\paragraph*{Statistical Complexity Required to Support the Claim.}
From the above arguments we want to approximate the disagreement rate of every hypothesis $h\in\Hh$ within $\varepsilon(1-2\eta)/2$ of their true value.
By Proposition~\ref{prop:Hoeffding} for $\alpha = 0, \beta = 1, \upxi = \varepsilon(1-2\eta)/2$, it follows that for every hypothesis $h_i\in\Hh$ it will hold that 
\begin{equation}\label{eq:RCN-ERM-proof-3}
%
%
%
\abs{\aemprisk{\sample}{h_i}-\arisk{\rcnoracle}{h_i}}\geq \varepsilon(1-2\eta)/2
\end{equation}
with probability
\begin{equation}\label{eq:RCN-ERM-proof-4}
p_i^{\textsl{fail}} \le 2\naturalexp{\frac{{-2m\left(\varepsilon(1-2\eta)/2\right)^{2}}}{(1-0)^{2}}} 
= 
2\naturalexp{-m\varepsilon^2(1-2\eta)^2/2}\,.
\end{equation}
We now want to require that this failure probability $p_i^{\textsl{fail}}$ in (\ref{eq:RCN-ERM-proof-4}) to be at most $\delta/\abs{\Hh}$.
Hence, 
it is enough if our sample size satisfies 
\begin{equation}\label{eq:RCN-ERM-proof-sample-size}
2\naturalexp{-m\varepsilon^2(1-2\eta)^2/2} \le \frac{\delta}{\abs{\Hh}}
\Longleftrightarrow
m \ge \frac{2}{\varepsilon^2(1-2\eta)^2}\ln\left(\frac{2\abs{\Hh}}{\delta}\right)\,.
\end{equation}


\paragraph*{Bounding the Failure Probability.}

Finally, we need to show that the overall failure probability is not more than $\delta$ so that Definition~\ref{def:pac-label-noise} is satisfied as required.

Indeed, for $i\in\set{1, \ldots, \abs{\Hh}}$, let $B_i$ be the bad event
\begin{displaymath}
B_i \equiv \mbox{Equation (\ref{eq:RCN-ERM-proof-3}) holds for hypothesis} \ h_i\,.
\end{displaymath}
Then, some bad event $B_i$ will be true with probability
\begin{eqnarray}
\probsub{\sample\sim\left(\rcnoracle\right)^m}{B_1 \vee \cdots \vee B_{\abs{\Hh}}}
&\le &
\sum_{i=1}^{\abs{\Hh}} \probsub{\sample\sim\left(\rcnoracle\right)^m}{B_i}\nonumber
\\
&=&
\sum_{i=1}^{\abs{\Hh}} p_i^{\textsl{fail}}\,.\label{eq:RCN-ERM-proof-6}
\end{eqnarray}
However, the selection of the sample size $m$ in (\ref{eq:RCN-ERM-proof-sample-size}) was done in such a way so that for every $i\in\set{1, \ldots, \abs{\Hh}}$, it holds 
\begin{displaymath}
p_i^{\textsl{fail}} \le \frac{\delta}{\abs{\Hh}}\,.
\end{displaymath}
Using this in (\ref{eq:RCN-ERM-proof-6}) it follows that the overall failure probability is upper bounded by $\delta$ as required.

\paragraph*{The Last Touch.}
Note that the analysis above is using the exact noise rate $\eta$ which is unknown to the learning algorithm.  However, the learning algorithm is assumed to have been given an upper bound $\eta_b$ of the true noise rate $\eta$. 
As $\eta \le \eta_b$, substituting $\eta$ with $\eta_b$ in the sample complexity calculated for $m$ in (\ref{eq:RCN-ERM-proof-sample-size}) only increases the sample size that we require and therefore the empirical estimates for the disagreement rates for all the hypotheses will be even better (or, if you prefer, we will be even more confident that they fall within their expected intervals). 
Therefore, all the above arguments still hold and this explains the sample size stated in this theorem.
\end{proof}

\begin{rem}
The finiteness assumption on \Hh in Theorem~\ref{thm:RCN-ERM-sample-complexity} is used to obtain uniform convergence by applying Hoeffding’s inequality separately to every hypothesis and then taking a union bound over \Hh. It is not essential to the underlying learning principle. For an infinite hypothesis space, one may instead control uniform convergence using an appropriate capacity measure, such as the VC dimension. In that case, the term $\ln\left(\abs{\Hh}\right)$ in the sample-complexity bound is replaced by a quantity depending on the complexity of \Hh, typically its VC dimension together with logarithmic factors in the accuracy parameter.
This remark is also made by~\citet{PAC:Noise:Angluin} a few lines after they prove Theorem 2 in their paper.

As a further clarifying comment, the dependence on the RCN rate continues to arise through the factor $(1-2\eta_b)^{-2}$, because RCN shrinks the separation between hypotheses by the multiplicative factor $(1-2\eta)$ as we saw in Proposition~\ref{prop:RCN-hypotheses-ordering}.
\end{rem}

\begin{rem}\label{rem:rcn:complexity}
Laird~\citep{Book:Laird}
has calculated a better upper bound than the one in Theorem~\ref{thm:RCN-ERM-sample-complexity}, as well as
a lower bound. 
In particular, the number of examples $m$ depends only on $\varepsilon^{-1}$, and not on $\varepsilon^{-2}$.
\citet{PAC:pConcepts:Simon:Bounds} has a lower bound similar to Laird's but with the term $\log\left(\abs{\Hh}\right)$ substituted by the VC dimension of \Hh. 
\citet{PAC:pConcepts:RCN:UpperBound} give a similar upper bound for the RCN model using the VC dimension of \Hh.  See also the work of~\citet{PAC:Noise:SampleComplexity} and~\citet{PAC:Noise:SampleEfficient} for additional discussion, results, and a more complete picture of the RCN model.
\end{rem}

\subsection{Computing an Upper Bound on the Random Classification Noise Rate}\label{sec:rcn-upper-bound}
Many results in noise frameworks rely on the idea that the learning algorithm has access to some non-trivial upper bound on the noise rate $\eta$ that affects the generated data. This was the situation in what was described in Theorem~\ref{thm:RCN-ERM-sample-complexity}, regarding the statistical complexity of ERM under RCN.

Having said that, it turns out that knowing an upper bound $\eta_b$ is not a prerequisite in the case of RCN when realizability holds~\citep{PAC:Noise:Angluin,Book:Laird}. Instead, 
one can calculate a non-trivial upper bound $\eta_b$ by merely requesting a sample of appropriate size from the noisy oracle \rcnoracle.
Following our discussion from Section~\ref{sec:RCN-ordering-of-hypotheses}, 
the idea is that since $h^\varheartsuit = c$ attains the minimum true disagreement rate $\arisk{\rcnoracle}{h^\varheartsuit}$ and in fact it holds that $\expected{\aemprisk{\sample}{h^\varheartsuit}} = \arisk{\rcnoracle}{h^\varheartsuit} = \eta$ on a sample \sample drawn i.i.d.~from \rcnoracle, it follows that if we have access to a sufficiently large dataset, then we can provide a non-trivial upper bound $\eta_b$ for the true noise rate $\eta$ that is governing the data-generating process inside the \rcnoracle oracle.

In order to provide such an upper bound we follow a guess-and-verify strategy that successively tries the values $1/4, 3/8, 7/16, \ldots$. 
In other words we look at the sequence 
\begin{displaymath}
    \eta_b^{(r)} = \frac{1}{2} - 2^{-(r+1)}\,,
\end{displaymath}
for $r = 1, 2, \ldots$, and when the smallest $r$ can provide us a candidate $\eta_b$ that can be justified as a safe upper bound for the noise rate $\eta$, we stop. 
Here the word ``safe'' is used in the sense that a certain relationship needs to hold (with high probability) given the number of examples that we have drawn i.i.d.~from the faulty oracle \rcnoracle in order to verify if our guess was correct.

Algorithm~\ref{algo:noise:upperbound} has the details for finding an upper bound $\eta_b$ within the RCN model.

\begin{algorithm}[ht]\caption{Calculating an upper bound $\eta_b$ on the RCN rate $\eta$}\label{algo:noise:upperbound}
Initialize $\hat{\eta}_b \assign 1/4$ and $r
\assign 1$;

\Repeat{the halt condition is satisfied}{
    Request a sample $\sample_r$ that has $m_r \ge 2^{2r + 3}\ln\left(\frac{\abs{\Hh}\cdot 2^{r+1}}{\delta}\right)$ examples;
    
    For each $h_i\in\Hh$, calculate the disagreement rate $\aemprisk{\sample_r}{h_i}$ of $h_i$ in the sample $\sample_r$
    
    Let $\widehat{R}_{\min} := \min_{h_{i}\in\Hh }\{\aemprisk{\sample_r}{h_i}\}$ be the minimum such value;
    
    \lIf{$\widehat{R}_{\min} < \hat{\eta}_b - 2^{-(r+2)}$}{
        halt and output $\hat{\eta}_b$\label{algoline:noise:upperbound:stopping-criterion}
    }
    \lElse{
        $r \assign r + 1$
        and 
        $\hat{\eta}_b \assign 1/2 - 2^{-(r+1)}$
    }
}
\end{algorithm}

\begin{theo}[\citet{PAC:Noise:Angluin}; \citet{Book:Laird}]\label{thm:upper-al:iterations-and-sample-size}
Let \Hh be a hypothesis class such that $\abs{\Hh} < \infty$ and $\Cc \subseteq \Hh$.
Let $\eta < 1/2$ be the true RCN rate.
With probability greater than $1-\delta$, 
Algorithm~\ref{algo:noise:upperbound} will terminate within $1 + \left\lceil \lg\left(1 - 2\eta\right)^{-1}\right\rceil$ iterations
and 
the total sample size for the entire process of estimating the upper bound $\eta_b$ is
$\OO{(1-2\eta)^{-2}\cdot\ln\left(\abs{\Hh}\delta^{-1}(1-2\eta)^{-1}\right)}$. 
The result is information-theoretic unless $\Hh$ has an efficient ERM routine.
\end{theo}
\begin{proof}
At iteration $r$, with $r\in\set{1, 2, \ldots }$, of Algorithm~\ref{algo:noise:upperbound} we 
are approximating the empirical disagreement rate of every hypothesis $h\in\Hh$ within $\upxi_{r}$ of their true value and the candidate upper bound for noise is set to be $\hat{\eta}_b^{(r)}$ as shown below:
\begin{equation}\label{eq:upper-al:parameters}
\left\{
\begin{array}{rcl}
    \hat{\eta}_b^{(r)} &=& 1/2 - 2^{-(r+1)} \\[0.1em]
    \upxi_{r} &=& 2^{-(r+2)}
\end{array}
\right.\,.
\end{equation}

\paragraph*{Correctness of the Claim.}

For a moment assume that a good event holds where all the empirical estimates $\aemprisk{\sample_{r}}{h_i}$ for all $h_i\in\Hh$ are calculated within $\upxi_r$ of their true values up to and including some iteration $r_{\max}$ which will be specified soon.

An important claim is that as long as $\hat{\eta}_b^{(r)} < \eta$, 
Algorithm~\ref{algo:noise:upperbound} is operating at some iteration $r < r_{\max}$ where, for all $h_i\in\Hh$, it holds that 
\begin{equation}\label{eq:upper-al:proof-1}
\aemprisk{\sample_r}{h_i} \ge \hat{\eta}_b^{(r)} - 2^{-(r+2)}\,.
\end{equation}

To see this, observe the following.
By assumption, we have that $\hat{\eta}_b^{(r)} < \eta$ and moreover $\aemprisk{\sample_r}{h_i}$ is calculated within $\upxi_r$ of its true value $\arisk{\rcnoracle}{h_i}$. 
Furthermore, recall by the discussion in Section~\ref{sec:RCN-ordering-of-hypotheses} that  regarding the hypothesis $h^{\varheartsuit} \in\arginf_{h\in\Hh}\arisk{\rcnoracle}{h}$, it holds $\arisk{\rcnoracle}{h^{\varheartsuit}} = \eta$. 
Hence, for all $h_i\in\Hh$, including $h^\varheartsuit$, the following holds:
\begin{equation}\label{eq:premature-upper-bound}
\aemprisk{\sample_r}{h_i} 
>  \arisk{\rcnoracle}{h_i} - \upxi_r 
\ge  \arisk{\rcnoracle}{h^{\varheartsuit}} - \upxi_r 
= \eta - 2^{-(r+2)}\,.
\end{equation}
In other words, as long as our guess $\hat{\eta}_b^{(r)}$ is less than the true noise rate $\eta$ and all the empirical estimates for the hypotheses in our hypothesis space are calculated within $\upxi_r = 2^{-(r+2)}$ of their true value, then 
by (\ref{eq:premature-upper-bound})
we have that 
$$\aemprisk{\sample_r}{h_i} > \eta - 2^{-(r+2)} > \hat{\eta}_b^{(r)} - 2^{-(r+2)}$$
and therefore
Line~\ref{algoline:noise:upperbound:stopping-criterion} in Algorithm~\ref{algo:noise:upperbound} will \emph{never be satisfied} and hence the algorithm will not stop prematurely.

Now we want to claim that the algorithm will terminate at some iteration $r_{\max}$ that requires fairly reasonable amount of time.
Note that Line~\ref{algoline:noise:upperbound:stopping-criterion} in Algorithm~\ref{algo:noise:upperbound} will be satisfied if there exists at least one hypothesis for which it holds
\begin{displaymath}
    \aemprisk{\sample_r}{h_i} < \hat{\eta}_b^{(r)} - \upxi_r\,.
\end{displaymath}
A simple upper bound on $r_{\max}$ can be obtained by arguing about the hypothesis $h^\varheartsuit = c$.
Assuming the good event holds where all the empirical estimates $\aemprisk{\sample_{r}}{h_i}$ for all $h_i\in\Hh$ are calculated within $\upxi_r$ of their true values up to and including some iteration $r_{\max}$, then for the hypothesis $h^\varheartsuit$ we have:
\begin{displaymath}
\aemprisk{\sample_r}{h^\varheartsuit} 
<
\arisk{\rcnoracle}{h^\varheartsuit} + \upxi_r = \eta + \upxi_r\,.
\end{displaymath}
Thus, it is enough if there exists some $r$ such that
\begin{equation}
\eta + \upxi_r \le \hat{\eta}_b^{(r)} - \upxi_r
\Longleftrightarrow
\eta + 2\upxi_r \le \hat{\eta}_b^{(r)}\,.
\end{equation}
Substituting the parameters with the values shown in (\ref{eq:upper-al:parameters}), it is equivalent if we require:
\begin{displaymath}
\eta + 2\cdot 2^{-(r+2)} \le 1/2 - 2^{-(r+1)}
\Longleftrightarrow
2^{-r} \le \frac{1-2\eta}{2}
\Longleftrightarrow
1-r \le \lg\left(1-2\eta\right)\,.
\end{displaymath}
In other words, assuming the good event holds where all the empirical estimates 
$\aemprisk{\sample_{r}}{h_i}$ 
for all $h_i\in\Hh$ are calculated within $\upxi_r$ of their true values for all iterations 
$r\in\set{1, 2, \ldots, r_{\max}}$, then  Algorithm~\ref{algo:noise:upperbound} will not run for more than 
\begin{equation}\label{eq:upper-al:iterations-upper-bound}
r_{\max} = 1 + \left\lceil\lg\left(\frac{1}{1-2\eta}\right)\right\rceil\,
\end{equation}
iterations, as required.

\paragraph*{Statistical Complexity Required to Support the Claim Per Iteration.}
The problem with (\ref{eq:upper-al:iterations-upper-bound}) is that since we do not know $\eta$, we cannot possibly calculate $r_{\max}$.  Hence, we need a way of distributing the overall failure probability $\delta$ for the bad events for arbitrarily many iterations $r_{\max}$.  The way out of these situations is to use a sequence $\{\upalpha_j\}_{j=1}^{\infty}$ of positive real values $\upalpha_j$ that add up to 1 so that we can allow failure probability at iteration $j$ equal to $\delta^{(j)} = \upalpha_j \delta$.  The two most popular sequences are
\begin{displaymath}
\upalpha_j = 2^{-j} 
\ \ \ \ \ \ \ \ \mbox{and}  \ \ \ \ \ \ \ \ 
\upalpha_j = \dfrac{6}{\pi^2 j^2},
\end{displaymath}
both defined for $j = 1, 2, \ldots$.
\citet{PAC:Noise:Angluin} as well as \citet{Book:Laird} are using the first sequence of the two and this is what we will also be using. 
In addition to using the sequence $\{\upalpha_j\}_{j=1}^{\infty}$ we will also be dividing the failure probability by $\abs{\Hh}$ in every iteration as we want the guarantee of the good approximation to hold for all the hypotheses in the hypothesis space \Hh.

\smallskip

In round $r$ we use Hoeffding's bound (Proposition~\ref{prop:Hoeffding}) with $\alpha = 0, \beta = 1, k = m_r$. 
Thus, it will be the case that
\begin{equation}\label{eq:upper-al:proof-bad-event}
\abs{\aemprisk{\sample_r}{h_i} - \arisk{\rcnoracle}{h_i}} \ge \upxi_r 
\end{equation}
holds for every hypothesis $h_i\in\Hh$ with probability 
\begin{equation}\label{eq:upper-al:proof-bad-event-failure}
p_{i, r}^{\textsl{fail}} \le 2\naturalexp{-2m_r\upxi_r^2}
\end{equation}
We now require for the per-hypothesis failure probability to hold 
$2\naturalexp{-2m_r\upxi_r^2} \le \dfrac{\delta^{(r)}}{\abs{\Hh}} = \dfrac{\delta}{2^r\abs{\Hh}}$. 
By (\ref{eq:upper-al:parameters}) we have $\upxi_r = 2^{-(r+2)}$ and therefore the last equation gives us
\begin{equation}\label{eq:upper-al:sample-complexity-round-r}
m_r \ge 
2^{2r+3}\ln\left(\dfrac{2^{r+1}\abs{\Hh}}{\delta}\right)\,.
\end{equation}

\paragraph*{Bounding the Failure Probability.}
We want to make sure that no bad event will happen.  That is, every hypothesis in any round up to and including $r_{\max}$ is approximated within $\upxi_r$ of its true value.

For $i\in\set{1, \ldots, \abs{\Hh}}$ and $r\in\set{1, \ldots, r_{\max}}$, let $B_{i, r}$ be the bad event 
\begin{displaymath}
    B_{i, r} \equiv \mbox{Equation (\ref{eq:upper-al:proof-bad-event}) holds for hypothesis $h_i$ at iteration $r$}\,.
\end{displaymath}

Then, some bad event $B_{i, r}$ will be true with probability
\begin{eqnarray}
\myprob{\bigcup_{r=1}^{r_{\max}} \bigcup_{i=1}^{\abs{\Hh}} B_{i, r}} 
&\le&  
\sum_{r=1}^{r_{\max}} \sum_{i=1}^{\abs{\Hh}} \myprob{\abs{\aemprisk{\sample_r}{h_i} - \arisk{\rcnoracle}{h_i}} \ge \upxi_r} \nonumber\\
&\le& 
\sum_{r=1}^{r_{\max}}\sum_{i=1}^{\abs{\Hh}} \dfrac{\delta^{(r)}}{\abs{\Hh}} \nonumber\\
&=& 
\sum_{r=1}^{r_{\max}}\sum_{i=1}^{\abs{\Hh}} \dfrac{\delta}{2^r\abs{\Hh}} \nonumber\\
&=& 
\sum_{r=1}^{r_{\max}} \dfrac{\delta}{2^r} \nonumber\\
&<& 
\delta\sum_{r=1}^{\infty} \dfrac{1}{2^r} 
\nonumber\\
&=& 
\delta\,.\nonumber
\end{eqnarray}
In other words, we have that 
\begin{displaymath}
\myprob{\bigcup_{r=1}^{r_{\max}} \bigcup_{i=1}^{\abs{\Hh}} B_{i, r}} < \delta\,.
\end{displaymath}

\paragraph*{Statistical Complexity for the Entire Process.}
Assuming all the empirical estimates are calculated within $\upxi_r$ of their true values for every hypothesis in every round, by (\ref{eq:upper-al:iterations-upper-bound}) then 
Algorithm~\ref{algo:noise:upperbound} will return a solution in at most $r_{\max} = 1 + \left\lceil\lg\left(\frac{1}{1-2\eta}\right)\right\rceil$ iterations.
Therefore it holds that
\begin{displaymath}
2^{r_{\max}} \le 2^{1 + \left\lceil \lg\left(\frac{1}{1-2\eta}\right)\right\rceil}
\le 2^{2 + \lg\left(\frac{1}{1-2\eta}\right)}
= \frac{4}{1 - 2\eta}\,.
\end{displaymath}
Furthermore, by (\ref{eq:upper-al:sample-complexity-round-r}) we have that $m_r \le \left\lceil 2^{2r+3}\ln\left(2^{r+1}\abs{\Hh}/\delta\right)\right\rceil$ samples are enough.
Hence, we have 
\begin{eqnarray}
\sum_{r=1}^{r_{\max}} m_r 
&\le& 
\sum_{r=1}^{r_{\max}}
\left\lceil 2^{2r+3}\ln\left(\dfrac{2^{r+1}\abs{\Hh}}{\delta}\right)\right\rceil \nonumber\\
&\le& 
\sum_{r=1}^{r_{\max}} \left[1 + 2^{2r+3}\ln\left(\dfrac{2^{r+1}\abs{\Hh}}{\delta}\right)\right] \nonumber\\
&\le & \sum_{r=1}^{r_{\max}} 1 + 
8\sum_{r=1}^{r_{\max}} 4^r\ln\left(\dfrac{2\cdot 2^r\cdot \abs{\Hh}}{\delta}\right)\nonumber\\
&\le & r_{\max} + 
8\ln\left(\dfrac{2\cdot 2^{r_{\max}}\cdot \abs{\Hh}}{\delta}\right)\sum_{r=1}^{r_{\max}} 4^r\nonumber\\
&\le & r_{\max} + 
8\ln\left(\dfrac{8\abs{\Hh}}{(1-2\eta)\delta}\right)\sum_{r=1}^{r_{\max}} 4^r\nonumber\\
&= & r_{\max} + 
8\ln\left(\dfrac{8\abs{\Hh}}{(1-2\eta)\delta}\right)\cdot\left(\dfrac{4^{1+r_{\max}}-4}{4-1}\right)\nonumber\\
&< & r_{\max} + 
8\ln\left(\dfrac{8\abs{\Hh}}{(1-2\eta)\delta}\right)\cdot\left(\dfrac{4^{1+r_{\max}}}{3}\right)\nonumber\\
&= & r_{\max} + 
\dfrac{32}{3}\ln\left(\dfrac{8\abs{\Hh}}{(1-2\eta)\delta}\right)\cdot 4^{r_{\max}}\nonumber\\
&= & r_{\max} + 
\dfrac{32}{3}\ln\left(\dfrac{8\abs{\Hh}}{(1-2\eta)\delta}\right)\cdot \left(2^{r_{\max}}\right)^2\nonumber\\
&\le & r_{\max} + 
\dfrac{32}{3}\ln\left(\dfrac{8\abs{\Hh}}{(1-2\eta)\delta}\right)\cdot \left(\dfrac{4}{1-2\eta}\right)^2\nonumber
\end{eqnarray}
Therefore, the total sample complexity of Algorithm~\ref{algo:noise:upperbound} when it terminates, which happens with probability greater than $1-\delta$, is 
$\OO{\dfrac{1}{(1-2\eta)^{2}}\ln\left(\dfrac{\abs{\Hh}}{(1-2\eta)\delta}\right)}$.
\end{proof}

\begin{rem}
Note that the algorithm may return an incorrect upper bound $\eta_b$ for the true noise rate $\eta$ if the disagreement rate of at least some hypothesis $h_i\in\Hh$ is grossly underestimated in some round that would falsely make the criterion of  Line~\ref{algoline:noise:upperbound:stopping-criterion} satisfied.
\end{rem}

\begin{rem}
The event that Algorithm~\ref{algo:noise:upperbound} runs forever has probability zero.
For example, because all the estimates for the true disagreement rates are overestimated in every round and hence the criterion of Line~\ref{algoline:noise:upperbound:stopping-criterion} is never satisfied so that the algorithm would terminate.
\end{rem}

Another remark that is subtle but is perhaps worth mentioning is the following.
\begin{rem} 
Regarding the sample complexity $m_r$ that we calculated in (\ref{eq:upper-al:sample-complexity-round-r}), 
in the proof of~\citet{PAC:Noise:Angluin} we find an additional factor of 2 inside the logarithm.  
This is done because in their proof they split evenly the total failure probability $\delta$ into two separate bad events that may happen:
\begin{enumerate}[label=(\arabic*),leftmargin=1.0cm]
    \item one bad event is about the probabilistic procedure that calculates the empirical estimates of the disagreement rates of all the hypotheses $h_i\in\Hh$ during rounds $r\in\set{1, \ldots, r_{\max}}$ no matter how large $r_{\max}$ is, and
    
    \item the other bad event is related to algorithmic reasoning regarding the validity of the upper bound $\hat{\eta}_b$ that is returned when the algorithm in fact terminates within the first $r_{\max}$ iterations.
\end{enumerate}
In other words, the second bad event is arguing that if the algorithm stops up until round $r_{\max}$, then with probability at least $1-\delta/2$ it holds that $\eta_b$ is indeed a valid upper bound for the noise rate $\eta$. 
However, if all the empirical estimates are calculated as expected within $\upxi_r$ of their true values in each round $r$, then as we argued earlier in order to derive (\ref{eq:upper-al:proof-1}), it can never logically be the case that the algorithm will stop and output an upper bound $\hat{\eta}_b$ for which it holds $\hat{\eta}_b < \eta$. 
Furthermore, under the same assumption, we also derived the maximum number of iterations $r_{\max}$ in (\ref{eq:upper-al:iterations-upper-bound}) that Algorithm~\ref{algo:noise:upperbound} would run in order to provide an upper bound.

As a summary, I understand that 
the proof of Angluin and Laird was arguing 
on one hand about the probabilistic portion of their proof (Hoeffding bound plus union bound) and on the other hand about the logical portion of their proof (termination criterion plus correctness of output), where each part would receive their own failure probability $\delta/2$.
However, 
the analysis that we provided here is pretty much standard and therefore nowadays 
I would not expect to see such a distinction in the argument anymore. 
Thus, in~(\ref{eq:upper-al:sample-complexity-round-r}) we calculate a slightly smaller sample size compared to their proof (but still asymptotically the same).
\end{rem}

Before we conclude this section we want to make one last remark that is not explicitly stated in~\citet{PAC:Noise:Angluin}.
Namely, one can also prove the following proposition.
\begin{prop}\label{prop:upper-al:ineq}
Upon successful execution of Algorithm~\ref{algo:noise:upperbound} (which will happen with probability greater than $1-\delta$), 
regarding the upper bound $\eta_b$ that is returned, it holds:
$\frac{1}{2}-\eta_{b} \geq \frac{1}{4} \left(\frac{1}{2}-\eta\right)$. 
\end{prop}
\begin{proof}
For $r \le 1 + \left\lceil \lg\left(1 - 2\eta\right)^{-1}\right\rceil$ we have that 
$2^r \le 2^{1 + \left\lceil \lg\left(\frac{1}{1-2\eta}\right)\right\rceil}
\le 2^{2 + \lg\left(\frac{1}{1-2\eta}\right)}
= \frac{4}{1 - 2\eta}$. 
By Algorithm~\ref{algo:noise:upperbound} we have $\frac{1}{2} - \eta_b = 2 ^{-(r+1)}$.
Hence, 
$\frac{1}{2} - \eta_b = 2 ^{-(r+1)} \geq \frac{1}{2}\cdot\frac{1 - 2\eta}{4} =\frac{1}{4}\left(\frac{1}{2} - \eta\right)$.
\end{proof}

\begin{rem}\label{rem:RCN-ERM-sample-complexity-using-AL-upper-bound}
The important point here is that for the particular bound $\eta_b$ that is computed by Algorithm~\ref{algo:noise:upperbound}, due to Proposition~\ref{prop:upper-al:ineq} it follows that $1-2\eta_b = \Omega\left(1 - 2\eta\right)$, where $\eta$ is the true noise rate $\eta$. 
Therefore, the total sample size required for calculating an ERM solution for Theorem~\ref{thm:RCN-ERM-sample-complexity} in Section~\ref{sec:RCN-ERM-sample-complexity} will now be upper bounded \emph{directly} by the unknown noise rate $\eta$! In other words, if we use the upper bound $\eta_b$ of Algorithm~\ref{algo:noise:upperbound}, the sample complexity of Theorem~\ref{thm:RCN-ERM-sample-complexity} becomes 
\OO{\varepsilon^{-2}(1-2\eta)^{-2}\ln{\left(\abs{\Hh}/\delta\right)}} even if we do not really know the exact noise rate $\eta$.
\end{rem}

\subsection{On the Computational Efficiency of  Proper Learning Algorithms that Rely on the Empirical Risk Minimization Principle}
\label{sec:ERM-Noise-Computation}

Regarding the computational efficiency of proper learning algorithms that rely on minimizing disagreements against a sample obtained by an \rcnoracle oracle, in turn below we see an easy and a difficult case.

\subsubsection{An Easy Case for ERM: Symmetric Boolean Functions}\label{sec:RCN-ERM-SymmetricBoolean}
We will now give a simple and efficient application of the ERM principle in order to learn the concept class of \emph{symmetric Boolean functions}.
This is the class of functions $\Cc$ such that 
for a function $c \colon \set{0, 1}^n \rightarrow \set{0, 1}$ from that class it holds that 
$c(\bm{x}) = c(\pi(\bm{x}))$, where $\pi(\bm{x})$ is a permutation of the attribute values in $\bm{x}$.
In other words, a Boolean function $c$ is \emph{symmetric} if its value is uniquely determined by the number of 1's in the input.  
Furthermore, 
let us assume that we want to learn such functions properly; i.e.,~$\Hh = \Cc$.

Note that any hypothesis $h\in \Hh$ can be represented by a vector $\left(y_0, y_1, \ldots, y_n\right)^{T}\in\set{0, 1}^{n+1}$,
where $y_j$ is the value of $h$ on examples having precisely $j$ 1's.  Assume that we are using this representation class for \Hh so that we can perform proper learning as promised.  Note that $\abs{\Hh} = 2^{n+1}$.

\begin{theo}[Efficient ERM of Symmetric Boolean Functions] 
The ERM principle can be applied in a computationally efficient manner on a sample obtained from a random classification oracle \rcnoracle using a proper representation for symmetric Boolean functions. 
Therefore, the class of symmetric Boolean functions is properly efficiently PAC-learnable under RCN.
\end{theo}
\begin{proof}
We split the proof into two parts.

\paragraph*{An Efficient Proper ERM Learning Algorithm.}
%
We provide the following learning algorithm that operates using the ERM principle.
First of all, partition the sample \sample into $n+1$ disjoint subsets $\sample_j$, with $j\in\set{0, \ldots, n}$,
such that $\sample_j$ is the subset of examples in \sample that have exactly $j$ 1's.
In other words, we want the following to hold:
\begin{equation}\label{eq:erm-symmetric-proof-1}
\sample = \bigcup_{j=0}^{n} \sample_j 
\quad
\mbox{and}
\quad
\sample_j \cap \sample_k = \emptyset 
\quad
\mbox{for all $j \neq k$}, \mbox{with } j, k \in \set{0, \ldots, n}\,.
\end{equation}
Furthermore, define $\#0(\sigma)$ and $\#1(\sigma)$ to count the number of labels that correspond to 0's and 1's respectively in some sample $\sigma$.
Then, for any sample \sample 
we formulate the hypothesis $h^{\star}$ 
using the vector representation mentioned above and in particular we set its 
$j$-th entry to be
$\charfunc{\#1(\sample_j) > \#0(\sample_j)}$. 
In other  words, by (\ref{eq:erm-symmetric-proof-1}), we formulate the hypothesis $h^{\star}$ such that:
\begin{equation}\label{eq:erm-symmetric-proof-2}
h^{\star} = 
\begin{pmatrix}
\charfunc{\#1(\sample_0) > \#0(\sample_0)}
\\
\vdots
\\
\charfunc{\#1(\sample_j) > \#0(\sample_j)}
\\
\vdots
\\
\charfunc{\#1(\sample_n) > \#0(\sample_n)}
\end{pmatrix}\,.
\end{equation}
The claim now is that 
no other function $h\in\Hh$ can have a smaller disagreement rate  $\aemprisk{\sample}{h}$ than $\aemprisk{\sample}{h^{\star}}$ for the hypothesis $h^{\star}$ defined above and hence $h^{\star}$ is the ERM solution for the sample $\sample$ that was given to the learning algorithm.
This claim is straightforward and is left as an exercise to the reader.
Furthermore, notice that one can perform one pass over $\sample$ and by maintaining $2(n+1)$ counters can calculate the required quantities in~(\ref{eq:erm-symmetric-proof-2}).

\paragraph*{Efficient PAC-Learnability.}
By the above, we can calculate the ERM solution on a sample \sample by performing a single pass over \sample.  Therefore, all that we really need to show is that we can have a sample of polynomial size so that the information-theoretic portion of Definition~\ref{def:pac-label-noise} is satisfied for the case of RCN and it will automatically follow that the above learning algorithm will calculate an efficient PAC solution by applying the ERM principle.

Given on one hand an upper bound $\eta_b$ on the noise rate $\eta$, and on the other hand that for this vector representation of the functions we have $\abs{\Hh} = 2^{n+1}$, by Theorem~\ref{thm:RCN-ERM-sample-complexity} it follows that having access to 
\begin{displaymath}
m = \left\lceil2\varepsilon^{-2}(1-2\eta_{b})^{-2}\ln{\left(2^{n+2}/\delta\right)}\right\rceil\in\OO{n\varepsilon^{-2}(1-2\eta_b)^{-2}\ln\left(1/\delta\right)}
\end{displaymath}
examples from the \rcnoracle oracle 
we can formulate efficiently a hypothesis $h^{\star}$ 
using the ERM principle such that $h^{\star}$ satisfies Definition~\ref{def:pac-label-noise} for RCN. 
This sample has size polynomial w.r.t.~the input parameters for our algorithm as required.

\smallskip

Of course, because of Theorem~\ref{thm:upper-al:iterations-and-sample-size} in Section~\ref{sec:rcn-upper-bound} the learning algorithm does not need to be given an upper bound $\eta_b$ as it can be calculated using 
$\OO{n(1-2\eta)^{-2}\cdot\ln\left(\delta^{-1}(1-2\eta)^{-1}\right)}$ examples and then apply the analysis above.
Therefore, if we also use Remark~\ref{rem:RCN-ERM-sample-complexity-using-AL-upper-bound}, it follows that the overall sample complexity in this case is
$\OO{n(1-2\eta)^{-2}\left(\varepsilon^{-2}\ln{\left(\delta^{-1}\right)} + \ln\left(\delta^{-1}(1-2\eta)^{-1}\right)\right)}$.
\end{proof}

\subsubsection{A Difficult Case for ERM: Monotone Conjunctions}
We will now see that applying the 
ERM
principle using a proper learning algorithm, can in general be computationally hard to compute.  In this direction we will use the fact that computing the \emph{vertex cover} (decision version) of a graph $G = (V, E)$ is in general NP-Complete~\citep{Book:NPComplete}.  In particular, we will reduce the decision version of the vertex cover problem of having a solution with at most $k$ vertices to the problem of finding a hypothesis $h\in\Hh$ that has at most $k$ disagreements with a given sample.  As a consequence, the ERM problem is NP-hard.

\begin{defi}[Decision Problem for Vertex Cover]
Let $k \le n$. Let $G = (V, E)$ be an undirected graph on $n$ vertices. The question is if there is a set of vertices $C$, such that $\abs{C} \le k$, in $G$, such that every edge from $E$ is incident to at least one vertex in $C$. Then, $C$ is called \emph{vertex cover}.
\end{defi}

We are now ready to prove the theorem.

\begin{theo}[\citet{PAC:Noise:Angluin}]
Finding monotone conjunctions which minimize disagreements with a given sample is NP-hard.
\end{theo}




\begin{proof}
The proof is split into the following components: the reduction that we will use, the claim that we will prove, as well as the two directions of the proof of the claim.

\paragraph*{Reduction.}
Given a graph $G = (V, E)$, that has $\abs{V} = n$ vertices, we will use the 
encoding:
\begin{itemize}[leftmargin=1.0cm]
    \item For every vertex of the graph we will create a positive example. In particular, vertex $v_i$ will be represented by the all 1's vector (in $n$ dimensions) that has a single 0 in its $i$-th entry.
    \item For every edge of the graph we will create $(n+1)$ copies of the same negative example. In particular, edge $e_{ij} = (v_i, v_j)$ will be represented by the all 1's vector (in $n$ dimensions) that has 0's in its $i$-th and $j$-th entry.
\end{itemize}
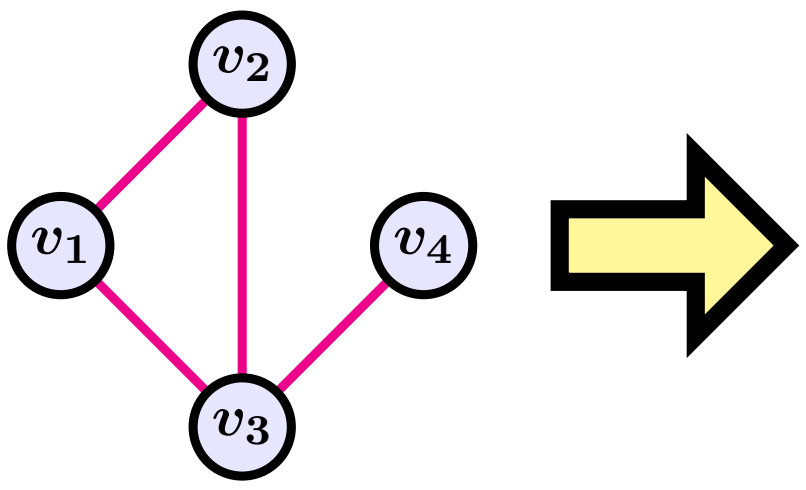
Figure~\ref{fig:ERM-Hardness:reduction} provides an example for a graph with four vertices and four edges.

\begin{figure}[ht]
\centering
\psset{unit=0.6cm}
\begin{pspicture*}(-0.7,-2.5)(23,6.7)
\psline[linecolor=magenta,linewidth=0.1]{-}(0,1.8)(2,3.8)
\psline[linecolor=magenta,linewidth=0.1]{-}(0,1.8)(2,-0.2)
\psline[linecolor=magenta,linewidth=0.1]{-}(2,-0.2)(2,3.8)
\psline[linecolor=magenta,linewidth=0.1]{-}(2,-0.2)(4,1.8)
%
\pscircle[linecolor=black,linewidth=0.1,fillcolor=blue!10](0,1.8){0.6}
\pscircle[linecolor=black,linewidth=0.1,fillcolor=blue!10](2,3.8){0.6}
\pscircle[linecolor=black,linewidth=0.1,fillcolor=blue!10](2,-0.2){0.6}
\pscircle[linecolor=black,linewidth=0.1,fillcolor=blue!10](4,1.8){0.6}
\rput(0,1.8){\textcolor{black}{\bm{$v_1$}}}
\rput(2,3.8){\textcolor{black}{\bm{$v_2$}}}
\rput(2,-0.2){\textcolor{black}{\bm{$v_3$}}}
\rput(4,1.8){\textcolor{black}{\bm{$v_4$}}}
\pspolygon[linecolor=black,linewidth=0.2,fillcolor=yellow!50](5.5,2.2)(7,2.2)(7,2.8)(8,1.8)(7,0.8)(7,1.4)(5.5,1.4)
\rput[l](8.5,6){\scriptsize{\textcolor{blue!70}{\bm{$\left((0, 1, 1, 1)^T, 1\right)$}}}}
\rput[l](8.5,5.2){\scriptsize{\textcolor{blue!70}{\bm{$\left((1, 0, 1, 1)^T, 1\right)$}}}}
\rput[l](8.5,4.4){\scriptsize{\textcolor{blue!70}{\bm{$\left((1, 1, 0, 1)^T, 1\right)$}}}}
\rput[l](8.5,3.6){\scriptsize{\textcolor{blue!70}{\bm{$\left((1, 1, 1, 0)^T, 1\right)$}}}}
%
\rput[l](14,6){\scriptsize{\textcolor{magenta}{\bm{$\left((0, 0, 1, 1)^T, 0\right)$}}}}
\rput[l](14,5.2){\scriptsize{\textcolor{magenta}{\bm{$\left((0, 0, 1, 1)^T, 0\right)$}}}}
\rput[l](14,4.4){\scriptsize{\textcolor{magenta}{\bm{$\left((0, 0, 1, 1)^T, 0\right)$}}}}
\rput[l](14,3.6){\scriptsize{\textcolor{magenta}{\bm{$\left((0, 0, 1, 1)^T, 0\right)$}}}}
\rput[l](14,2.8){\scriptsize{\textcolor{magenta}{\bm{$\left((0, 0, 1, 1)^T, 0\right)$}}}}
\rput[l](19,6){\scriptsize{\textcolor{magenta}{\bm{$\left((0, 1, 0, 1)^T, 0\right)$}}}}
\rput[l](19,5.2){\scriptsize{\textcolor{magenta}{\bm{$\left((0, 1, 0, 1)^T, 0\right)$}}}}
\rput[l](19,4.4){\scriptsize{\textcolor{magenta}{\bm{$\left((0, 1, 0, 1)^T, 0\right)$}}}}
\rput[l](19,3.6){\scriptsize{\textcolor{magenta}{\bm{$\left((0, 1, 0, 1)^T, 0\right)$}}}}
\rput[l](19,2.8){\scriptsize{\textcolor{magenta}{\bm{$\left((0, 1, 0, 1)^T, 0\right)$}}}}
\rput[l](14,1.5){\scriptsize{\textcolor{magenta}{\bm{$\left((1, 0, 0, 1)^T, 0\right)$}}}}
\rput[l](14,0.7){\scriptsize{\textcolor{magenta}{\bm{$\left((1, 0, 0, 1)^T, 0\right)$}}}}
\rput[l](14,-0.1){\scriptsize{\textcolor{magenta}{\bm{$\left((1, 0, 0, 1)^T, 0\right)$}}}}
\rput[l](14,-0.9){\scriptsize{\textcolor{magenta}{\bm{$\left((1, 0, 0, 1)^T, 0\right)$}}}}
\rput[l](14,-1.7){\scriptsize{\textcolor{magenta}{\bm{$\left((1, 0, 0, 1)^T, 0\right)$}}}}
\rput[l](19,1.5){\scriptsize{\textcolor{magenta}{\bm{$\left((1, 1, 0, 0)^T, 0\right)$}}}}
\rput[l](19,0.7){\scriptsize{\textcolor{magenta}{\bm{$\left((1, 1, 0, 0)^T, 0\right)$}}}}
\rput[l](19,-0.1){\scriptsize{\textcolor{magenta}{\bm{$\left((1, 1, 0, 0)^T, 0\right)$}}}}
\rput[l](19,-0.9){\scriptsize{\textcolor{magenta}{\bm{$\left((1, 1, 0, 0)^T, 0\right)$}}}}
\rput[l](19,-1.7){\scriptsize{\textcolor{magenta}{\bm{$\left((1, 1, 0, 0)^T, 0\right)$}}}}
\end{pspicture*}
\caption{Example of reduction from a graph $G=(V, E)$ to a sample \sample of positive and negative training examples over which we would like to find a monotone conjunction that has as few disagreements as possible. The sample \sample that we construct has $n$ positive and $\abs{E}(n+1)$ negative examples.}\label{fig:ERM-Hardness:reduction}
\end{figure}

\paragraph*{Claim.}
The claim that we want to prove now is the following.
\begin{cla}\label{claim:ERM-Hardness:claim-for-proof}
Let $k \le n$.
Then, a graph $G$ has a vertex cover of size at most $k$ \emph{if and only if} there is a monotone conjunction with at most $k$ disagreements in the sample~\sample obtained by the reduction.
\end{cla}

\paragraph*{Direction \ensuremath{(\Longrightarrow)}.}
Let $G$ have a vertex cover $C$ of size at most $k$ and consider the hypothesis
\begin{displaymath}
h^{\blacktriangleright} = \bigwedge_{v_j\in C} x_j
\end{displaymath}

\begin{rem}
As a specific example you can think of the graph shown in Figure~\ref{fig:ERM-Hardness:reduction} where for $k=2$ we have the cover $C = \set{v_1, v_3}$. The above transformation is suggesting that we should consider the hypothesis $h^{\blacktriangleright} = x_1 \wedge x_3$.
\end{rem}
We want to show that this particular hypothesis $h^{\blacktriangleright}$ makes at most $k$ mistakes among the training examples. This can be verified with the following observations.

\smallskip

Since every \emph{positive} example $(\bm{x_i}, \posex)$ has exactly one zero in the representation of its instance $\bm{x_i}$ it follows that each such $\bm{x_i}$ can falsify at most one variable in \emph{any} hypothesis $h\in\Hh$. Furthermore, all the instances among the positive examples have a zero at different coordinates with each other.  As a consequence, any hypothesis with $\uplambda$ variables can make at most $\uplambda$ mistakes. Therefore, in particular, $h^{\blacktriangleright}$ can make at most $k$ mistakes among the positive examples; one for each variable corresponding to a node in the cover $C$.

Among the negative examples, the hypothesis $h^{\blacktriangleright}$ makes no mistakes. To see this, we can argue in the following way.  For every node $v_j$ in the cover, $h^{\blacktriangleright}$ includes the corresponding variable $x_j$ among its constraints.  
However, every edge $e_{ij}\in E$ is incident to a node in $C$ and therefore at least one of the two zeros in the corresponding instance of each negative example will falsify one of the constraints in $h^{\blacktriangleright}$.

\smallskip

Hence, if a graph $G$ has a vertex cover of size at most $k$, then there is a monotone conjunction $h\in\Hh$ that disagrees with at most $k$ examples.

\paragraph*{Direction \ensuremath{(\Longleftarrow)}.}
Now we want to show that if there is a monotone conjunction $h^{\blacktriangleleft}$ that disagrees with at most $k \le n$ training examples obtained from the reduction, then we can easily construct a cover of size $k$ in the original graph $G$.

We will do that in the following way.
\begin{itemize}[leftmargin=1.0cm] 
    \item First, we will show that any hypothesis $h\in\Hh$ that disagrees with at most $k\le n$ examples obtained from the reduction, has to have size at most $k$.
    \item Second, we will define a set $C$ of nodes in the original graph $G$ that correspond to the respective variables that appear in $h$ and moreover show that this construction of $C$ makes it a vertex cover in the original graph $G$.
\end{itemize}

   
Since the negative examples in our reduction contain $\abs{E}$ blocks of $(n+1)$ repetitions of the same negative example (that corresponds to some edge in the original graph $G$), it follows that $h^{\blacktriangleleft}$ has to agree with all such negative examples as otherwise it would disagree with at least $n+1 > k$ examples contradicting our assumption that it disagrees with \sample at most $k\le n$ times.
Therefore, $h^{\blacktriangleleft}$ may disagree only with the positive examples in the sample \sample obtained from the reduction.  At the same time, since for every attribute in the instances that correspond to the positive class in \sample there is a unique zero, it follows that \emph{that} particular zero can falsify at most one variable in any given $h\in\Hh$. Therefore, when we consider a hypothesis $h\in\Hh$ that may disagree only with the positive examples, it follows that for every such disagreement there is a \emph{bijection} between a variable that is falsified in $h$ and a particular instance that appears in the examples of the positive class.
Hence, if $h^{\blacktriangleleft}$ makes at most $k\le n$ mistakes it follows that each variable in the representation of $h^{\blacktriangleleft}$ is falsified by exactly one of the instances that appear among the positive examples. As a consequence, $h^{\blacktriangleleft}$ contains at most $k$ variables $x_i$ and this proves the first part that we wanted to prove.

\smallskip

Regarding the second part of the proof, we  
define $C$ to be the set of all those vertices $v_i$ such that $x_i$ appears in $h^{\blacktriangleleft}$ and we want to show that $C$ is a cover in the graph $G$ that has size at most $k$.

By the above argument, $h^{\blacktriangleleft}$ has at most $k$ variables and therefore $\abs{C} \le k$. 

To see that $C$ is indeed a cover note that if $G$ has an edge $e_{ij} = (v_i, v_j)$, then by the construction of the reduction, in the sample \sample there is a block of $(n+1)$ instances that are classified as negative and since $h^{\blacktriangleleft}$ makes at most $k$ mistakes it has to predict correctly that the instances corresponding to that edge are negative.
However, because $h^{\blacktriangleleft}$ makes a correct prediction in such negative examples, it follows that some of the variables $x_i$ or $x_j$ has to appear in $h^{\blacktriangleleft}$ so that it is falsified by one of the two zeros that appear in the instance and thus allow $h^{\blacktriangleleft}$ to make a correct prediction.
But then this means that the set $C$ that we defined contains at least one of the variables $v_i$ or $v_j$ and therefore covers the edge in the original graph $G$.  As a consequence, every edge in the original graph $G$ has to be covered by at least one node that can be found inside $C$.  Put it differently, $C$ is a vertex cover of the original graph as desired.
\end{proof}

\begin{rem}\label{rem:RCN-conjunctions}
The above result indicates that monotone conjunctions are hard to learn properly using the ERM principle on arbitrary labeled instances.  However, one can in fact bypass the application of the ERM principle and learn this class of functions in a computationally efficient manner. 
The proof is a variant of the algorithm that is given by~\citet[Section~5.2]{Book:KearnsVazirani} where it is argued that we can PAC learn general conjunctions by including all those literals that are significant but not harmful when predicting a label using our hypothesis.  In that context, a literal is \emph{significant} if it can be falsified with non-negligible probability and \emph{harmful} if in addition in an example drawn from \oracle, the correct label is positive, thus indicating that such a literal is not present in the representation of the ground truth function and hence should we include it in our hypothesis we may introduce significant excess risk, ergo \emph{harmful}. 
In fact the proof relies on the fact that even if all the literals that do not appear in the ground truth $c$ that are insignificant are retained in $h$, then $h$ will still have low risk.  

Hence, the crucial component in the proof is the identification (with high probability) of all the harmful literals, so that they are removed from $h$.  In turn, this is a variant of the \textsc{Find-S} algorithm discussed by~\citet[Section 2.4]{Book:Mitchell}\footnote{The \textsc{Find-S} algorithm probably appeared first by~\citet{Book:Thinking}. It is also discussed by~\citet[Section 1.3]{Book:KearnsVazirani}.} that PAC learns conjunctions by starting from the conjunction that has all the literals in its representation and then iterates through all the positive examples and removes from the representation of $h$ all those literals that are falsified in such truth assignments that correspond to instances classified as positive.  
However, under RCN such an approach would be erratic as we may start deleting literals that are significant and actually appear in $c$, thus creating a hypothesis in the end that has prohibitively large risk.  
Therefore, the idea is that instead of removing literals based on information that we obtain from single examples, we rather draw a large enough sample and by using statistics on that sample we can decide which literals (or, in the case of monotone conjunctions, variables) need not be included in our hypothesis $h$. 
Hence, with high probability we will remove all the harmful literals and we will formulate a hypothesis that will satisfy the PAC criterion against examples coming from the honest oracle \oracle.
The reader is urged to first read the algorithm from~\citet[Section~5.2]{Book:KearnsVazirani} and then try 
to prove such a result.
\end{rem}

\begin{rem}
As a brief takeaway of the above, NP-hardness of proper ERM is not an impossibility result for PAC learning. A concept class may remain efficiently learnable otherwise; for example, through a non-ERM or an improper learning procedure.
\end{rem}



\begin{rem}
As a closely related remark, agnostically learning monomials using halfspaces is also NP-hard~\citep{Hardness:Agnostic:Monomials}.
\end{rem}

\subsection{How Far Can Empirical Risk Minimization Take Us?}
\label{sec:ERM-Beyond}
We conclude this section with an investigation of learning algorithms that rely on the ERM principle and by making some remarks on poisoning attacks.

\begin{theo}[\citet{PAC:Noise:FourTypes:Sloan}]\label{thm:MCN-ERM-Sloan}
Let $\Hh$ be a finite hypothesis space 
and 
$\Cc$ be a concept class, both over $\XX$, such that $\Cc\subseteq\Hh$. 
Let $0 < \eta_b < 1/2$ be an upper bound on the malicious classification noise rate $\eta$.
If 
\begin{displaymath}
m\geq 2\varepsilon^{-2}(1-2\eta_{b})^{-2}\ln{\left(2\abs{\Hh}/\delta\right)}\,,
\end{displaymath}
then 
for any $h\in\Hh$ that minimizes the disagreements (i.e., $h(\bm{x_{i}})\neq y_i$) with the sequence 
of examples $\sample=\{(\bm{x_{i}},y_i)\}_{i=1}^{m}$ obtained by the noisy oracle \mcnoracle in a single draw (as explained in Definition~\ref{def:pac-label-noise}), it holds that 
\begin{displaymath}
\probsub{\sample\getsample\mcnoracle}{\arisk{\oracle}{h} < \varepsilon} \geq 1-\delta\,.
\end{displaymath}
That is, whenever $\log\abs{\Hh}$ is polynomially bounded in the representation parameters, Definition~\ref{def:pac-label-noise} is satisfied and therefore $\Cc$ is PAC-learnable by $\Hh$ over $\XX$ 
having access to examples from a maliciously noisy oracle \mcnoracle via a method that minimizes disagreements with the sequence of training examples \sample.
\end{theo}
\begin{proof}
We want to argue that the same number of examples that are sufficient for PAC learning under RCN are sufficient also for MCN.

For the proof it is easier to argue if we think in terms of the actual count of the disagreements that the different hypotheses have and for the moment assume that all the examples are generated using an \rcnoracle oracle.
In such a scenario, for a hypothesis $h_i\in\Hh$ the expected number of disagreements that $h_i$ has against such a sample \sample is
\begin{displaymath}
    \expected{\text{\# disagreements of \ensuremath{h_i}}} = 
    m\cdot\arisk{\rcnoracle}{h_i}\,.
\end{displaymath}    
Figure~\ref{fig:ERM-MCN-proof} provides a pictorial representation for this situation.

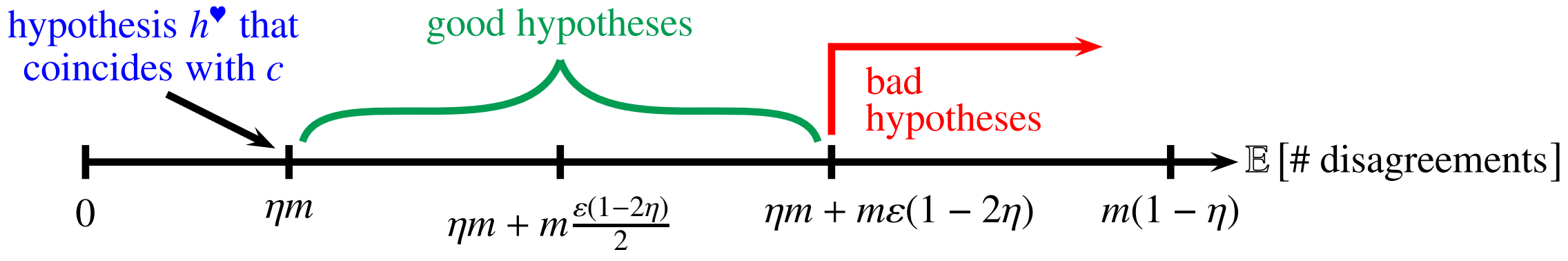
\begin{figure}[ht]
\begin{center}
\psset{unit=0.65cm}
\begin{pspicture*}(-1.2,-1.5)(22,2.4)
%
%
%
%
\psline[linecolor=black,linewidth=2pt]{->}(0,0)(17,0)
\psline[linecolor=black,linewidth=2pt]{-}(0,-0.25)(0,0.25)
\psline[linecolor=black,linewidth=2pt]{-}(16,-0.25)(16,0.25)
\psline[linecolor=black,linewidth=2pt]{-}(3,-0.25)(3,0.25)
\psline[linecolor=black,linewidth=2pt]{-}(7,-0.25)(7,0.25)
\psline[linecolor=black,linewidth=2pt]{-}(11,-0.25)(11,0.25)
%
\rput[l](17.1,0){\textcolor{black}{$\expected{\text{\# disagreements}}$}}
\rput[c](0,-0.75){\textcolor{black}{0}}
\rput[c](16,-0.75){\textcolor{black}{$m(1-\eta)$}}
\rput[c](3,-0.75){\textcolor{black}{$\eta$}$m$}
\rput[c](7,-0.9){\textcolor{black}{$\eta m + m\frac{\varepsilon(1-2\eta)}{2}$}}
\rput[c](12,-0.75){\textcolor{black}{$\eta m + m\varepsilon(1-2\eta)$}}
%
\rput[c](1,2){\textcolor{blue}{hypothesis $h^\varheartsuit$ that}}
\rput[c](1,1.4){\textcolor{blue}{coincides with $c$}}
\rput[c](7,2){\textcolor{MyGreen}{good hypotheses}}
\psline[linecolor=black,linewidth=2pt]{->}(1.2,1.0)(2.8,0.2)
\psbezier[linecolor=MyGreen,linewidth=2pt,showpoints=false]{-}(3.2,0.3)(3.5,1.3)(6.5,0.1)(7,1.5)
\psbezier[linecolor=MyGreen,linewidth=2pt,showpoints=false]{-}(7,1.5)(7.5,0.1)(10.5,1.3)(10.8,0.3)
\psline[linewidth=2pt,linecolor=red]{->}(11,0.4)(11,1.7)(15,1.7)
\rput[l](11.5,1.2){\textcolor{red}{bad}}
\rput[l](11.5,0.6){\textcolor{red}{hypotheses}}
\end{pspicture*}
\end{center}
\caption{Schematic representation regarding the relationship of the different hypotheses as far as their expected disagreements are concerned, with respect to $m$ examples obtained from the noisy oracle \rcnoracle that has noise rate $\eta$. 
Note that for the hypothesis $h^{\varheartsuit} = c$ we have 
$\expected{\text{\# disagreements of \ensuremath{h^{\varheartsuit}}}} = \eta m$
under RCN.  As a consequence, should the examples be coming from a malicious oracle, then under MCN it holds that  
$\expected{\text{\# disagreements of \ensuremath{h^{\varheartsuit}}}} \le \eta m$. A hypothesis $h\in\Hh$ is called \emph{bad} if $\arisk{\oracle}{h} \ge \varepsilon$; otherwise, it is called \emph{good}.}\label{fig:ERM-MCN-proof}
\end{figure}

As it was explained in Figure~\ref{fig:ex:noise-application}, in the case of malicious classification noise the adversary is allowed to assign any label on an instance  whenever they have a chance 
(i.e., a tampering opportunity occurs for the respective example; see also Figure~\ref{fig:noise:mechanism}).
The important part in the proof of Theorem~\ref{thm:RCN-ERM-sample-complexity} is that given a sample \sample obtained from the \rcnoracle oracle, with high probability we look at the good event $\mathcal{E}$ where all the empirical estimates 
$\aemprisk{\sample}{h_i}$
of the hypotheses $h_i$ are within $\varepsilon(1-2\eta)/2$ of their true values.  Hence, for the hypothesis $h^{\varheartsuit} = c$ in particular it holds that 
\begin{displaymath}
\eta m - m\frac{\varepsilon(1-2\eta)}{2}
<
\text{\# disagreements of \ensuremath{h^{\varheartsuit}}} 
<
\eta m + m\dfrac{\varepsilon(1-2\eta)}{2}\,.
\end{displaymath}
Moreover, when this good event $\mathcal{E}$ holds, for every  hypothesis $h^{\text{bad}}\in\Hh$ that has true risk $\varepsilon$ or higher (i.e., $\arisk{\oracle}{h^{\text{bad}}} \ge \varepsilon$), we have that 
\begin{displaymath}
\text{\# disagreements of \ensuremath{h^{\text{bad}}}} 
> 
\eta m + m\dfrac{\varepsilon(1-2\eta)}{2}\,.
\end{displaymath}
Therefore, $h^{\varheartsuit}$ makes, with high probability, at least one mistake less than any bad hypothesis $h^{\text{bad}}$ and as a consequence any learning algorithm that relies on the ERM principle will never select such a bad hypothesis $h^{\text{bad}}$ that has prohibitively large risk over the hypothesis $h^{\varheartsuit}$.

\medskip

We can now turn to the case of malicious classification noise.  In the case of malicious classification noise, we can think of the sequence of examples $\sample'$ obtained by the \mcnoracle as if it was the case that first the sequence of examples \sample was generated from the \rcnoracle oracle and then some of the labels that had to be flipped under RCN are now rectified in $\sample'$ to their original clean labels since the malicious adversary chose not to flip these even if they had the opportunity to do so.  
%
However, for every such example in $\sample'$ whose label the adversary chose \emph{not to flip}, the hypothesis $h^{\varheartsuit} = c$ will make one mistake less. 
At the same time, whenever such a ``rectification'' occurs in a label, any bad hypothesis $h^{\text{bad}}\in\Hh$ cannot improve the number of disagreements with the modified sample by more than one. 
As a further consequence, since $h^{\varheartsuit}$ is a better choice than any $h^{\text{bad}}$ under that sequence of examples $\sample$ generated by \rcnoracle, it will also be better under the sequence of examples $\sample'$ generated by \mcnoracle as any one label-flip rectification cannot alter the relationship between $h^{\varheartsuit}$ and any bad hypothesis $h^{\text{bad}}$.
This is because for every rectified label, the difference
$$\text{\# disagreements of \ensuremath{h^{\text{bad}}}} - \text{\# disagreements of \ensuremath{h^{\varheartsuit}}}$$
either remains unchanged or increases by two. 
Hence, the disagreement advantage of $h^{\varheartsuit} = c$ over any bad hypothesis cannot decrease.
%
%
As a further consequence, any learning algorithm that relies on the ERM principle, will never select a bad hypothesis $h^{\text{bad}}$ over $h^{\varheartsuit}$ and as a final conclusion, we can infer that the PAC criterion is satisfied, since an $\varepsilon$-good hypothesis will eventually be selected.
\end{proof}

\begin{rem}
Judging from the number of mislabeled examples in a sample, at a first glance it appears that there is some extent of a \emph{naming paradox} when one compares the RCN and MCN noise models.  This can be justified from the fact that when a learning algorithm has access to two samples with the same number of examples in each one of them, one sample obtained from an RCN oracle and one sample obtained from an MCN oracle, then under RCN the learning algorithm has to face, in general, a larger amount of mislabelings compared to the MCN case, even if both noise models are governed by the same noise rate $\eta$ when the two samples of the same size are generated.
In other words, the term \emph{malicious} seems a bit paradoxical judging from this comparison between RCN and MCN alone.

Going one step further, in the above analysis of Theorem~\ref{thm:MCN-ERM-Sloan} where one assumes a finite \Hh, $\Cc \subseteq \Hh$, and ERM hypothesis-testing comparison for the learning algorithm, we observe that the learning algorithm needs to obtain a sample from an MCN oracle that has size not larger than the equivalent sample that would be obtained from an RCN oracle.
\end{rem}

\medskip

While this survey is not meant to cover any results on poisoning attacks, perhaps the following example can illustrate how ERM fails in the case of poisoning attacks and thus it can be in sharp contrast to positive results that we can obtain for learnability in the presence of noise, as in Theorems~\ref{thm:RCN-ERM-sample-complexity} and~\ref{thm:MCN-ERM-Sloan}.

\begin{examp}[Label-Poisoning Can Be Catastrophic for ERM Learning Algorithms]
\label{ex:poisoning-symmetric-Boolean}
Let \Hh be the class of symmetric Boolean functions over
$\XX=\{0,1\}^n$, and suppose that examples are drawn from the uniform distribution $U$.  For $j\in\set{0,\ldots,n}$, write
$L_j=\set{\bm{x}\in\{0,1\}^n : \norm{\bm{x}}_1=j}$
for the $j$-th Hamming-weight layer. 
The probability mass of this layer $L_j$ is
$p_j = U(L_j) = \binom{n}{j}\cdot 2^{-n}$. 

Let $\varepsilon < 1/2$ be the PAC bound on the accuracy and suppose that the adversary has a small poisoning budget $b$ that is slightly larger than $\varepsilon/2$.
Write $b=\varepsilon/2+\upxi$, for some appropriate $\upxi > 0$. 
Let 
\begin{displaymath}
p_{\max} = \max_{0 \le j \le n} p_j = p_{\lfloor n/2\rfloor} = \binom{n}{\lfloor n/2\rfloor}\cdot 2^{-n}    
\end{displaymath}
and moreover let $n$ be large enough so that $p_{\max} < 2\upxi$.
This is possible because $p_{\max}= \binom{n}{\lfloor n/2\rfloor}\cdot 2^{-n} \in \Theta\left(1/\sqrt n\right)$.

Now order the Hamming-weight layers in any fixed order, for example in
nonincreasing order of their true masses $p_j$.  
Let $r^\star$ be the minimum index in this sequence such that 
$\sum_{i=1}^{r^\star} p_{j_i} \ge \varepsilon$. 
%
Let the collection of these indices be 
$\mathcal{J}^\star=\{j_1,\ldots,j_{r^\star}\}$
and define the 
set
\begin{displaymath}
A^\star
=
\bigcup_{j\in \mathcal{J}^\star} L_j
=
\{\bm{x}\in\{0,1\}^n:\norm{\bm{x}}_1\in \mathcal{J}^\star\}\,,
\end{displaymath}
which has probability mass
$\alpha := U(A^\star) = \sum_{j\in \mathcal{J}^\star} p_j$.
Furthermore, let
\begin{displaymath}
p_{\min}^{\mathcal{J}^\star} = \min_{j\in\mathcal{J}^\star}p_j > 0\,.
\end{displaymath}

Any layer $L_j$ has probability mass $p_j \le p_{\max}$ and therefore the last layer added in the formation of $A^\star$ contributed not more than $p_{\max}$.
Therefore,
\begin{equation}\label{eq:poisoning-alpha-vs-budget}
\varepsilon \le \alpha < \varepsilon+p_{\max}
< \varepsilon+2\upxi
= 2b.    
\end{equation}

The argument now is that any adversary with a budget $b = \varepsilon/2 + \upxi$ 
should be able to change the labels of more than half of the examples that correspond to instances 
in $A^\star$ in a clean sample \sample obtained from the \customoracle{U} oracle, provided that the sample \sample is sufficiently large.  We will come to the calculation of this sample size in a minute.
The critical comparison will be on the performance between the hypothesis
\begin{displaymath}
h^\varheartsuit = c
\end{displaymath}
and the poisoned hypothesis $h^{\text{\faSkullCrossbones}}$ that is defined as 
\begin{equation}
h^{\text{\faSkullCrossbones}}(\bm{x}) = 
\left\{
\begin{array}{rcl}
    1 - h^{\varheartsuit}(\bm{x}) &,& \text{if } \norm{\bm{x}}_1\in \mathcal{J}^\star,\\[3pt]
    h^{\varheartsuit}(\bm{x}) &,& \mbox{otherwise.}
\end{array}
\right.
\end{equation}

Consider a clean sample $\sample$ of size $m$ drawn from the \customoracle{U} oracle and suppose that every layer $L_j$ with
$j\in\mathcal{J}^\star$ is represented in $\sample$.  For each such layer, the adversary flips the labels of a strict majority of
the examples in $\sample$ whose instances belong to $L_j$, and it does not modify any other examples, creating a poisoned sample $\sample^{\text{\faSkullCrossbones}}$.
%
%
%
Hence, after poisoning, within every layer \(L_j\) with $j\in \mathcal{J}^\star$,
the empirical majority label is the opposite of the clean label.  On all other layers, the empirical labels remain clean.  
Since a symmetric hypothesis chooses one label for each Hamming-weight layer,
empirical risk minimization over \Hh is performed layer by layer.
Consequently, every ERM hypothesis against
$\sample^{\text{\faSkullCrossbones}}$ must agree with
$h^{\text{\faSkullCrossbones}}$ on every layer
$L_j$ with $j\in\mathcal{J}^\star$.
Ties are possible only on layers outside $\mathcal{J}^\star$ for which no
examples were obtained in \sample.
%
%
Therefore, under the original clean distribution $U$, any ERM hypothesis
$h^{\star}$ against the poisoned sample satisfies
\begin{displaymath}
\arisk{\customoracle{U}}{h^{\star}}
\ge
\probsub{\bm{x}\sim U}{
\norm{\bm{x}}_1\in\mathcal{J}^\star
}
=
U(A^\star)
=
\alpha
\ge
\varepsilon.    
\end{displaymath}
Thus any hypothesis $h^\star$ selected by ERM on the poisoned sample has true risk at least $\varepsilon$, which is prohibitive for a strict PAC error criterion.  Consequently, even a poisoning budget slightly larger than $\varepsilon/2$ may force ERM to output a hypothesis that fails the PAC criterion for accuracy.

Let us now discuss a bit more the required sample size for such an attack to be successful.
Due to (\ref{eq:poisoning-alpha-vs-budget}) 
we have 
$\alpha<2b$.
Now, define $\Delta$ to be the difference 
\begin{displaymath}
\Delta = 2b - \alpha
\end{displaymath}
which is therefore positive due to (\ref{eq:poisoning-alpha-vs-budget}).
For each $j\in \mathcal{J}^\star$, let $m_j$ denote the number of examples in the clean sample 
$\sample$ whose instances belong to $L_j$.  Also let
$M=\sum_{j\in\mathcal{J}^\star} m_j$
be the total number of sampled examples whose instances lie in $A^\star$.

We now want to make sure that from every layer $L_j$ with $j\in\mathcal{J}^\star$ at least one example will be obtained in the sample \sample that we draw. 
Therefore, for $j\in\mathcal{J}^\star$ consider the bad event
\begin{displaymath}
B^\star_{j} \equiv \mbox{no instance with $\norm{\bm{x}}_1 = j$ appears in \sample; i.e., $m_j = 0$ in \sample}\,.
\end{displaymath}
Then, by a union bound, some bad event $B^\star_j$ will be true with probability
\begin{displaymath}
p_{\text{bad}} = \myprob{\bigcup_{j\in\mathcal{J}^\star} B^\star_j} \le \sum_{j\in\mathcal{J}^\star} \myprob{B^\star_j} 
=
\sum_{j\in\mathcal{J}^\star} \probsub{\sample\sim U^m}{m_j = 0}
= 
\sum_{j\in\mathcal{J}^\star} \left(1 - p_j\right)^m\,.
\end{displaymath}
Using the facts that $p_j \ge p_{\min}^{\mathcal{J}^\star}$ and $(1-z)^{m} \le \exp(-m z)$, we have
\begin{displaymath}
p_{\text{bad}} 
\le 
\sum_{j\in\mathcal{J}^\star} \left(1 - p_{\min}^{\mathcal{J}^\star}\right)^m
\le 
\abs{\mathcal{J}^\star}\exp\left(-mp_{\min}^{\mathcal{J}^\star}\right)\,.
\end{displaymath}
%
Therefore, if
\begin{equation}\label{eq:poisoning-layer-coverage}
m
\ge
\frac{1}{p_{\min}^{\mathcal{J}^\star}}
\ln\left(
\frac{2\abs{\mathcal{J}^\star}}{\delta}
\right),
\end{equation}
then every selected layer is represented in $\sample$ with probability at least
$1-\delta/2$.

Assuming no bad event $B^\star_j$ happens, to flip a strict majority of the examples in every selected layer, the adversary needs to perform $C$ corruptions such that
\begin{equation}\label{eq:poisoning-corruptions-upper-bound}
C =
\sum_{j\in\mathcal{J}^\star}
\left(
\left\lfloor \frac{m_j}{2}\right\rfloor+1
\right)
\le
\frac{1}{2}\sum_{j\in\mathcal{J}^\star}m_j
+\abs{\mathcal{J}^\star}
=
\frac{1}{2}M+\abs{\mathcal{J}^\star}
\le
\frac{1}{2}M+(n+1)\,.
\end{equation}
Therefore, it is enough to guarantee
$M/2+(n+1) \le bm$.

Since each sampled instance in \sample lies in $A^\star$ with probability $\alpha$,
we have that the quantity $M$ is binomially distributed with parameters $\abs{\sample} = m$ and $\alpha$; in other words,~
$M\sim \operatorname{Binomial}(m,\alpha)$.
By Hoeffding's inequality (Proposition~\ref{prop:Hoeffding}), we have
\begin{displaymath}
\probsub{\sample\sim U^m}{
\abs{\frac{M}{m}-\alpha} \ge \frac{\Delta}{2}
}
\le
2\exp\left(-2m\left(\Delta/2\right)^2\right)
=
2\exp\left(-m\Delta^2/2\right).
\end{displaymath}
In other words, with probability at least $1-\delta/2$, 
assuming 
\begin{displaymath}
m\ge \frac{2}{\Delta^2}\ln\frac{4}{\delta},    
\end{displaymath}
we have that 
\begin{equation}\label{eq:poisoning-ratio1}
\frac{M}{m} < \alpha  +  \frac{\Delta}{2}\,.    
\end{equation}
Moreover, if
$m\ge \frac{4(n+1)}{\Delta}$,
then
\begin{equation}\label{eq:poisoning-ratio2}
\frac{n+1}{m}\le \frac{\Delta}{4}.
\end{equation}
Combining the inequalities (\ref{eq:poisoning-ratio1}) and~(\ref{eq:poisoning-ratio2}), due to~(\ref{eq:poisoning-corruptions-upper-bound}) we get
\begin{displaymath}
\frac{C}{m}
\le
\frac{1}{2}\frac{M}{m}+\frac{n+1}{m}
<
\frac{1}{2}\left(\alpha+\frac{\Delta}{2}\right)
+
\frac{\Delta}{4}
=
\frac{\alpha}{2}+\frac{\Delta}{2} = b.
\end{displaymath}
In other words, we have that 
\begin{displaymath}
C < bm\,.    
\end{displaymath}
Hence, by a union bound, if the clean sample \sample drawn from
\customoracle{U} has size 
\begin{equation}
m
\ge
\max\left\{
\frac{4(n+1)}{2b-\alpha},
\;
\frac{2}{(2b-\alpha)^2}\ln\frac{4}{\delta},
\;
\frac{1}{p_{\min}^{\mathcal{J}^\star}}
\ln\left(
\frac{2\abs{\mathcal{J}^\star}}{\delta}
\right)
\right\},
\end{equation}
then, with probability at least $1-\delta$, every selected layer is represented in \sample and the adversary can flip a strict majority of the examples in every selected layer while corrupting strictly fewer examples than a $b$-fraction of the sample \sample.
Consequently, every ERM hypothesis against the resulting poisoned sample 
has true risk at least $\varepsilon$.
\end{examp}

\begin{rem}
In Example~\ref{ex:poisoning-symmetric-Boolean} above, the adversary was able to mislead the learning algorithm through carefully selected label corruptions so that the learning algorithm will select a bad hypothesis in the output assuming it relies on the ERM principle.  Similar results have been proved in other cases for poisoning.  For example, in~\citep{PAC:Noise:Clean:Diochnos} 
a similar idea is used so that a certain part of the domain is hidden from the learning algorithm in what is called a 
\emph{strong $p$-budget poisoning attack}, 
so that 
essentially the output hypothesis of the learning algorithm has to be a guess for that hidden region, and in turn this results in a hypothesis that may have prohibitively large risk both in that region, as well as overall.
\end{rem}

\section{Related Work and a Taxonomy of Noise Models}\label{sec:related}

We have seen some important results for 
label-noise models, 
and we have also discussed how label-poisoning attacks can be catastrophic for ERM learners.
We also saw a classic result on arbitrary malicious noise early in our discussion.
Many additional noise models have been proposed over time, with different assumptions regarding how corruptions arise, what information the corruption mechanism may use, and whether only the label or the entire example may be modified.  This section briefly reviews representative models, including some models that are not restricted to label noise.  We first discuss stochastic and structured label-noise models, then consider other label-generation mechanisms and adversarial label noise, and finally review models that may
modify the instance itself.

\subsection{Classical and Structured Label-Noise Models}
\label{sec:related:structured-label-noise}

At the beginning of Section~\ref{sec:classification-noise}, and in particular in
Definition~\ref{def:pac-label-noise}, we introduced three models of noise:
the \emph{random classification noise (RCN)} model of
\citet{PAC:Noise:Angluin}, the \emph{malicious classification noise (MCN)}
model studied by Sloan~\citep{PAC:Noise:FourTypes:Sloan}, and
\emph{Massart noise}~\citep{PAC:Noise:Massart}.
Under RCN, every label is independently flipped with the same probability.
Massart noise is instead an instance-dependent model, since the probability
of a label flip may vary with the instance $\bm{x}\in\XX$.
MCN is more adversarial: whenever an opportunity for corruption occurs,
the adversary may choose whether to change the label and which label to
return.  We postpone further discussion of MCN until
Section~\ref{sec:related:other-label-mechanisms}.

The RCN framework has a natural appeal as a benchmark adversarial
environment.  For this reason, it has been used extensively in investigations
of the learnability of particular concept classes and in robust extensions of
learning models.  As an important example, parity functions are PAC-learnable
in the absence of noise, but they are not learnable in the statistical query
model~\citep{PAC:Noise:SQ,PAC:Fixed:Blum}, which was explicitly designed to
support learning in the presence of noise.  One natural relaxation is to reduce
the number of variables on which the target parity depends, and this restriction
has also received attention~\citep{Parities:Reyzin,Parities:Blum}.
The interested reader is referred to~\citep{PAC:SQ:Survey} and the references
therein for further results on statistical queries.
Outside the statistical query model, efficiently learning parity functions under
RCN remains a central open problem.  The problem is widely believed to be
computationally hard and forms the basis of several cryptographic
constructions~\citep{LPN:CryptoConnect}.

A natural extension of RCN allows different portions of the instance space to
have different mislabeling rates, leading to the Massart noise model.
For example, instances near a decision boundary may have very similar
attribute values despite belonging to different classes.  Such instances may
be especially difficult for a human expert or another labeling mechanism to
classify correctly, which motivates an instance-dependent probability of error.
Massart noise can pose significant learnability challenges, and several
important lines of work have investigated learning under this condition;
see, for example,
\citep{PAC:Noise:Massart,
PAC:Noise:Label:LossFunctions:RCNandMassart,
PAC:Noise:Massart:Awasthi,
PAC:Noise:Massart:Balcan,
PAC:Noise:Label:UniformNonuniformEarly}.
A particularly notable example is the problem of learning halfspaces under
Massart noise.  
A major breakthrough was obtained by~\citet{PAC:Halfspaces:Massart}, who gave a distribution-independent polynomial-time algorithm for halfspaces under Massart noise with misclassification error at most $\eta+\epsilon$ against the noisy labeled distribution, where $\eta<1/2$ bounds the flip probabilities. This guarantee differs from the clean-target error criterion in Definition~\ref{def:pac-label-noise}.

Although Massart noise provides one way to model the intuition that points near
a decision boundary may have higher noise rates, a more refined model for this
purpose is \emph{Tsybakov noise}%
~\citep{PAC:Noise:TsybakovOld,PAC:Noise:Tsybakov}.
Under Tsybakov noise, the probability of corruption is controlled through a
margin-dependent condition.  In one common formulation, the model is governed
by parameters $(\alpha,A)$ and satisfies
\begin{displaymath}
\probsub{\bm{x}\sim\DD}{
   \eta(\bm{x})\geq \frac{1}{2}-t
}
 \leq
 A t^{\alpha/(1-\alpha)}.
\end{displaymath}
Here $\alpha\in (0, 1)$ and $A > 0$ are fixed parameters, and the inequality holds for every $t \in (0,1/2]$.
Massart noise is bounded in the sense that
\begin{displaymath}
  \eta(\bm{x})\leq \eta_b<\frac{1}{2}
  \qquad
  \text{for every }\bm{x}\in\XX.
\end{displaymath}
By contrast, under Tsybakov noise the instance-dependent noise rate may be
arbitrarily close to $1/2$, provided that the probability mass of such
high-noise instances is appropriately controlled.
Tsybakov noise has attracted considerable attention despite being more general
than Massart noise; see, for example,
\citep{PAC:Noise:Tsybakov,
PAC:Noise:Tsybakov:Diakonikolas:STOC,
PAC:Noise:Tsybakov:Gentile,
PAC:Noise:Tsybakov:UPM,
PAC:Noise:Tsybakov:Enhanced}.
Margin-dependent models beyond the standard Tsybakov condition have also been
studied~\citep{PAC:Noise:TsybakovBeyond}.

Another structured model is
\emph{constant-partition classification noise (CPCN)}%
~\citep{PAC:Noise:Decatur:ConstantPartition}.
In this framework, the noise rate may depend not only on the instance but also
on its true label.  More precisely, the product space
$\XX\times\YY$ is partitioned into a constant number $k$ of disjoint regions.
Within each region of the partition $\Pi$, labels are flipped at a common rate,
although different regions may have different rates.
A formal investigation of the relationship between CPCN and RCN is available
in~\citep{PAC:Noise:CNequalsCPCN}.

CPCN can model many interesting situations.  In particular, the example space
may be divided into two regions according to the true class defined by the
ground-truth concept $c$.  The resulting faulty oracle can then use different
noise rates for false-positive and false-negative labels.
This two-region setting is closely related to the
\emph{class-conditional noise (CCN)} model.  CCN is controlled by two
parameters $(\eta_-,\eta_+)$, where $\eta_-$ and $\eta_+$ describe the rates
of false-positive and false-negative labels, respectively.  These parameters
are typically required to satisfy
\begin{displaymath}
  \eta_- + \eta_+ < 1\,.
\end{displaymath}
Consequently, one of the two class-conditional rates may equal or exceed $1/2$, as long as their sum remains below one.
This possibility is particularly relevant in semi-supervised learning. Many semi-supervised algorithms construct hypotheses using a combination of correctly labeled and pseudo-labeled data~\citep{PseudolabelingSurvey}.  Because pseudo-labels are generated by a previously trained hypothesis, the hypothesis may be systematically biased on some parts of the domain and may produce incorrect labels there at a rate exceeding $1/2$.
The two-rate model appeared in the original analysis of
co-training~\citep{SSL:Co-Training}; see also
\citep{SSL:MCT,SSL:CoTraining:TRiCo} for more recent variants of co-training. CCN has subsequently received substantial attention; see, for example, \citep{PAC:Noise:Label:Menon,
PAC:Noise:Label:Menon2,
PAC:Noise:Unconditional:Cost-Sensitive:Natarajan,
PAC:Noise:Unconditional:Natarajan,
PAC:Noise:Label:Asymmetric,
SSL:Co-Training,
PAC:Noise:Label:DeepLearning}.

A special case arises when the faulty oracle makes errors in only one direction, resulting in \emph{one-sided classification noise}~\citep{PAC:Label:OneSided:Simon,PAC:Label:OneSided:MIL}.  One-sided classification noise is also connected to
\emph{multiple-instance learning}~\citep{PAC:Label:OneSided:MIL}.  Multiple-instance learning is outside the scope of this chapter, but the interested reader may consult~\citep{MIL:Survey,Book:MIL} and the references therein.

\smallskip

The models discussed so far modify only the label of an example, but they
differ in the information on which the corruption probability may depend.
RCN uses a single instance- and class-independent rate; Massart and Tsybakov
noise permit dependence on the instance or its margin; and CPCN and CCN permit
dependence on a region of the example space or on the true class.
Table~\ref{tbl:stochastic-label-noise} summarizes these distinctions.
\begin{table}[tbp]
\centering
\small
\caption{Representative stochastic and structured label-noise models.
Only the label may be modified in these models.}
\label{tbl:stochastic-label-noise}

\setlength{\tabcolsep}{4pt}
\renewcommand{\arraystretch}{1.15}

\begin{tabularx}{\textwidth}{
    @{}
    p{3.0cm}
    Y
    Y
    @{}
}
\toprule
\textbf{Model}
&
\textbf{Dependence of the corruption mechanism}
&
\textbf{Main rate condition}
\\
\midrule

Random classification noise (RCN)
&
Independent of the instance $\bm{x}$, the true label
$y=c(\bm{x})$, and the distribution $\DD$.
&
Uniform rate $\eta \le \eta_b < 1/2$.
\\

\addlinespace

Massart noise
&
Depends on the instance $\bm{x}$ through an instance-dependent rate
$\eta(\bm{x})$.
&
$\eta(\bm{x})\leq \eta_b<1/2$ for every $\bm{x}\in\XX$.
\\

\addlinespace

Tsybakov noise
&
Depends on the instance, commonly through its proximity to the
decision boundary or through a margin quantity.
&
The rate may approach $1/2$, but the probability mass of high-noise
points is controlled by a Tsybakov condition.
\\

\addlinespace

Constant-partition classification noise (CPCN)
&
Depends on the region containing $(\bm{x},y)$ in a fixed partition
of $\XX\times\YY$.
&
A constant number of regions, usually with each regional rate
strictly below $1/2$.
\\

\addlinespace

Class-conditional noise (CCN)
&
Depends on the true class $y=c(\bm{x})$; in the binary case, separate
rates govern flips from the negative and positive classes.
&
Typically $\eta_-+\eta_+<1$; either individual rate may be at least
$1/2$.
\\

\bottomrule
\end{tabularx}
\end{table}
The progression in Table~\ref{tbl:stochastic-label-noise} may be viewed as an
increase in the information available to the corruption mechanism.
RCN is uniform throughout the domain, whereas the remaining models allow the
corruption rate to vary according to increasingly detailed information about
the example.  Nevertheless, these stochastic and structured models do not give
an adversary the same freedom as MCN to choose an arbitrary returned label
whenever an opportunity for corruption occurs.

\subsection{Other Label-Generation Mechanisms and Adversarial Label Noise}
\label{sec:related:other-label-mechanisms}

Label noise may also arise because the available teachers or annotation sources
have different levels of reliability.  One approach models different teachers as
different sources with different mislabeling rates%
~\citep{PAC:Noise:MultipleSources}.
A related approach formalizes such sources through the notion of
\emph{weak teachers}~\citep{PAC:Noise:WeakTeachers}.
These situations arise naturally when labels are crowdsourced, for example
through Amazon Mechanical Turk%
~\citep{PAC:Noise:WeakTeachers:WhoseVote,
PAC:Noise:WeakTeachers:Crowd}, obtained through remote sensing%
~\citep{WeakLabeling:RemoteSensing}, or generated in other environments where
annotators have varying levels of expertise.
Models of noisy teachers often describe a data-generation paradigm rather than
a single formal noise process.  They may therefore be more naturally viewed
within the framework of \emph{weak supervision}%
~\citep{Book:WeakSupervision}.
For example, labels may be generated through crowdsourcing, heuristic labeling
rules, pseudo-labeling, or automated systems such as
Snorkel~\citep{WeakSupervision:Snorkel}.
Such labeling mechanisms may be both noisy and systematically biased.

Another general approach is to model the target itself as a
\emph{probabilistic concept}.
Probabilistic concepts provide a natural way to describe label-generation
mechanisms that may depend on the instance space $\XX$, the label space $\YY$,
and the underlying distribution $\DD$.
General upper and lower bounds for learning under noise, including results
formulated in terms of probabilistic concepts, can be found in
\citep{PAC:pConcepts:Simon:Bounds,
PAC:Noise:SampleComplexity,
PAC:Noise:Bounds:Gentile,
PAC:Noise:Label:pconcepts}.
A related general framework has been investigated by
\citet{PAC:Noise:GeneralClassNoise}.

The malicious classification noise model introduced earlier differs from these source-dependent and probabilistic mechanisms because the corruption is explicitly adversarial.  Under MCN, the instance is still generated honestly, but whenever an opportunity for corruption occurs, the adversary may choose the label returned to the learner.  The adversary may therefore base its decision on information available during the learning process, including choices made for earlier examples.

Table~\ref{tbl:other-label-noise} summarizes these additional label-only
noise frameworks.
\begin{table}[tbp]
\centering
\small
\caption{Other label-generation mechanisms and adversarial models in
which only the label may be modified.}
\label{tbl:other-label-noise}

\setlength{\tabcolsep}{4pt}
\renewcommand{\arraystretch}{1.15}

\begin{tabularx}{\textwidth}{
    @{}
    p{3.2cm}
    Y
    p{2.7cm}
    Y
    @{}
}
\toprule
\textbf{Model}
&
\textbf{Dependence of the mechanism}
&
\textbf{Adversarial choice?}
&
\textbf{Main restriction}
\\
\midrule

Multiple or variable-quality labeling sources
&
Depends on the labeling source and may also depend on the instance
and the underlying distribution.
&
Usually no.
&
Source-specific assumptions; there is no single universal rate
condition.
\\

\addlinespace

Probabilistic concepts
&
May depend jointly on the instance, label, and distribution through
the probabilistic label-generation mechanism.
&
No explicit adversary is required.
&
Controlled by the assumptions imposed on the probabilistic concept.
\\

\addlinespace

Malicious classification noise (MCN)
&
May depend on prior observations and on information available to the
adversary.
&
Yes; the adversary chooses the returned label whenever an
opportunity for corruption occurs.
&
The corruption-opportunity rate satisfies $\eta<1/2$ in the
framework considered in this chapter.
\\

\bottomrule
\end{tabularx}
\end{table}
The models in Table~\ref{tbl:other-label-noise} illustrate that label noise need
not be described solely through a fixed flip-rate function.  It may arise from
heterogeneous labeling sources, from an intrinsically probabilistic
label-generation rule, or from an adversary that actively selects the observed
label.  We next consider stronger frameworks in which the instance itself may
also be modified.

\subsection{Noise Models That May Modify Instances}
\label{sec:related:instance-noise}

A stronger form of noise is obtained when the adversary may manipulate both the
instance and the label whenever an opportunity for corruption occurs, for
example when $\mathcal{T}_i = 1$ 
in Figure~\ref{fig:noise:mechanism}.
When this modification may be arbitrary, we obtain the
\emph{malicious noise} model introduced by
\citet{PAC:Noise:Valiant}.
Several fundamental results for this model were subsequently established by
\citet{PAC:Noise:Malicious}.

MCN is a restriction of arbitrary malicious noise: under MCN the adversary may
modify only the label, whereas under arbitrary malicious noise the adversary
may replace the entire example.
In Section~\ref{sec:arbitrary-malicious}, and specifically in
Theorem~\ref{thm:KearnsLi}, we examined the classic result of Kearns and Li
showing that no learning algorithm can tolerate arbitrary malicious noise at
rate $\varepsilon/(1+\varepsilon)$ or higher under the strict PAC objective.
This result contrasts sharply with Theorem~\ref{thm:MCN-ERM-Sloan}, where an
ERM learner supplied with an upper bound $\eta_b$ may tolerate malicious
classification noise at a rate arbitrarily close to $1/2$.

A related framework allows the adversary to choose a noisy distribution and
labeling rule ahead of time.  Each generated example is then honest with
probability $1-\eta$ and is drawn from the adversarially chosen noisy
distribution with probability $\eta$%
~\citep{PAC:Fixed:Mansour:Conjunctions:Product}.
Unlike an adaptive malicious adversary, the corrupting distribution in this
framework is fixed before the sample is generated.

Another important model is \emph{nasty sample noise (NSN)}, introduced by
\citet{PAC:Noise:Nasty}.
Historically, this model was developed alongside several classical noise
frameworks.  Under the convention adopted in
Section~\ref{sec:poisoning-vs-noise}, however, we view NSN as a form of
poisoning: the adversary observes the training dataset and chooses which
datapoints to replace, subject to a budget of permitted replacements.
A label-only restriction of this model is sometimes referred to as
\emph{nasty classification noise (NCN)}.  Since the adversary still selects
which training examples are modified, we likewise view this restriction as a
poisoning attack.

The interested reader may consult
\citep{PAC:Poisoning:Certified,
Surveys:Poisoning:DL,
Surveys:Poisoning:Tsipras,
Surveys:Poisoning:WildPatterns}
for further ideas and directions concerning poisoning attacks.
Recent work by \citet{PAC:Noise:NastyVSMalicious} studies the relationship
between nasty noise and malicious noise in both distribution-independent and
distribution-specific learning, and establishes a separation between the two
settings.

Another framework in which instances may be modified is
\emph{clean-label noise}%
~\citep{PAC:Noise:CleanFrogs,
PAC:Noise:Clean:Diochnos,
PAC:CleanLabel}.
Here the adversary modifies the attributes of an example, but the label
returned with the perturbed instance must remain consistent with the underlying
ground-truth concept.  Thus, if the perturbation moves the instance across the
ground-truth decision boundary, the label must also change accordingly.
This line of work is also known as
\emph{$p$-tampering}~\citep{PAC:Noise:Clean:Diochnos}.

Under \emph{attribute noise}, the corruption mechanism modifies only the
attributes of an example%
~\citep{PAC:Noise:Attribute:SV,
PAC:Noise:Attribute:Sloan,
PAC:Noise:FourTypes:Sloan}.
This framework is closely related to learning with
\emph{corrupted inputs}~\citep{CorruptedInputs}, where the attributes may be
modified but the resulting example must satisfy the label requirements imposed
by the particular model.

Noise models may also be combined to obtain \emph{hybrid} frameworks.
Decatur introduced the
\emph{classification and malicious (CAM)} noise model%
~\citep{PAC:Noise:Decatur:CAM}, which combines RCN with arbitrary malicious
noise.  With one probability, the label is flipped as under RCN; with a smaller
probability, the adversary may replace the entire example.
Other combinations are possible, although, to the best of our knowledge,
hybrid noise models have not been investigated as extensively beyond
Decatur's CAM framework.

Unlike the label-only models summarized earlier, the frameworks in this
subsection may modify the instance itself, or both the instance and its label.
They also differ in whether the adversary selects particular training points,
whether the modified label must remain consistent with the ground-truth
concept, and whether stochastic and adversarial corruption mechanisms are
combined.  Table~\ref{tbl:instance-corruption-models} summarizes these
distinctions.
\begin{table}[ht]
\centering
\small
\caption{Noise, tampering, and poisoning frameworks in which the
instance, or both the instance and its label, may be modified.}
\label{tbl:instance-corruption-models}

\setlength{\tabcolsep}{4pt}
\renewcommand{\arraystretch}{1.15}

\begin{tabularx}{\textwidth}{
    @{}
    p{3.0cm}
    Y
    p{2.7cm}
    Y
    @{}
}
\toprule
\textbf{Model}
&
\textbf{What may be modified?}
&
\textbf{Selection mechanism}
&
\textbf{Main restriction}
\\
\midrule

Arbitrary malicious noise
&
Both the instance and the label.
&
An adversary acts whenever a corruption opportunity occurs.
&
In the general distribution-free setting, the tolerable rate is
strictly below $\varepsilon/(1+\varepsilon)$ under the strict PAC
objective.
\\

\addlinespace

Nasty sample noise
&
Selected examples may be replaced entirely.
&
The adversary chooses which points in the completed training sample
to replace.
&
A fixed replacement budget; this chapter treats the model as a
poisoning framework.
\\

\addlinespace

Clean-label noise or $p$-tampering
&
The instance, and when necessary its associated label.
&
The adversary selects a modified instance.
&
The returned label must be correct for the modified instance under
the ground-truth concept.
\\

\addlinespace

Attribute noise or corrupted inputs
&
The attributes of the instance.
&
Depends on the particular corruption model.
&
A model-dependent perturbation constraint and a specified rule for
the associated label.
\\

\addlinespace

Classification and malicious (CAM) noise
&
The label under the RCN component; both the instance and label under
the malicious component.
&
Combines stochastic label flips with adversarial example
replacement.
&
Separate rates govern the random-classification and malicious
components.
\\

\bottomrule
\end{tabularx}
\end{table}
These models lie partly outside the central label-noise focus of this chapter,
but including them clarifies the boundary between label corruption, input
corruption, malicious example replacement, and poisoning.  This distinction
will also be useful for the taxonomy developed below, since some of its
categories may contain mechanisms that alter an entire example rather than
only its label.

\subsection{Algorithmic and Complexity Perspectives}
\label{sec:related:algorithmic-perspectives}

As the earlier sections of this chapter have illustrated, one basic defense
against noisy examples is to increase the sample size sufficiently to overcome
the statistical uncertainty introduced by the corruption process.
This is the approach emphasized by the classic ERM results developed in
Section~\ref{sec:classification-noise}.

An important alternative is to design algorithms that operate through
\emph{statistical queries}~\citep{PAC:Noise:SQ}.
Rather than accessing individual labeled examples directly, a statistical query
learner obtains approximate expectations of bounded functions of the data.
This abstraction can make algorithms naturally robust to several forms of
classification noise, although it may exclude certain efficiently PAC-learnable
classes, with unrestricted parity functions being the only notable known such class.
Our focus in this chapter has been on classic results in which the learner requests a sufficiently large sample from a faulty oracle, but for a broader discussion of the statistical query model the interested reader is referred to the survey of \citet{PAC:SQ:Survey} which appears as a separate book chapter in this book that is marking the 35th anniversary of ISAIM.

More generally, the effect of noise can be studied through upper and lower bounds on sample complexity, computational complexity, and achievable error.  The results cited throughout this section as well as in Remark~\ref{rem:rcn:complexity} demonstrate that the answer depends not only on the corruption rate, but also on the structure of the noise, the expressiveness of the concept and hypothesis classes, the distributional assumptions, and whether proper or improper learning is permitted.


The variety of models reviewed above motivates a broader organizational framework.  We next relate these classical theoretical models to the NCAR, NAR, and NNAR terminology that is often used in more applied discussions of label noise.
We note that we use a slightly different NAR/NNAR taxonomy compared to Fr{\'{e}}nay and Verleysen~\citep{Surveys:LabelNoise:Taxonomy} as explained below.


\subsection{A Taxonomy}
\label{subsec:ncar-nar-nnar}

A different notion of robustness for machine learning algorithms 
relies on the development of learning algorithms that may operate even if some of the attribute values in the data are missing.  For example, sometimes people may refuse to provide answers on some question surveys (e.g., for privacy concerns) and thus one needs to deal with data that have incomplete instances. While such approaches are outside of our scope, nevertheless, in the context of missing data a popular taxonomy is due to \citet{RUBINTAXONOMY} where the nature of the missingness belongs to three different categories: \emph{missing completely at random (MCAR)}, \emph{missing at random (MAR)}, and \emph{missing not at random (MNAR)}.  One can 
borrow this idea and view the nature of noise in data in a similar way.  This way, and in particular for label noise, we can have
labels that are \emph{noisy completely at random (NCAR)}, labels that are \emph{noisy at random (NAR)}, and labels that are \emph{noisy not at random (NNAR)}~\citep{Surveys:LabelNoise:Taxonomy}.  Therefore, in this chapter we will try to provide a taxonomy of the theoretical work that was mentioned above in the related work section, using the lens of this NCAR-NAR-NNAR separation.

In this analogy, NCAR corresponds to label corruption that is independent of both the instance and the true label. This aligns directly with RCN, where each label is flipped independently with fixed probability. 

The term NAR is associated with corruption mechanisms that have some dependence on some property of the data but are otherwise random.  For example, we can have a dependence 
\begin{inparaenum}[\itshape (i)\upshape]
\item on the attributes, 
\item the attributes and the label, or 
\item the attributes, the label, and the underlying distribution.
\end{inparaenum}
In turn these three different cases would correspond to instance-dependent noise models such as Massart and Tsybakov noise, structured noise variants such as CPCN and CCN, connections to labelers of varied quality including contamination at different rates, and connections to probabilistic concepts.
Furthermore, one can envision some hybrid noise models that combine different NAR models into one new noise model, or NAR models together with RCN which is probably a more benign noise model. Having said that we are not aware of such formal hybrid noise model investigations.
We also note that our NAR/NNAR convention differs conceptually from that of Fr{\'{e}}nay and Verleysen~\citep{Surveys:LabelNoise:Taxonomy}. In their taxonomy, NNAR includes stochastic label noise whose mislabeling probability depends on instance features beyond its dependence on the true class. In this chapter, we place such structured stochastic noise under NAR and reserve NNAR for adversarial corruption models, including hybrid models with at least one adversarial corruption component.

The term 
NNAR refers here to corruption mechanisms that 
may be chosen adversarially, potentially with knowledge of the learning algorithm used for the task. 
This regime includes malicious noise~\cite{PAC:Noise:Malicious,PAC:Noise:Attribute:Sloan} and nasty noise~\cite{PAC:Noise:Nasty}, under which an adversary may arbitrarily corrupt or replace a bounded fraction of the examples. 
As explained in Section~\ref{sec:poisoning-vs-noise}, we reserve the term \emph{``noise model''} for settings in which the adversary does not choose the sample locations at which corruption opportunities arise; otherwise, we speak of a \emph{poisoning attack}.
Accordingly, we view nasty noise as a poisoning attack rather than a separate noise model. Malicious noise admits both an online and an offline, budgeted formulation: only the online formulation is treated here as a noise model, whereas the budgeted formulation is classified as poisoning.
As in the NAR case, hybrid NNAR models may combine several mechanisms, provided that at least one component is NNAR. One example is Decatur's hybrid model, which combines RCN with arbitrary malicious noise, including corruption beyond the labels. Although it is therefore not purely a label-noise model, we include it in the taxonomy for completeness. More generally, modern data-poisoning frameworks~\cite{PAC:Poisoning:Certified,Surveys:Poisoning:DL,Surveys:Poisoning:WildPatterns,Surveys:Poisoning:Tsipras,Poisoning:ArmsRace} can be viewed as algorithm-aware refinements of the NNAR regime. Identifiability and learnability in such settings typically require stronger assumptions, robust estimators, or algorithmic defenses, often tailored to particular domains or data distributions.

Finally, 
note that unlike Rubin’s missing-data taxonomy, 
the NCAR/NAR/NNAR terminology does not define a universally accepted three-way partition of label-noise models.
Nevertheless, the NCAR/NAR/NNAR viewpoint provides a useful high-level lens for organizing these established models according to whether corruption is independent, feature/structure-dependent, or adversarially driven.  We hope that the readers will find this taxonomy interesting for a conceptual separation of various theoretical models and how these relate to some approaches that are taken in more applied machine learning research.

\smallskip

\begin{figure}[!t]
\centering
\resizebox{0.75\textwidth}{!}{
\begin{forest}
for tree={
  grow'=east,
  parent anchor=east,
  child anchor=west,
  anchor=west,
  align=left,
  edge={-Latex},
  rounded corners,
  draw,
  inner sep=4pt,
  s sep=8pt,
  l sep=12pt,
  font=\small,
}
[
{\textbf{Label}\\ \textbf{Noise}}
  [
    {\textbf{NCAR:}\\ 
    \textbf{Noisy Completely At Random}\\
    Independent of $\XX, \YY$, and $\DD$}
    [
      {Random classification noise (RCN)\\
       Same flip rate $\eta < 1/2$
       for all $(\bm{x}, y)$\\
       \citep{PAC:Noise:Angluin,Book:Laird,PAC:Noise:Label:LossFunctions:RCNandMassart,Diochnos:PACI-RCN,PAC:Fixed:Blum,Parities:Reyzin,Parities:Blum}}
    ]
  ]
  [
    {\textbf{NAR:}\\ \textbf{Noisy At Random}\\Depends on\\ observed structure}
    [
        {Noise rate $\eta$\\ depends on $\XX$}
        [
            {Massart noise\\
            Rate \emph{bounded} for all $\bm{x}\in\XX$\\
            $\eta(x)\le \eta < 1/2$\\
            \citep{PAC:Noise:Massart,PAC:Halfspaces:Massart,PAC:Noise:Label:LossFunctions:RCNandMassart,PAC:Noise:Massart:Awasthi,PAC:Noise:Massart:Balcan,PAC:Noise:Label:UniformNonuniformEarly}}
        ]
        [
            {Tsybakov noise\\ 
            Instance-dependent rate $\eta(\bm{x})$, subject to\\
            $\myprob{\eta(\bm{x}) \ge 1/2 -t} \le At^{\alpha/(1-\alpha)}$\\
            \citep{PAC:Noise:Tsybakov,PAC:Noise:Tsybakov:Diakonikolas:STOC,PAC:Noise:Tsybakov:Gentile,PAC:Noise:Tsybakov:UPM,PAC:Noise:Tsybakov:Enhanced}\\
            or other margin-based noise models~\citep{PAC:Noise:TsybakovBeyond}}
        ]
    ]
    [
        {Noise rate $\eta$\\ depends on\\
        partition of\\ $\XX$ and $\YY$}
        [
            {Class-conditional noise (CCN)\\
            Partition of size 2 where  
            flip rates $\eta_-, \eta_+$\\ 
            depend on class label $y$ 
            s.t.~$\eta_- + \eta_+ < 1$\\
            \citep{PAC:Noise:Label:Menon,PAC:Noise:Label:Menon2,PAC:Noise:Unconditional:Cost-Sensitive:Natarajan,PAC:Noise:Unconditional:Natarajan,PAC:Noise:Label:Asymmetric,SSL:Co-Training,PAC:Noise:Label:DeepLearning}
            \\
            may also be one-sided~\citep{PAC:Label:OneSided:Simon,PAC:Label:OneSided:MIL}
            }
        ]
        [
            {Constant-partition classification noise (CPCN)\\
            Non-overlapping partition of size $k\ge 2$ with\\ flip rate
            $\eta_k < 1/2$ in region $\pi_k$\\
            \citep{PAC:Noise:Decatur:ConstantPartition,PAC:Noise:CNequalsCPCN}}
        ]
    ]
    [
        {Noise rate $\eta$\\ depends on the\\ source ($\XX$ and $\DD$)}
        [
            {Variable quality, multiple sources\\
            \citep{PAC:Noise:WeakTeachers,PAC:Noise:WeakTeachers:Crowd,PAC:Noise:WeakTeachers:WhoseVote,PAC:Noise:MultipleSources,PAC:Fixed:Mansour:Conjunctions:Product}}
        ]
    ]
    [
        {Noise rate $\eta$\\ depends on\\
        $\XX, \YY$, and $\DD$}
        [
            {Conversion of deterministic\\
            to probabilistic concepts\\
            \citep{PAC:Noise:GeneralClassNoise,PAC:Noise:Label:pconcepts}}
        ]
    ]
  ]
  [
    {\textbf{NNAR:}\\ \textbf{Noisy Not At Random}\\ Adversarial noise}
    [
      {Malicious classification noise (MCN)\\
       Adversary chooses label with rate $\eta < 1/2$\\
       \citep{PAC:Noise:Malicious,PAC:Noise:FourTypes:Sloan}}
    ]
    [
        {Choose labels and distribution\\ 
        adversarially ahead of time\\
        \citep{PAC:Fixed:Mansour:Conjunctions:Product}}
    ]
    [
      {Hybrid noise environments\\
        Mixture mechanisms involving NNAR\\
       \citep{PAC:Noise:Decatur:CAM}}
    ]
    [
      {\textbf{Outside of our scope}\\
       Data poisoning / replace data\\
       Training-set manipulation with budget\\
       including nasty noise\\
       \citep{PAC:Poisoning:Certified,PAC:Noise:Nasty,Surveys:Poisoning:DL,Surveys:Poisoning:Tsipras,Surveys:Poisoning:WildPatterns,Poisoning:ArmsRace}}
    ]
  ]
]
\end{forest}
}
\caption{The label-noise taxonomy adopted in this chapter.}
\label{fig:noise-taxonomy-forest}
\end{figure}
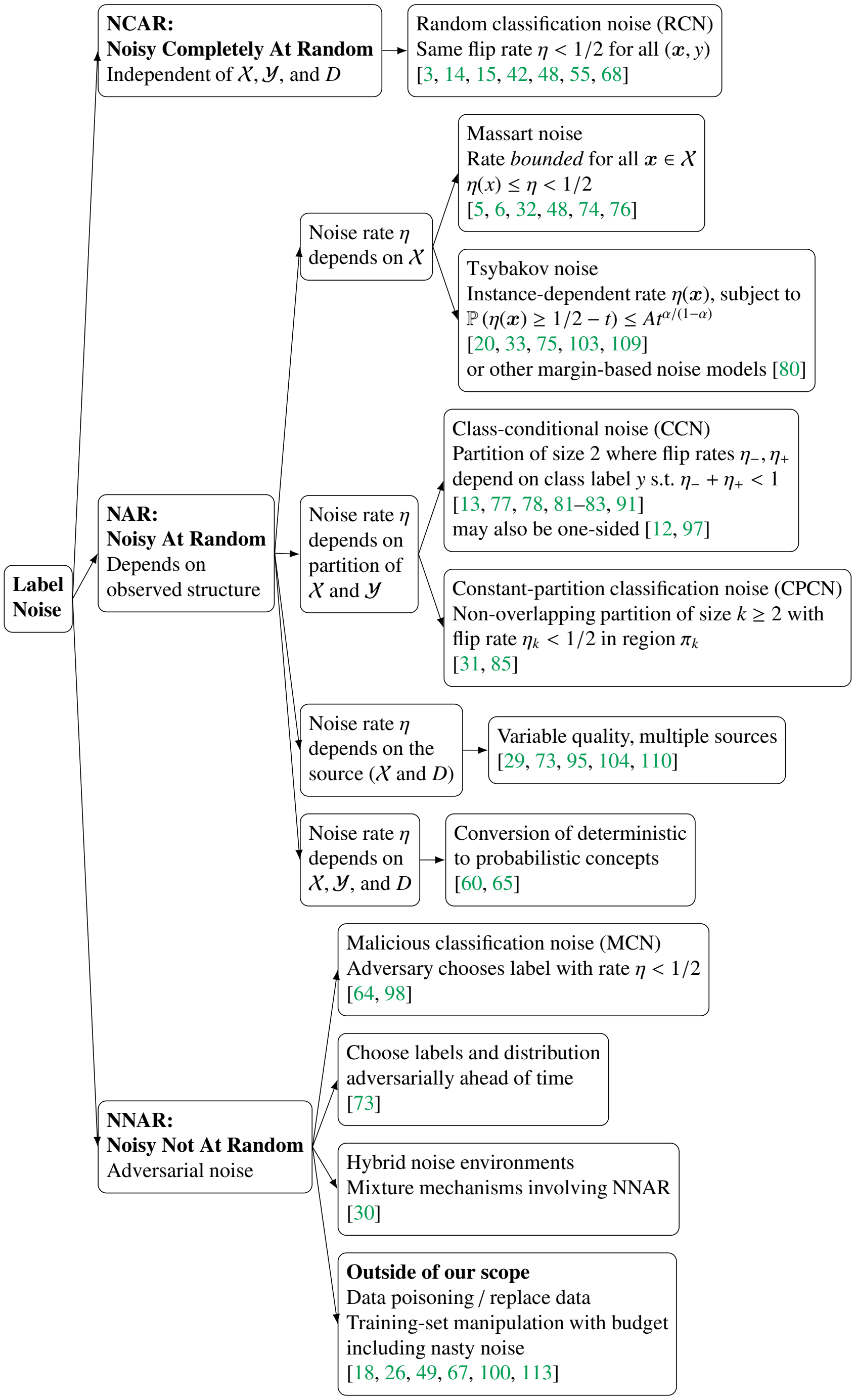

Taking all 
the above information 
into account, Figure~\ref{fig:noise-taxonomy-forest} provides a taxonomy for label-noise models.


\section{Conclusion}\label{sec:conclusion}
We discussed fundamental results for binary classification under label noise. In the process, we also touched upon the classic result of~\citet{PAC:Noise:Malicious} on arbitrary malicious noise, and we saw an example near the end (Example~\ref{ex:poisoning-symmetric-Boolean}) illustrating the catastrophic effect that poisoning attacks can have on datasets. 
In Section~\ref{sec:related} we surveyed label-noise models as well as models of noise that were not necessarily restricted to modifying the labels of the generated examples and we provided brief summaries in Tables~\ref{tbl:stochastic-label-noise}, \ref{tbl:other-label-noise}, and \ref{tbl:instance-corruption-models}.  We also provided a taxonomy in Figure~\ref{fig:noise-taxonomy-forest} that relates fine-grained label-noise models to the broader categories NCAR, NAR, and NNAR, mirroring Rubin's taxonomy for missing data~\citep{RUBINTAXONOMY}.  
We hope that this distinction can make it easier to understand where the different noise models stand in various cases and the relevant references can serve as useful starting points for further research.

This actually brings us to the fact that noise models have been studied for quite some time and at least the classical noise models possess a mature theory.
However, other directions in machine learning could inspire new tools and techniques, or provide new insights, for such well-established results on noise models.

One example is the situation where one is trying to accomplish not just low error rate for the learned hypothesis but also \emph{high recall} and/or \emph{high precision}.  This situation arises in a variant of the traditional PAC model that is relevant to \emph{class imbalance}~\citep{Diochnos:PACI,Imbalanced:margin,Imbalanced:ThrowingAway}, including situations that subsume class imbalance~\citep{Imbalanced:Mansour}.  Recent work~\citep{Diochnos:PACI-RCN} has been able to provide such guarantees under random classification noise, but in doing so,
\begin{inparaenum}[\itshape (i)\upshape]
\item on one hand a non-trivial lower bound on the noise rate may have to be established by the learning algorithm, and
\item well-known techniques may be brought under new light.
\end{inparaenum}
As an example of this latter case, a new noise-isolation algorithm has been developed in~\citep{Diochnos:PACI-RCN} 
for class-imbalanced classification.
Clearly there are many more noise models that await to be investigated.

A different direction could follow the lines of \emph{replicability}~\citep{Replicability,Replicability:Chase} 
where 
the goal 
is to produce the same hypothesis (with high probability) given access to two different i.i.d.~samples from the same oracle, 
using the same internal randomness in both runs.
An interesting question is how the structure of particular noise models can be exploited so that we can obtain sample complexity guarantees and computationally efficient replicable learners.

{
\small
\paragraph*{Acknowledgements.}
I want to thank Fred Hoffman and Marty Golumbic for all the trust that they have shown me ever since I was a PhD student.  I had a tremendous opportunity to learn from them as I have been part of the organizing committee for the International Symposium on Artificial Intelligence and Mathematics (ISAIM) since 2012.  I was Chair for Publications and Publicity for ISAIM during the years 2012--2024 and I had the privilege of being a Program Chair for ISAIM 2026 together with J\"org Rothe. 
I am also grateful to Maria Provost, who has also helped me throughout these years with the organization of ISAIM.  Maria and Fred have been working silently in the background so that everything would run smoothly for ISAIM and I cannot think of anyone else doing a better job than them!

Marty also graciously invited me to serve as an Associate Editor for the journal of Annals of Mathematics and Artificial Intelligence (AMAI) in 2022 and I also served as Associate Managing Editor for the calendar year of 2024.  Now that Marty has stepped down from leading AMAI, I continue to have a happy collaboration with the current Editor-in-Chief, J\"urgen Dix. 
Actually J\"urgen, together with Guillermo Simari, were the Program Chairs for ISAIM 2018 and together with the two of them I had the privilege of being a Guest Editor for a special issue that was devoted to selected papers from ISAIM 2018~\citep{Diochnos:ISAIM:2018}. This was valuable experience as it happened before I became an Associate Editor for AMAI and I started realizing the effort that it sometimes takes to obtain reviews for certain papers.  Speaking about forewords for ISAIM conferences, I am very happy that Fred, Marty, and I are also bonded by a foreword for ISAIM~\citep{Diochnos:ISAIM:2022}; namely the one that happened in the aftermath of the COVID-19 pandemic, in 2022.  That is the only occurrence of ISAIM that was held completely online and provided a cheerful note that better days were coming not only in the realm of Mathematics and AI with all the excellent contributed talks, but also for humanity in general.

Last, but certainly not least, I am also grateful to Gyuri Tur\'an who introduced me to the ISAIM community in 2010 when I gave a talk regarding our work in evolvability~\citep{Evolvability:SAGA}.  Without him, perhaps none of the above would have happened and I am so happy that they did happen!

\paragraph*{Competing Interests.}
The author has no conflicts of interest to declare that are relevant to the content of this chapter.
}

\normalsize

%
%
\bibliographystyle{plainnat}
\bibliography{references_short}

@book{Book:NPComplete,
  author       = {Michael R. Garey and David S. Johnson},
  title        = "{Computers and Intractability: {A} Guide to the Theory of NP-Completeness}",
  publisher    = {W. H. Freeman},
  year         = {1979},
  isbn         = {0-7167-1044-7},
  bibsource    = {dblp computer science bibliography, https://dblp.org}
}

@book{Book:Understanding,
  author       = {Shai Shalev{-}Shwartz and
                  Shai Ben{-}David},
  title        = "{Understanding Machine Learning - From Theory to Algorithms}",
  publisher    = {Cambridge University Press},
  year         = {2014},
  isbn         = {978-1-10-705713-5},
  bibsource    = {dblp computer science bibliography, https://dblp.org}
}

@book{Book:Mitchell,
  author       = {Tom M. Mitchell},
  title        = {Machine learning},
  series       = {McGraw-Hill Series in Computer Science},
  publisher    = {McGraw-Hill},
  year         = {1997},
  isbn         = {978-0-07-042807-2},
  bibsource    = {dblp computer science bibliography, https://dblp.org}
}

@book{Book:KearnsVazirani,
  author       = {Michael J. Kearns and Umesh V. Vazirani},
  title        = "{An Introduction to Computational Learning Theory}",
  publisher    = {{MIT} Press},
  year         = {1994},
  isbn         = {978-0-262-11193-5},
  bibsource    = {dblp computer science bibliography, https://dblp.org}
}

@book{Book:Trustworthy,
    author="Kush R. Varshney",
    title="Trustworthy Machine Learning",
    publisher="Independently Published",
    address="Chappaqua, NY, USA",
    year="2022"
}

@book{Book:Laird, 
    author = {Laird, Philip D.}, 
    title = "{Learning from Good and Bad Data}", 
    year = {1988}, 
    isbn = {0898382637}, 
    publisher = {Kluwer Academic Publishers}, 
    address = {USA}
}

@book{Book:AdvML1,
  author       = {Yevgeniy Vorobeychik and Murat Kantarcioglu},
  title        = "{Adversarial Machine Learning}",
  series       = {Synthesis Lectures on Artificial Intelligence and Machine Learning},
  publisher    = {Morgan {\&} Claypool Publishers},
  year         = {2018},
  bibsource    = {dblp computer science bibliography, https://dblp.org}
}

@book{Book:AdvML2, 
    author = {Joseph, Anthony D. and Nelson, Blaine and Rubinstein, Benjamin I. P. and Tygar, J. D.}, 
    title = "{Adversarial Machine Learning}", 
    year = {2019}, isbn = {1107043468}, 
    publisher = {Cambridge University Press}, 
    address = {USA}, 
    edition = {1st}
}

@book{Book:DatasetShift,
    author = {Qui\~{n}onero-Candela, Joaquin and Sugiyama, Masashi and Schwaighofer, Anton and Lawrence, Neil D.},
    title = "{Dataset Shift in Machine Learning}",
    publisher = {The MIT Press},
    year = {2008},
    isbn = {9780262170055}
}

@book{Book:Thinking,
 author = {Bruner, Jerome S. and Goodnow, Jacqueline J. and Austin, George A.},
 title = "{A study of thinking}",
 year = {1957},
 isbn = {0471114154},
 publisher = {John Wiley \& Sons},
 address = {New York, NY, USA}
}

@book{Book:WeakSupervision,
  author    = {Sugiyama, Masashi and Bao, Han and Ishida, Takashi and Lu, Nan and Sakai, Tomoya and Niu, Gang},
  title     = "{Machine Learning from Weak Supervision: {A}n Empirical Risk Minimization Approach}",
  publisher = {{MIT} Press},
  year      = {2022},
  pages     = {320},
  isbn      = {978-0262047074}
}

@Article{Basics:Hoeffding,
author = {Wassily Hoeffding},
title = "{Probability Inequalities for Sums of Bounded Random Variables}",
journal = {Journal of the American Statistical Association},
year = {1963},
volume = {58},
number = {301},
pages = {13-30},
month = {March}
}

@article{PAC:Valiant,
  author    = {Leslie G. Valiant},
  title     = {A Theory of the Learnable},
  journal   = {Commun. {ACM}},
  volume    = {27},
  number    = {11},
  pages     = {1134--1142},
  year      = {1984},
  bibsource = {dblp computer science bibliography, https://dblp.org}
}

@article{PAC:Fixed:BenedekItai,
  author       = {Gyora M. Benedek and Alon Itai},
  title        = "{Learnability with Respect to Fixed Distributions}",
  journal      = {Theor. Comput. Sci.},
  volume       = {86},
  number       = {2},
  pages        = {377--390},
  year         = {1991},
  bibsource    = {dblp computer science bibliography, https://dblp.org}
}

@inproceedings{PAC:Fixed:Blum,
  author       = {Avrim Blum and Merrick L. Furst and Jeffrey C. Jackson and Michael J. Kearns and Yishay Mansour and Steven Rudich},
  title        = "{Weakly learning {DNF} and characterizing statistical query learning using Fourier analysis}",
  booktitle    = {Proceedings of the Twenty-Sixth Annual {ACM} Symposium on Theory of Computing (STOC)},
  pages        = {253--262},
  publisher    = {{ACM}},
  year         = {1994},
  bibsource    = {dblp computer science bibliography, https://dblp.org}
}

@article{PAC:Fixed:Mansour:Conjunctions:Product,
  author       = {Yishay Mansour and Michal Parnas},
  title        = "{Learning Conjunctions with Noise under Product Distributions}",
  journal      = {Inf. Process. Lett.},
  volume       = {68},
  number       = {4},
  pages        = {189--196},
  year         = {1998},
  bibsource    = {dblp computer science bibliography, https://dblp.org}
}

@inproceedings{PAC:Halfspaces:Massart,
  author       = {Ilias Diakonikolas and
                  Themis Gouleakis and
                  Christos Tzamos},
  title        = "{Distribution-Independent {PAC} Learning of Halfspaces with Massart Noise}",
  booktitle    = {Advances in Neural Information Processing Systems 32 (NeurIPS)},
  pages        = {4751--4762},
  year         = {2019},
  bibsource    = {dblp computer science bibliography, https://dblp.org}
}

@inproceedings{PAC:Noise:Massart:Awasthi,
  author       = {Pranjal Awasthi and
                  Maria{-}Florina Balcan and
                  Nika Haghtalab and
                  Hongyang Zhang},
  title        = "{Learning and 1-bit Compressed Sensing under Asymmetric Noise}",
  booktitle    = {Proceedings of the 29th Conference on Learning Theory, {COLT} 2016},
  series       = {{JMLR} Workshop and Conference Proceedings},
  volume       = {49},
  pages        = {152--192},
  publisher    = {JMLR.org},
  year         = {2016},
  bibsource    = {dblp computer science bibliography, https://dblp.org}
}

@inproceedings{PAC:Noise:Massart:Balcan,
  author       = {Pranjal Awasthi and
                  Maria{-}Florina Balcan and
                  Nika Haghtalab and
                  Ruth Urner},
  title        = "{Efficient Learning of Linear Separators under Bounded Noise}",
  booktitle    = {Proceedings of The 28th Conference on Learning Theory, {COLT} 2015},
  series       = {{JMLR} Workshop and Conference Proceedings},
  volume       = {40},
  pages        = {167--190},
  publisher    = {JMLR.org},
  year         = {2015},
  bibsource    = {dblp computer science bibliography, https://dblp.org}
}

@article{PAC:Noise:Label:UniformNonuniformEarly,
  author       = {Naresh Manwani and
                  P. S. Sastry},
  title        = "{Noise Tolerance Under Risk Minimization}",
  journal      = {{IEEE} Trans. Cybern.},
  volume       = {43},
  number       = {3},
  pages        = {1146--1151},
  year         = {2013},
  bibsource    = {dblp computer science bibliography, https://dblp.org}
}

@article{PAC:Noise:SampleComplexity,
  author       = {Javed A. Aslam and
                  Scott E. Decatur},
  title        = "{On the Sample Complexity of Noise-Tolerant Learning}",
  journal      = {Inf. Process. Lett.},
  volume       = {57},
  number       = {4},
  pages        = {189--195},
  year         = {1996},
  bibsource    = {dblp computer science bibliography, https://dblp.org}
}

@article{PAC:pConcepts:Simon:Bounds,
  author       = {Hans Ulrich Simon},
  title        = "{General Bounds on the Number of Examples Needed for Learning Probabilistic Concepts}",
  journal      = {J. Comput. Syst. Sci.},
  volume       = {52},
  number       = {2},
  pages        = {239--254},
  year         = {1996},
  bibsource    = {dblp computer science bibliography, https://dblp.org}
}

@inproceedings{PAC:pConcepts:RCN:UpperBound,
  author       = {Nicol{\`{o}} Cesa{-}Bianchi and Paul Fischer and Eli Shamir and Hans Ulrich Simon},
  title        = "{Randomized Hypotheses and Minimum Disagreement Hypotheses for Learning with Noise}",
  booktitle    = {Computational Learning Theory, Third European Conference, EuroCOLT '97},
  series       = {Lecture Notes in Computer Science},
  volume       = {1208},
  pages        = {119--133},
  publisher    = {Springer},
  year         = {1997},
  doi          = {10.1007/3-540-62685-9\_11},
  bibsource    = {dblp computer science bibliography, https://dblp.org}
}

@article{PAC:Noise:NastyVSMalicious,
  author       = {Guy Blanc and
                  Yizhi Huang and
                  Tal Malkin and
                  Rocco A. Servedio},
  title        = "{Is nasty noise actually harder than malicious noise?}",
  journal      = {CoRR},
  volume       = {abs/2511.09763},
  year         = {2025},
  eprinttype    = {arXiv},
  eprint       = {2511.09763},
  bibsource    = {dblp computer science bibliography, https://dblp.org}
}

@inproceedings{PAC:Noise:Tsybakov:Gentile,
  author       = {Giovanni Cavallanti and
                  Nicol{\`{o}} Cesa{-}Bianchi and
                  Claudio Gentile},
  title        = "{Linear Classification and Selective Sampling Under Low Noise Conditions}",
  booktitle    = {Proceedings of the Twenty-Second Annual Conference on Neural Information Processing Systems (NeurIPS)},
  pages        = {249--256},
  publisher    = {Curran Associates, Inc.},
  year         = {2008},
  bibsource    = {dblp computer science bibliography, https://dblp.org}
}

@article{PAC:Noise:Bounds:Gentile,
  author       = {Claudio Gentile and
                  David P. Helmbold},
  title        = "{Improved Lower Bounds for Learning from Noisy Examples: An Information-Theoretic Approach}",
  journal      = {Inf. Comput.},
  volume       = {166},
  number       = {2},
  pages        = {133--155},
  year         = {2001},
  bibsource    = {dblp computer science bibliography, https://dblp.org}
}

@inproceedings{PAC:Noise:Tsybakov:UPM,
  author       = {Qizhou Wang and
                  Bo Han and
                  Tongliang Liu and
                  Gang Niu and
                  Jian Yang and
                  Chen Gong},
  title        = "{Tackling Instance-Dependent Label Noise via a Universal Probabilistic Model}",
  booktitle    = {Thirty-Fifth {AAAI} Conference on Artificial Intelligence, {AAAI} 2021},
  pages        = {10183--10191},
  publisher    = {{AAAI} Press},
  year         = {2021},
  bibsource    = {dblp computer science bibliography, https://dblp.org}
}

@article{PAC:Noise:TsybakovBeyond,
  author       = {Mehryar Mohri and Yutao Zhong},
  title        = "{Beyond Tsybakov: Model Margin Noise and {\(\mathcal{H}\)}-Consistency Bounds}",
  journal      = {CoRR},
  volume       = {abs/2511.15816},
  year         = {2025},
  note         = "{Presented at the International Symposium on Artificial Intelligence and Mathematics (ISAIM) 2026}",
  eprinttype    = {arXiv},
  eprint       = {2511.15816},
  bibsource    = {dblp computer science bibliography, https://dblp.org}
}

@inproceedings{PAC:Noise:Tsybakov:Enhanced,
  author       = {Anqi Mao and
                  Mehryar Mohri and
                  Yutao Zhong},
  title        = "{Enhanced H-Consistency Bounds}",
  booktitle    = {International Conference on Algorithmic Learning Theory},
  series       = {Proceedings of Machine Learning Research},
  volume       = {272},
  pages        = {772--813},
  publisher    = {{PMLR}},
  year         = {2025},
  bibsource    = {dblp computer science bibliography, https://dblp.org}
}

@inproceedings{PAC:Noise:Valiant,
  author       = {Leslie G. Valiant},
  editor       = {Aravind K. Joshi},
  title        = "{Learning Disjunctions of Conjunctions}",
  booktitle    = {Proceedings of the 9th International Joint Conference on Artificial Intelligence (IJCAI)},
  pages        = {560--566},
  publisher    = {Morgan Kaufmann},
  year         = {1985},
  bibsource    = {dblp computer science bibliography, https://dblp.org}
}

@article{PAC:Noise:Angluin,
  author       = {Dana Angluin and Philip D. Laird},
  title        = "{Learning From Noisy Examples}",
  journal      = {Mach. Learn.},
  volume       = {2},
  number       = {4},
  pages        = {343--370},
  year         = {1988},
  bibsource    = {dblp computer science bibliography, https://dblp.org}
}

@inproceedings{PAC:Noise:Attribute:SV,
  author       = {George Shackelford and Dennis Volper},
  title        = "{Learning k-DNF with Noise in the Attributes}",
  booktitle    = {Proceedings of the First Annual Workshop on Computational Learning Theory, {COLT} '88},
  pages        = {97--103},
  publisher    = {{ACM/MIT}},
  year         = {1988},
  bibsource    = {dblp computer science bibliography, https://dblp.org}
}

@article{PAC:Noise:Malicious,
  author       = {Michael J. Kearns and Ming Li},
  title        = "{Learning in the Presence of Malicious Errors}",
  journal      = {{SIAM} J. Comput.},
  volume       = {22},
  number       = {4},
  pages        = {807--837},
  year         = {1993},
  bibsource    = {dblp computer science bibliography, https://dblp.org}
}

@article{PAC:Noise:Attribute:Sloan,
  author       = {Sally A. Goldman and Robert H. Sloan},
  title        = "{Can {PAC} Learning Algorithms Tolerate Random Attribute Noise?}",
  journal      = {Algorithmica},
  volume       = {14},
  number       = {1},
  pages        = {70--84},
  year         = {1995},
  bibsource    = {dblp computer science bibliography, https://dblp.org}
}

@article{PAC:Noise:FourTypes:Sloan,
  author       = {Robert H. Sloan},
  title        = "{Four Types of Noise in Data for {PAC} Learning}",
  journal      = {Inf. Process. Lett.},
  volume       = {54},
  number       = {3},
  pages        = {157--162},
  year         = {1995},
  bibsource    = {dblp computer science bibliography, https://dblp.org}
}

@article{PAC:Noise:SQ,
  author       = {Michael J. Kearns},
  title        = "{Efficient Noise-Tolerant Learning from Statistical Queries}",
  journal      = {J. {ACM}},
  volume       = {45},
  number       = {6},
  pages        = {983--1006},
  year         = {1998},
  bibsource    = {dblp computer science bibliography, https://dblp.org}
}

@article{PAC:SQ:Survey,
  author       = {Lev Reyzin},
  title        = "{Statistical Queries and Statistical Algorithms: Foundations and Applications}",
  journal      = {CoRR},
  volume       = {abs/2004.00557},
  year         = {2020},
  eprinttype    = {arXiv},
  eprint       = {2004.00557},
  bibsource    = {dblp computer science bibliography, https://dblp.org}
}

@article{PAC:Noise:Nasty,
  author       = {Nader H. Bshouty and Nadav Eiron and Eyal Kushilevitz},
  title        = "{{PAC} learning with nasty noise}",
  journal      = {Theor. Comput. Sci.},
  volume       = {288},
  number       = {2},
  pages        = {255--275},
  year         = {2002},
  bibsource    = {dblp computer science bibliography, https://dblp.org}
}

@article{PAC:Noise:SampleEfficient,
  author       = {Nicol{\`{o}} Cesa{-}Bianchi and Eli Dichterman and Paul Fischer and Eli Shamir and Hans Ulrich Simon},
  title        = "{Sample-Efficient Strategies for Learning in the Presence of Noise}",
  journal      = {J. {ACM}},
  volume       = {46},
  number       = {5},
  pages        = {684--719},
  year         = {1999},
  doi          = {10.1145/324133.324221},
  bibsource    = {dblp computer science bibliography, https://dblp.org}
}

@article{PAC:Noise:Massart,
author = {Pascal Massart and {\'E}lodie N{\'e}d{\'e}lec},
title = "{Risk Bounds for Statistical Learning}",
volume = {34},
journal = {The Annals of Statistics},
number = {5},
publisher = {Institute of Mathematical Statistics},
pages = {2326 -- 2366},
year = {2006}
}

@inproceedings{PAC:Noise:CleanFrogs,
  author       = {Ali Shafahi and W. Ronny Huang and Mahyar Najibi and Octavian Suciu and Christoph Studer and Tudor Dumitras and Tom Goldstein},
  title        = "{Poison Frogs! Targeted Clean-Label Poisoning Attacks on Neural Networks}",
  booktitle    = {Advances in Neural Information Processing Systems 31: Annual Conference on Neural Information Processing Systems 2018, NeurIPS 2018},
  pages        = {6106--6116},
  year         = {2018},
  bibsource    = {dblp computer science bibliography, https://dblp.org}
}

@inproceedings{PAC:Noise:Decatur:ConstantPartition,
  author       = {Scott E. Decatur},
  title        = "{{PAC} Learning with Constant-Partition Classification Noise and Applications to Decision Tree Induction}",
  booktitle    = {Proceedings of the Fourteenth International Conference on Machine Learning (ICML) 1997},
  pages        = {83--91},
  publisher    = {Morgan Kaufmann},
  year         = {1997},
  bibsource    = {dblp computer science bibliography, https://dblp.org}
}

@inproceedings{PAC:Noise:Decatur:CAM,
  author       = {Scott E. Decatur},
  title        = "{Learning in Hybrid Noise Environments Using Statistical Queries}",
  booktitle    = {Learning from Data - Fifth International Workshop on Artificial Intelligence and Statistics, {AISTATS} 1995. Proceedings},
  pages        = {259--270},
  publisher    = {Springer},
  year         = {1995},
  bibsource    = {dblp computer science bibliography, https://dblp.org}
}

@article{PAC:Noise:Clean:Diochnos,
  author       = {Saeed Mahloujifar and Dimitrios I. Diochnos and Mohammad Mahmoody},
  title        = "{Learning under p-tampering poisoning attacks}",
  journal      = {Ann. Math. Artif. Intell.},
  volume       = {88},
  number       = {7},
  pages        = {759--792},
  year         = {2020},
  bibsource    = {dblp computer science bibliography, https://dblp.org}
}

@article{PAC:Noise:TsybakovOld,
  title="{Smooth Discrimination Analysis}",
  author={Mammen, Enno and Tsybakov, Alexandre B},
  journal={The Annals of Statistics},
  volume={27},
  number={6},
  pages={1808--1829},
  year={1999},
  publisher={Institute of Mathematical Statistics}
}

@article{PAC:Noise:Tsybakov,
author = {Alexander B. Tsybakov},
title = "{Optimal Aggregation of Classifiers in Statistical Learning}",
volume = {32},
journal = {The Annals of Statistics},
number = {1},
publisher = {Institute of Mathematical Statistics},
pages = {135 -- 166},
year = {2004}
}

@inproceedings{PAC:Poisoning:Certified,
  author       = {Jacob Steinhardt and
                  Pang Wei Koh and
                  Percy Liang},
  title        = "{Certified Defenses for Data Poisoning Attacks}",
  booktitle    = {Advances in Neural Information Processing Systems 30 (NeurIPS)},
  pages        = {3517--3529},
  year         = {2017},
  bibsource    = {dblp computer science bibliography, https://dblp.org}
}

@inproceedings{PAC:Noise:Unconditional:Natarajan,
  author       = {Nagarajan Natarajan and
                  Inderjit S. Dhillon and
                  Pradeep Ravikumar and
                  Ambuj Tewari},
  title        = "{Learning with Noisy Labels}",
  booktitle    = {Advances in Neural Information Processing Systems 26: 27th Annual
                  Conference on Neural Information Processing Systems (NeurIPS) 2013. Proceedings},
  pages        = {1196--1204},
  year         = {2013},
  bibsource    = {dblp computer science bibliography, https://dblp.org}
}

@article{PAC:Noise:Unconditional:Cost-Sensitive:Natarajan,
  author       = {Nagarajan Natarajan and
                  Inderjit S. Dhillon and
                  Pradeep Ravikumar and
                  Ambuj Tewari},
  title        = "{Cost-Sensitive Learning with Noisy Labels}",
  journal      = {J. Mach. Learn. Res.},
  volume       = {18},
  pages        = {155:1--155:33},
  year         = {2017},
  bibsource    = {dblp computer science bibliography, https://dblp.org}
}

@inproceedings{PAC:Noise:Label:Asymmetric,
  author       = {Clayton Scott and
                  Gilles Blanchard and
                  Gregory Handy},
  title        = "{Classification with Asymmetric Label Noise: Consistency and Maximal Denoising}",
  booktitle    = {{COLT} 2013 - The 26th Annual Conference on Learning Theory},
  series       = {{JMLR} Workshop and Conference Proceedings},
  volume       = {30},
  pages        = {489--511},
  publisher    = {JMLR.org},
  year         = {2013},
  bibsource    = {dblp computer science bibliography, https://dblp.org}
}

@inproceedings{PAC:Noise:Label:Menon,
  author       = {Aditya Krishna Menon and
                  Brendan van Rooyen and
                  Cheng Soon Ong and
                  Bob Williamson},
  title        = "{Learning from Corrupted Binary Labels via Class-Probability Estimation}",
  booktitle    = {Proceedings of the 32nd International Conference on Machine Learning, {ICML} 2015},
  series       = {{JMLR} Workshop and Conference Proceedings},
  volume       = {37},
  pages        = {125--134},
  publisher    = {JMLR.org},
  year         = {2015},
  bibsource    = {dblp computer science bibliography, https://dblp.org}
}

@article{PAC:Noise:Label:Menon2,
  author       = {Aditya Krishna Menon and
                  Brendan van Rooyen and
                  Nagarajan Natarajan},
  title        = "{Learning from binary labels with instance-dependent noise}",
  journal      = {Mach. Learn.},
  volume       = {107},
  number       = {8-10},
  pages        = {1561--1595},
  year         = {2018},
  bibsource    = {dblp computer science bibliography, https://dblp.org}
}

@inproceedings{PAC:Noise:Label:DeepLearning,
  author       = {Giorgio Patrini and
                  Alessandro Rozza and
                  Aditya Krishna Menon and
                  Richard Nock and
                  Lizhen Qu},
  title        = "{Making Deep Neural Networks Robust to Label Noise: {A} Loss Correction Approach}",
  booktitle    = {2017 {IEEE} Conference on Computer Vision and Pattern Recognition, {CVPR} 2017},
  pages        = {2233--2241},
  publisher    = {{IEEE} Computer Society},
  year         = {2017},
  bibsource    = {dblp computer science bibliography, https://dblp.org}
}

@inproceedings{PAC:Noise:CNequalsCPCN,
  author       = {Liva Ralaivola and
                  Fran{\c{c}}ois Denis and
                  Christophe Nicolas Magnan},
  title        = "{{CN} = {CPCN}}",
  booktitle    = {Machine Learning, Proceedings of the Twenty-Third International Conference (ICML) 2006},
  series       = {{ACM} International Conference Proceeding Series},
  volume       = {148},
  pages        = {721--728},
  publisher    = {{ACM}},
  year         = {2006},
  bibsource    = {dblp computer science bibliography, https://dblp.org}
}

@inproceedings{PAC:Noise:GeneralClassNoise,
  author       = {Shahin Jabbari and
                  Robert C. Holte and
                  Sandra Zilles},
  title        = "{PAC-Learning with General Class Noise Models}",
  booktitle    = {{KI} 2012: Advances in Artificial Intelligence - 35th Annual German Conference on AI. Proceedings},
  series       = {Lecture Notes in Computer Science},
  volume       = {7526},
  pages        = {73--84},
  publisher    = {Springer},
  year         = {2012},
  bibsource    = {dblp computer science bibliography, https://dblp.org}
}

@inproceedings{PAC:Noise:Tsybakov:Diakonikolas:STOC,
  author       = {Ilias Diakonikolas and
                  Daniel M. Kane and
                  Vasilis Kontonis and
                  Christos Tzamos and
                  Nikos Zarifis},
  title        = "{Efficiently learning halfspaces with Tsybakov noise}",
  booktitle    = {{STOC} '21: 53rd Annual {ACM} {SIGACT} Symposium on Theory of Computing},
  pages        = {88--101},
  publisher    = {{ACM}},
  year         = {2021},
  bibsource    = {dblp computer science bibliography, https://dblp.org}
}

@article{PAC:Noise:MultipleSources,
  author       = {Koby Crammer and
                  Michael J. Kearns and
                  Jennifer Wortman},
  title        = "{Learning from Multiple Sources}",
  journal      = {J. Mach. Learn. Res.},
  volume       = {9},
  pages        = {1757--1774},
  year         = {2008},
  bibsource    = {dblp computer science bibliography, https://dblp.org}
}

@article{PAC:Noise:Label:LossFunctions:RCNandMassart,
  author       = {Aritra Ghosh and
                  Naresh Manwani and
                  P. S. Sastry},
  title        = "{Making risk minimization tolerant to label noise}",
  journal      = {Neurocomputing},
  volume       = {160},
  pages        = {93--107},
  year         = {2015},
  bibsource    = {dblp computer science bibliography, https://dblp.org}
}

@article{PAC:Noise:Label:pconcepts,
  author       = {Michael J. Kearns and
                  Robert E. Schapire},
  title        = "{Efficient Distribution-Free Learning of Probabilistic Concepts}",
  journal      = {J. Comput. Syst. Sci.},
  volume       = {48},
  number       = {3},
  pages        = {464--497},
  year         = {1994},
  bibsource    = {dblp computer science bibliography, https://dblp.org}
}

@article{PAC:Label:OneSided:Simon,
  author       = {Hans Ulrich Simon},
  title        = "{PAC-learning in the presence of one-sided classification noise}",
  journal      = {Ann. Math. Artif. Intell.},
  volume       = {71},
  number       = {4},
  pages        = {283--300},
  year         = {2014},
  bibsource    = {dblp computer science bibliography, https://dblp.org}
}

@article{PAC:Label:OneSided:MIL,
  author       = {Avrim Blum and
                  Adam Kalai},
  title        = "{A Note on Learning from Multiple-Instance Examples}",
  journal      = {Mach. Learn.},
  volume       = {30},
  number       = {1},
  pages        = {23--29},
  year         = {1998},
  bibsource    = {dblp computer science bibliography, https://dblp.org}
}

@article{AdvEgs:Survey,
  author       = {Ian J. Goodfellow and
                  Patrick D. McDaniel and
                  Nicolas Papernot},
  title        = "{Making machine learning robust against adversarial inputs}",
  journal      = {Commun. {ACM}},
  volume       = {61},
  number       = {7},
  pages        = {56--66},
  year         = {2018},
  bibsource    = {dblp computer science bibliography, https://dblp.org}
}

@inproceedings{Adversarial:EvasionBackdoor:Diochnos,
  author       = {Dimitrios I. Diochnos and Saeed Mahloujifar and Mohammad Mahmoody},
  title        = "{Lower Bounds for Adversarially Robust {PAC} Learning under Evasion and Hybrid Attacks}",
  booktitle    = {19th {IEEE} International Conference on Machine Learning and Applications, {ICMLA} 2020},
  pages        = {717--722},
  publisher    = {{IEEE}},
  year         = {2020},
  bibsource    = {dblp computer science bibliography, https://dblp.org}
}

@inproceedings{AdvEgs:Diochnos:Neurips,
  author       = {Dimitrios I. Diochnos and Saeed Mahloujifar and Mohammad Mahmoody},
  title        = "{Adversarial Risk and Robustness: General Definitions and Implications for the Uniform Distribution}",
  booktitle    = {Advances in Neural Information Processing Systems 31, NeurIPS 2018},
  pages        = {10380--10389},
  year         = {2018},
  bibsource    = {dblp computer science bibliography, https://dblp.org}
}

@inproceedings{AdvEgs:Diochnos:AAAI,
  author       = {Saeed Mahloujifar and Dimitrios I. Diochnos and Mohammad Mahmoody},
  title        = "{The Curse of Concentration in Robust Learning: Evasion and Poisoning Attacks from Concentration of Measure}",
  booktitle    = {The Thirty-Third {AAAI} Conference on Artificial Intelligence, {AAAI} 2019},
  pages        = {4536--4543},
  publisher    = {{AAAI} Press},
  year         = {2019},
  bibsource    = {dblp computer science bibliography, https://dblp.org}
}

@inproceedings{Evolvability:SAGA,
  author       = {Dimitrios I. Diochnos and
                  Gy{\"{o}}rgy Tur{\'{a}}n},
  title        = "{On Evolvability: The Swapping Algorithm, Product Distributions, and Covariance}",
  booktitle    = {Stochastic Algorithms: Foundations and Applications, 5th International Symposium, {SAGA} 2009. Proceedings},
  series       = {Lecture Notes in Computer Science},
  volume       = {5792},
  pages        = {74--88},
  publisher    = {Springer},
  year         = {2009},
  bibsource    = {dblp computer science bibliography, https://dblp.org}
}

@inproceedings{Evolvability:ALT,
  author       = {Dimitrios I. Diochnos},
  title        = "{On the Evolution of Monotone Conjunctions: Drilling for Best Approximations}",
  booktitle    = {Algorithmic Learning Theory - 27th International Conference, {ALT} 2016, Proceedings},
  series       = {Lecture Notes in Computer Science},
  volume       = {9925},
  pages        = {98--112},
  year         = {2016},
  bibsource    = {dblp computer science bibliography, https://dblp.org}
}

@article{Evolvability:JAIR,
  author       = {Dimitrios I. Diochnos},
  title        = "{On the Evolvability of Monotone Conjunctions with an Evolutionary Mutation Mechanism}",
  journal      = {J. Artif. Intell. Res.},
  volume       = {70},
  pages        = {891--921},
  year         = {2021},
  bibsource    = {dblp computer science bibliography, https://dblp.org}
}

@inproceedings{Evolvability:EuroGP,
  author       = {Pantia{-}Marina Alchirch and
                  Dimitrios I. Diochnos and
                  Katia Papakonstantinopoulou},
  title        = "{Evolving Monotone Conjunctions in Regimes Beyond Proved Convergence}",
  booktitle    = {Genetic Programming - 25th European Conference, EuroGP 2022},
  series       = {Lecture Notes in Computer Science},
  volume       = {13223},
  pages        = {228--244},
  publisher    = {Springer},
  year         = {2022},
  bibsource    = {dblp computer science bibliography, https://dblp.org}
}

@article{PseudolabelingSurvey,
  author       = {Patrick Kage and Jay C. Rothenberger and Pavlos Andreadis and Dimitrios I. Diochnos},
  title        = "{A Review of Pseudo-Labeling for Computer Vision}",
  journal      = {J. Artif. Intell. Res},
  volume       = {85},
  year         = {2026},
  bibsource    = {dblp computer science bibliography, https://dblp.org}
}

@inproceedings{SSL:Co-Training,
  author       = {Avrim Blum and Tom M. Mitchell},
  title        = "{Combining Labeled and Unlabeled Data with Co-Training}",
  booktitle    = {Proceedings of the Eleventh Annual Conference on Computational Learning Theory, {COLT} 1998},
  pages        = {92--100},
  publisher    = {{ACM}},
  year         = {1998},
  bibsource    = {dblp computer science bibliography, https://dblp.org}
}

@inproceedings{SSL:MCT,
  author       = {Jay C. Rothenberger and Dimitrios I. Diochnos},
  title        = "{Meta Co-Training: Two Views are Better than One}",
  booktitle    = {{ECAI} 2025 - 28th European Conference on Artificial Intelligence},
  series       = {Frontiers in Artificial Intelligence and Applications},
  volume       = {413},
  pages        = {3355 - 3362},
  publisher    = {IOS Press Ebooks},
  year         = {2025}
}

@article{Surveys:Poisoning:Tsipras,
  author       = {Micah Goldblum and
                  Dimitris Tsipras and
                  Chulin Xie and
                  Xinyun Chen and
                  Avi Schwarzschild and
                  Dawn Song and
                  Aleksander Madry and
                  Bo Li and
                  Tom Goldstein},
  title        = "{Dataset Security for Machine Learning: Data Poisoning, Backdoor Attacks, and Defenses}",
  journal      = {{IEEE} Trans. Pattern Anal. Mach. Intell.},
  volume       = {45},
  number       = {2},
  pages        = {1563--1580},
  year         = {2023},
  bibsource    = {dblp computer science bibliography, https://dblp.org}
}

@article{Surveys:Poisoning:DL,
  author       = {Pinlong Zhao and
                  Weiyao Zhu and
                  Pengfei Jiao and
                  Di Gao and
                  Ou Wu},
  title        = "{Data Poisoning in Deep Learning: {A} Survey}",
  journal      = {CoRR},
  volume       = {abs/2503.22759},
  year         = {2025},
  eprinttype    = {arXiv},
  eprint       = {2503.22759},
  bibsource    = {dblp computer science bibliography, https://dblp.org}
}

@article{Surveys:Poisoning:WildPatterns,
  author       = {Antonio Emanuele Cin{\`{a}} and
                  Kathrin Grosse and
                  Ambra Demontis and
                  Sebastiano Vascon and
                  Werner Zellinger and
                  Bernhard Alois Moser and
                  Alina Oprea and
                  Battista Biggio and
                  Marcello Pelillo and
                  Fabio Roli},
  title        = "{Wild Patterns Reloaded: {A} Survey of Machine Learning Security against Training Data Poisoning}",
  journal      = {{ACM} Comput. Surv.},
  volume       = {55},
  number       = {13s},
  pages        = {294:1--294:39},
  year         = {2023},
  bibsource    = {dblp computer science bibliography, https://dblp.org}
}

@article{Surveys:AdvEgs:Opportunities,
  author       = {Jiliang Zhang and
                  Chen Li},
  title        = "{Adversarial Examples: Opportunities and Challenges}",
  journal      = {{IEEE} Trans. Neural Networks Learn. Syst.},
  volume       = {31},
  number       = {7},
  pages        = {2578--2593},
  year         = {2020},
  bibsource    = {dblp computer science bibliography, https://dblp.org}
}

@article{Surveys:AdvEgs:LongLists,
  author       = {Rey Reza Wiyatno and
                  Anqi Xu and
                  Ousmane Dia and
                  Archy de Berker},
  title        = "{Adversarial Examples in Modern Machine Learning: {A} Review}",
  journal      = {CoRR},
  volume       = {abs/1911.05268},
  year         = {2019},
  eprinttype    = {arXiv},
  eprint       = {1911.05268},
  bibsource    = {dblp computer science bibliography, https://dblp.org}
}

@inproceedings{AdvEgs:Intriguing,
  author       = {Christian Szegedy and
                  Wojciech Zaremba and
                  Ilya Sutskever and
                  Joan Bruna and
                  Dumitru Erhan and
                  Ian J. Goodfellow and
                  Rob Fergus},
  title        = "{Intriguing properties of neural networks}",
  booktitle    = {2nd International Conference on Learning Representations, {ICLR} 2014},
  year         = {2014},
  bibsource    = {dblp computer science bibliography, https://dblp.org}
}

@article{Diochnos:ISAIM:2022,
  author       = {Dimitrios I. Diochnos and Martin Charles Golumbic and Frederick Hoffman},
  title        = "{{ISAIM-2022:} international symposium on artificial intelligence and mathematics}",
  journal      = {Ann. Math. Artif. Intell.},
  volume       = {92},
  number       = {1},
  pages        = {1--4},
  year         = {2024},
  bibsource    = {dblp computer science bibliography, https://dblp.org}
}

@article{Diochnos:ISAIM:2018,
  author       = {Dimitrios I. Diochnos and
                  J{\"{u}}rgen Dix and
                  Guillermo Ricardo Simari},
  title        = "{Foreword to special issue for {ISAIM} 2018}",
  journal      = {Ann. Math. Artif. Intell.},
  volume       = {88},
  number       = {7},
  pages        = {687--689},
  year         = {2020},
  bibsource    = {dblp computer science bibliography, https://dblp.org}
}

@article{Gourdeau:RobustHardness,
  author       = {Pascale Gourdeau and Varun Kanade and Marta Kwiatkowska and James Worrell},
  title        = "{On the Hardness of Robust Classification}",
  journal      = {J. Mach. Learn. Res.},
  volume       = {22},
  pages        = {273:1--273:29},
  year         = {2021},
  bibsource    = {dblp computer science bibliography, https://dblp.org}
}

@inproceedings{Gourdeau:Robustness:DL,
  author       = {Pascale Gourdeau and
                  Varun Kanade and
                  Marta Kwiatkowska and
                  James Worrell},
  title        = "{Sample Complexity Bounds for Robustly Learning Decision Lists against Evasion Attacks}",
  booktitle    = {Proceedings of the Thirty-First International Joint Conference on
                  Artificial Intelligence, {IJCAI} 2022},
  pages        = {3022--3028},
  publisher    = {ijcai.org},
  year         = {2022},
  bibsource    = {dblp computer science bibliography, https://dblp.org}
}

@inproceedings{Urner:UAI,
  author       = {Nivasini Ananthakrishnan and
                  Shai Ben{-}David and
                  Tosca Lechner and
                  Ruth Urner},
  title        = "{Identifying regions of trusted predictions}",
  booktitle    = {Proceedings of the Thirty-Seventh Conference on Uncertainty in Artificial Intelligence (UAI)},
  series       = {Proceedings of Machine Learning Research},
  volume       = {161},
  pages        = {2125--2134},
  publisher    = {{AUAI} Press},
  year         = {2021},
  bibsource    = {dblp computer science bibliography, https://dblp.org}
}

@inproceedings{Urner:Robustness:NotAtOdds,
  author       = {Sadia Chowdhury and
                  Ruth Urner},
  title        = "{Robustness Should Not Be at Odds with Accuracy}",
  booktitle    = {3rd Symposium on Foundations of Responsible Computing, {FORC} 2022},
  series       = {LIPIcs},
  volume       = {218},
  pages        = {5:1--5:20},
  publisher    = {Schloss Dagstuhl - Leibniz-Zentrum f{\"{u}}r Informatik},
  year         = {2022},
  bibsource    = {dblp computer science bibliography, https://dblp.org}
}

@article{Adversarial:Barreno,
  author       = {Marco Barreno and Blaine Nelson and Anthony D. Joseph and J. D. Tygar},
  title        = "{The security of machine learning}",
  journal      = {Mach. Learn.},
  volume       = {81},
  number       = {2},
  pages        = {121--148},
  year         = {2010},
  bibsource    = {dblp computer science bibliography, https://dblp.org}
}

@inproceedings{Adversarial:Evasion:Biggio,
  author       = {Battista Biggio and
                  Igino Corona and
                  Davide Maiorca and
                  Blaine Nelson and
                  Nedim Srndic and
                  Pavel Laskov and
                  Giorgio Giacinto and
                  Fabio Roli},
  title        = "{Evasion Attacks against Machine Learning at Test Time}",
  booktitle    = {Machine Learning and Knowledge Discovery in Databases - European Conference, {ECML} {PKDD} 2013},
  series       = {Lecture Notes in Computer Science},
  volume       = {8190},
  pages        = {387--402},
  publisher    = {Springer},
  year         = {2013},
  bibsource    = {dblp computer science bibliography, https://dblp.org}
}

@inproceedings{AdvEgs:ExplainHarness,
  author       = {Ian J. Goodfellow and
                  Jonathon Shlens and
                  Christian Szegedy},
  title        = "{Explaining and Harnessing Adversarial Examples}",
  booktitle    = {3rd International Conference on Learning Representations, {ICLR} 2015},
  year         = {2015},
  bibsource    = {dblp computer science bibliography, https://dblp.org}
}

@inproceedings{AdvEgs:Madry,
  author       = {Aleksander Madry and
                  Aleksandar Makelov and
                  Ludwig Schmidt and
                  Dimitris Tsipras and
                  Adrian Vladu},
  title        = "{Towards Deep Learning Models Resistant to Adversarial Attacks}",
  booktitle    = {6th International Conference on Learning Representations, {ICLR} 2018},
  publisher    = {OpenReview.net},
  year         = {2018},
  bibsource    = {dblp computer science bibliography, https://dblp.org}
}

@inproceedings{Diochnos:PACI,
  author       = {Dimitrios I. Diochnos and Theodore B. Trafalis},
  title        = "{Learning Reliable Rules under Class Imbalance}",
  booktitle    = {{SIAM} International Conference on Data Mining ({SDM}) 2021},
  pages        = {28--36},
  publisher    = {{SIAM}},
  year         = {2021},
  bibsource    = {dblp computer science bibliography, https://dblp.org}
}

@unpublished{Diochnos:PACI-RCN,
  author       = {Dimitrios I.~Diochnos and Keng-Hung Steven Lin and Ting-Hung Yu},
  title        = "{Binary Imbalanced Classification under Random Classification Noise}",
  year   = {2026},
  note   = {In preparation}
}

@inproceedings{Imbalanced:margin,
  author       = {Corinna Cortes and
                  Anqi Mao and
                  Mehryar Mohri and
                  Yutao Zhong},
  title        = "{Balancing the Scales: {A} Theoretical and Algorithmic Framework for Learning from Imbalanced Data}",
  booktitle    = {Forty-second International Conference on Machine Learning, {ICML} 2025},
  series       = {Proceedings of Machine Learning Research},
  volume       = {267},
  publisher    = {{PMLR} / OpenReview.net},
  year         = {2025},
  bibsource    = {dblp computer science bibliography, https://dblp.org}
}

@article{Imbalanced:Mansour,
  author       = {Lee Cohen and
                  Yishay Mansour and
                  Shay Moran and
                  Han Shao},
  title        = "{Probably Approximately Precision and Recall Learning}",
  journal      = {CoRR},
  volume       = {abs/2411.13029},
  year         = {2024},
  eprinttype    = {arXiv},
  eprint       = {2411.13029},
  bibsource    = {dblp computer science bibliography, https://dblp.org}
}

@inproceedings{Imbalanced:ThrowingAway,
  author       = {Kamalika Chaudhuri and
                  Kartik Ahuja and
                  Mart{\'{\i}}n Arjovsky and
                  David Lopez{-}Paz},
  title        = "{Why does Throwing Away Data Improve Worst-Group Error?}",
  booktitle    = {International Conference on Machine Learning, {ICML} 2023},
  series       = {Proceedings of Machine Learning Research},
  volume       = {202},
  pages        = {4144--4188},
  publisher    = {{PMLR}},
  year         = {2023},
  bibsource    = {dblp computer science bibliography, https://dblp.org}
}

@inproceedings{Replicability:Chase,
  author       = {Zachary Chase and
                  Shay Moran and
                  Amir Yehudayoff},
  title        = "{Stability and Replicability in Learning}",
  booktitle    = {64th {IEEE} Annual Symposium on Foundations of Computer Science ({FOCS})},
  pages        = {2430--2439},
  publisher    = {{IEEE}},
  year         = {2023},
  bibsource    = {dblp computer science bibliography, https://dblp.org}
}

@inproceedings{Replicability,
  author       = {Russell Impagliazzo and
                  Rex Lei and
                  Toniann Pitassi and
                  Jessica Sorrell},
  title        = "{Reproducibility in learning}",
  booktitle    = {{STOC} '22: 54th Annual {ACM} {SIGACT} Symposium on Theory of Computing},
  pages        = {818--831},
  publisher    = {{ACM}},
  year         = {2022},
  bibsource    = {dblp computer science bibliography, https://dblp.org}
}

@inproceedings{PAC:Robustness:Mohri,
  author       = {Pranjal Awasthi and
                  Anqi Mao and
                  Mehryar Mohri and
                  Yutao Zhong},
  title        = "{Theoretically Grounded Loss Functions and Algorithms for Adversarial Robustness}",
  booktitle    = {International Conference on Artificial Intelligence and Statistics},
  series       = {Proceedings of Machine Learning Research},
  volume       = {206},
  pages        = {10077--10094},
  publisher    = {{PMLR}},
  year         = {2023},
  bibsource    = {dblp computer science bibliography, https://dblp.org}
}

@inproceedings{PAC:CleanLabel,
  author       = {Avrim Blum and
                  Steve Hanneke and
                  Jian Qian and
                  Han Shao},
  title        = "{Robust learning under clean-label attack}",
  booktitle    = {Conference on Learning Theory, {COLT} 2021},
  series       = {Proceedings of Machine Learning Research},
  volume       = {134},
  pages        = {591--634},
  publisher    = {{PMLR}},
  year         = {2021},
  bibsource    = {dblp computer science bibliography, https://dblp.org}
}

@inproceedings{CorruptedInputs,
  author       = {Uriel Feige and
                  Yishay Mansour and
                  Robert E. Schapire},
  title        = "{Learning and inference in the presence of corrupted inputs}",
  booktitle    = {Proceedings of The 28th Conference on Learning Theory, {COLT} 2015},
  series       = {{JMLR} Workshop and Conference Proceedings},
  volume       = {40},
  pages        = {637--657},
  publisher    = {JMLR.org},
  year         = {2015},
  bibsource    = {dblp computer science bibliography, https://dblp.org}
}

@article{MIL:Survey,
  author       = {Marc{-}Andr{\'{e}} Carbonneau and
                  Veronika Cheplygina and
                  Eric Granger and
                  Ghyslain Gagnon},
  title        = "{Multiple instance learning: {A} survey of problem characteristics and applications}",
  journal      = {Pattern Recognit.},
  volume       = {77},
  pages        = {329--353},
  year         = {2018},
  bibsource    = {dblp computer science bibliography, https://dblp.org}
}

@article{RUBINTAXONOMY,
  title="{Inference and Missing Data}",
  author={Rubin, Donald B},
  journal={Biometrika},
  volume={63},
  number={3},
  pages={581--592},
  year={1976},
  publisher={Oxford University Press}
}

@article{Surveys:LabelNoise:Taxonomy,
  author       = {Beno{\^{\i}}t Fr{\'{e}}nay and
                  Michel Verleysen},
  title        = "{Classification in the Presence of Label Noise: {A} Survey}",
  journal      = {{IEEE} Trans. Neural Networks Learn. Syst.},
  volume       = {25},
  number       = {5},
  pages        = {845--869},
  year         = {2014},
  bibsource    = {dblp computer science bibliography, https://dblp.org}
}

@article{Poisoning:ArmsRace,
  author       = {Pang Wei Koh and
                  Jacob Steinhardt and
                  Percy Liang},
  title        = "{Stronger data poisoning attacks break data sanitization defenses}",
  journal      = {Mach. Learn.},
  volume       = {111},
  number       = {1},
  pages        = {1--47},
  year         = {2022},
  bibsource    = {dblp computer science bibliography, https://dblp.org}
}

@inproceedings{PAC:Noise:WeakTeachers,
  author       = {Ruth Urner and
                  Shai Ben{-}David and
                  Ohad Shamir},
  title        = "{Learning from Weak Teachers}",
  booktitle    = {Proceedings of the Fifteenth International Conference on Artificial Intelligence and Statistics, {AISTATS} 2012},
  series       = {{JMLR} Proceedings},
  volume       = {22},
  pages        = {1252--1260},
  publisher    = {JMLR.org},
  year         = {2012},
  bibsource    = {dblp computer science bibliography, https://dblp.org}
}

@inproceedings{PAC:Noise:WeakTeachers:Crowd,
  author       = {Victor S. Sheng and
                  Foster J. Provost and
                  Panagiotis G. Ipeirotis},
  title        = "{Get another label? improving data quality and data mining using multiple, noisy labelers}",
  booktitle    = {Proceedings of the 14th {ACM} {SIGKDD} International Conference on Knowledge Discovery and Data Mining},
  pages        = {614--622},
  publisher    = {{ACM}},
  year         = {2008},
  bibsource    = {dblp computer science bibliography, https://dblp.org}
}

@inproceedings{PAC:Noise:WeakTeachers:WhoseVote,
  author       = {Jacob Whitehill and
                  Paul Ruvolo and
                  Tingfan Wu and
                  Jacob Bergsma and
                  Javier R. Movellan},
  title        = "{Whose Vote Should Count More: Optimal Integration of Labels from Labelers of Unknown Expertise}",
  booktitle    = {Advances in Neural Information Processing Systems 22 (NeurIPS)},
  pages        = {2035--2043},
  publisher    = {Curran Associates, Inc.},
  year         = {2009},
  bibsource    = {dblp computer science bibliography, https://dblp.org}
}

@article{SSL:CoTraining:TRiCo,
  author       = {Hongyang He and
                  Xinyuan Song and
                  Yangfan He and
                  Zeyu Zhang and
                  Yanshu Li and
                  Haochen You and
                  Lifan Sun and
                  Wenqiao Zhang},
  title        = "{TRiCo: Triadic Game-Theoretic Co-Training for Robust Semi-Supervised Learning}",
  journal      = {CoRR},
  note = "{Accepted to NeurIPS 2025}",
  volume       = {abs/2509.21526},
  year         = {2025},
  eprinttype    = {arXiv},
  eprint       = {2509.21526},
  bibsource    = {dblp computer science bibliography, https://dblp.org}
}

@inproceedings{WeakLabeling:RemoteSensing,
  author       = {Padhraic Smyth and
                  Usama M. Fayyad and
                  Michael C. Burl and
                  Pietro Perona and
                  Pierre Baldi},
  title        = "{Inferring Ground Truth from Subjective Labelling of Venus Images}",
  booktitle    = {Advances in Neural Information Processing Systems 7 (NeurIPS)},
  pages        = {1085--1092},
  publisher    = {{MIT} Press},
  year         = {1994},
  bibsource    = {dblp computer science bibliography, https://dblp.org}
}

@article{WeakSupervision:Snorkel,
  author       = {Alexander Ratner and
                  Stephen H. Bach and
                  Henry R. Ehrenberg and
                  Jason A. Fries and
                  Sen Wu and
                  Christopher R{\'{e}}},
  title        = "{Snorkel: rapid training data creation with weak supervision}",
  journal      = {{VLDB} J.},
  volume       = {29},
  number       = {2-3},
  pages        = {709--730},
  year         = {2020},
  bibsource    = {dblp computer science bibliography, https://dblp.org}
}

@article{Hardness:Agnostic:Monomials,
  author       = {Vitaly Feldman and
                  Venkatesan Guruswami and
                  Prasad Raghavendra and
                  Yi Wu},
  title        = "{Agnostic Learning of Monomials by Halfspaces Is Hard}",
  journal      = {{SIAM} J. Comput.},
  volume       = {41},
  number       = {6},
  pages        = {1558--1590},
  year         = {2012},
  bibsource    = {dblp computer science bibliography, https://dblp.org}
}

@inproceedings{Parities:Reyzin,
  author       = {Elena Grigorescu and
                  Lev Reyzin and
                  Santosh S. Vempala},
  title        = "{On Noise-Tolerant Learning of Sparse Parities and Related Problems}",
  booktitle    = {Algorithmic Learning Theory - 22nd International Conference, {ALT} 2011. Proceedings},
  series       = {Lecture Notes in Computer Science},
  volume       = {6925},
  pages        = {413--424},
  publisher    = {Springer},
  year         = {2011},
  bibsource    = {dblp computer science bibliography, https://dblp.org}
}

@article{Parities:Blum,
  author       = {Avrim Blum and
                  Adam Kalai and
                  Hal Wasserman},
  title        = "{Noise-tolerant learning, the parity problem, and the statistical query model}",
  journal      = {J. {ACM}},
  volume       = {50},
  number       = {4},
  pages        = {506--519},
  year         = {2003},
  bibsource    = {dblp computer science bibliography, https://dblp.org}
}

@article{LPN:CryptoConnect,
  author       = {Oded Regev},
  title        = "{On lattices, learning with errors, random linear codes, and cryptography}",
  journal      = {J. {ACM}},
  volume       = {56},
  number       = {6},
  pages        = {34:1--34:40},
  year         = {2009},
  bibsource    = {dblp computer science bibliography, https://dblp.org}
}

@book{Book:MIL,
  author       = {Francisco Herrera and
                  Sebasti{\'{a}}n Ventura and
                  Rafael Bello and
                  Chris Cornelis and
                  Amelia Zafra and
                  D{\'{a}}nel S{\'{a}}nchez Tarrag{\'{o}} and
                  Sarah Vluymans},
  title        = "{Multiple Instance Learning - Foundations and Algorithms}",
  publisher    = {Springer},
  year         = {2016},
  isbn         = {978-3-319-47758-9},
  bibsource    = {dblp computer science bibliography, https://dblp.org}
}

\end{document}